\documentclass[prb,amsmath,amssymb,floatfix,citeautoscript,preprint]{revtex4}
\usepackage{bm}
\usepackage{epsfig}
\usepackage{subfigure}
\usepackage[usenames]{color}

\begin{document}

\title{Topological Superconductivity in Metal/Quantum-Spin-Ice Heterostructures}

\author{Jian-Huang She$^1$, Choong H. Kim$^2$, Craig J. Fennie$^2$, Michael J. Lawler$^{1, 3}$, Eun-Ah Kim$^1$}

\affiliation{$^1$Department of Physics, Cornell University, Ithaca, New York 14853, USA \\
$^2$School of Applied and Engineering Physics, Cornell University, Ithaca, NY 14853, USA\\$^3$Department of physics, Binghamton University, Vestal NY 13850, USA}

\begin{abstract}
The original proposal to achieve superconductivity by starting from a quantum spin-liquid (QSL) and doping it with charge carriers, as proposed by Anderson in 1987, has yet to be realized. 
Here we propose an alternative strategy: use a QSL as a substrate for heterostructure growth of metallic films to design exotic superconductors. By spatially separating the two key ingredients of  superconductivity, i.e., charge carriers (metal) and pairing interaction (QSL), the proposed setup naturally lands on the parameter regime conducive to a controlled theoretical prediction. 
 Moreover, the proposed setup allows us to ``customize'' electron-electron interaction imprinted on the metallic layer. The QSL material of our choice is quantum spin ice well-known for its emergent gauge-field description of spin frustration. Assuming the metallic layer forms an isotropic single Fermi pocket,  we predict that the coupling between the emergent gauge-field and the electrons of the metallic layer will drive topological odd-parity pairing. We further present guiding principles for materializing the suitable heterostructure using ab initio calculations and describe the band structure we predict for the case of
Y$_2$Sn$_{2-x}$Sb$_x$O$_7$ grown on the (111) surface of Pr$_2$Zr$_2$O$_7$. Using this microscopic information, we predict topological odd-parity superconductivity at a few Kelvin in this heterostructure, which is comparable to the $T_c$ of the only other confirmed odd-parity superconductor Sr$_2$RuO$_4$.

\end{abstract}

\date{\today \ [file: \jobname]}
 \maketitle

 An intimate connection between the quantum spin liquid (QSL) state and superconductivity has long been suspected. Anderson conjectured that a QSL could turn into a superconductor upon doping holes in 1987\cite{Anderson87}. His idea is based on the resonating valence bond (RVB) description of a QSL\cite{Anderson73} which involves a quantum superposition of singlet configurations in which all spins form a singlet with a partner (See Fig1A). Such spins simultaenously point in many directions due to quantum fluctuation effects and hence show no sign of magnetic order. Nevertheless widely separated spins in a QSL maintain a high degree of entanglement driven by the exchange interaction $J_{\rm ex}$\cite{Balents10,Lee08}. Anderson conjectured that a RVB state can turn into a superconducting state by removing spins (doping holes) and allowing singlets to move around, which promote spin singlets to Cooper pairs (See Fig 1A).  However so far no QSL has been successfully doped into becoming a superconductor to the best of our knowledge. Here we propose a conceptually new framework for using a QSL to drive superconductivity without doping: grow a heterostructure consisting of a QSL and a metal (See Fig 1B). 

 We propose to ``borrow'' the spin correlation of a QSL without destroying QSL phase. This is conceptually distinct from Anderson's proposal and it has several advantages. Firstly, the superconductor need not be a singlet 
superconductor. Instead the pairing symmetry now depends on the 
 dynamic spin-spin correlation function and the structure of the interlayer coupling and hence it can be ``chosen'' at will, through the choice of the QSL layer. Specifically, we will show that the quantum spin ice\cite{Gingras14} as a QSL material will drive topological triplet pairing at the interface. Secondly the two distinct characteristic energy scales of each layers, namely the Fermi energy of the metal $E_F$ (or equivalently $N(0)^{-1}$, the inverse of the density of states at the Fermi level) and the spin-spin exchange interaction of the QSL $J_{\rm ex}$, 
enables us to be in the regime where $J_{\rm ex}/E_F\ll 1$. Theoretically this small parameter can play the role  of $\omega_{\rm D}/E_F\ll1$ (where $\omega_{\rm D}$ represents characteristic phonon frequency) in the Migdal-Eliashberg theory which justifies ignoring a certain set of diagrams and in turn serves as the key to their essentially exact treatment of 
phonon-mediated superconductivity \cite{Migdal58}. Finally, the coupling between the spins and the itinerant electrons in the form of a Kondo-like coupling $J_K$ \cite{Hewson} across the interface is expected to be small, i.e., $J_KN(0)\ll1$, making a perturbative treatment in this parameter reliable.  These advantages in concert with advances in the atomically precise preparation of relevant heterostructures \cite{Ohtomo04, Reyren07, Schlom10} present an unusual opportunity for a theoretically guided ``design'' of a new topological superconductor.

Interestingly the problem of coupling between local spin-moments and itinerant electrons has a long and celebrated history itself, especially in the context of heavy-fermion systems\cite{Coleman}.
In the strong coupling limit of $J_K N(0)\gg 1$, the conduction electrons hybridize with the local moments to form Kondo singlets resulting in a heavy Fermi liquid ground state (the gray phase labeled HFL in Fig 1C).  
On the other hand, in the weak coupling limit $J_K N(0)\ll 1$ of our interest, the spins are asymptotically free and there are many more possibilities depending on the strength of the spin-spin exchange interaction $J_{\rm ex}$.  When $J_{\rm ex}=0$, the RKKY interaction mediated by itinerant electrons, which is a perturbative effect of the coupling to the local moments with the characteristic interaction strength $J_{\rm RKKY}\sim J_K^2N(0)$, typically drives an antiferromagnetic order\cite{Doniach77} (see Supplemental Material (SM) Figure 1). 
 However when $J_{\rm ex}\neq0$ \cite{Coleman89} and furthermore frustrated \cite{Senthil03} as it is expected for the QSL, such antiferromagnetic ordering will be suppressed. Further, for sufficiently strong $J_K N(0)$, the Kondo singlet, the RVB singlet and Cooper pairs may all cooperate to form an exotic superconductor (the purple phase labeled SC$\times$QSL in Fig.1C).\cite{Coleman89, DHLee89, Senthil03} However, the coupling through the interface would naturally put the proposed heterostructure in the small $J_K N(0)$ region which has not 
received much attention to-date. 
 
 For $J_KN(0)\ll1$, the effect of the interfacial coupling on the metal can be treated perturbatively. Moreover when the QSL has a gapped spectrum with the gap scale $\omega_{\rm sf}\sim 2J_{\rm ex}$, we anticipate the QSL to stay intact on the insulator side  (see Fig 1C) as long as $J_{\rm RKKY}\sim J_K^2N(0)<J_{\rm ex}$. 
Under these conditions, one can safely ``integrate out'' the local spin degrees of freedom to arrive at an effective interaction for the itinerant electrons. In the absence of the $J_K$ coupling, the ground state for the heterostructure will consist of decoupled and coexisting Fermi liquid and QSL for a trivial reason (labeled FL$|$QSL in Fig.1C).
\footnote{This phase corresponds to the so-called FL* phase of Refs.\cite{Senthil03, Senthil04}.}
However the Fermi liquid state of the metal may be unstable against ordering once the effective electronic interaction due to the coupling $J_K$ is taken into account.  In the absence of Fermi surface nesting, the only such instability that is accessible at infinitesimal coupling strength is a
 superconducting instability\cite{Shankar94}. Hence as long as $J_{\rm RKKY}<J_{\rm ex}$, we anticipate the ground state in the $J_KN(0)\ll1$ regime to consist of superconducting itinerant electrons from the metallic side coexisting with the QSL. This interfacial superconductor, which we dubbed a SC$|$QSL  phase (the yellow phase in Fig1C), will be the focus of the rest of this paper.

 In order to materialize the SC$|$QSL
phase, we propose to grow a metallic layer on a QSL substrate. The goal will be for 
each sides of the heterostructure to be individually well-understood and to provide one of two essential ingredients of superconductivity: the charge carriers and the pairing interaction. For this, we will focus on a class of QSL materials known as the quantum spin ice (QSI) family.  
 The QSI materials are frustrated pyrochlore magnets that not only show no sign of order down to low temperatures, but also exhibit quantum dynamics\cite{Hermele04, Onoda10, Balents11, Balents12, Gardner99, Thompson11, Ross11, Kimura13, Gingras14}.
For our purpose, the advantage of the QSI materials over other spin-$1/2$ QSL materials\footnote{See SM section SII for a list of candidate QSL materials.} is that the QSI's appear to be quantum fluctuating relatives of the well-understood classical spin ice \cite{Kimura13}. Specifically, the classical spin ice materials obey the ice rule which amounts to the divergence-free constraint, i.e., $\nabla\cdot {\vec S}({\bm r})=0$ for the coarse-grained spin field ${\vec S}({\bm r})$. This constraint, which can be elegantly expressed in terms of an emergent gauge field ${\vec A}({\bm r})$ as ${\vec S}({\bm r})=\nabla\times{\vec A}({\bm r})$ \cite{Henley05, Isakov04}, appears to simply gain relaxational dynamics in QSI\cite{Gingras14}. 
Specifically we will focus on Pr$_2$Zr$_2$O$_7$ for concreteness and for its appealing properties. Experimentally, Pr$_2$Zr$_2$O$_7$ exhibits QSL phenomenology over a large temperature window ($T<1.4$K). Inelastic neutron scattering results on single crystals of Pr$_2$Zr$_2$O$_7$  reveals a gapped spectrum with a single frequency scale $\omega_{\rm sf}\sim 0.17{\rm meV}$\cite{Kimura13} and a peak around ${\bm Q}=0$. Theoretically, the magnetic degree of freedom is a non-Kramers doublet governed in the absence of disorder by a simple Hamiltonian\cite{Gingras14}. Armed with these facts we can construct a reliable phenomenological model for the heterostructure.

We first consider the relevant low energy effective theory $H = H_c + H_s + H_K$.  $H_c$ describes the metallic layer with an isotropic Fermi surface:
\begin{equation}
H_c=\sum_{{\bm k}\alpha}\left(\frac{\hslash^2k^2}{2m}-E_F\right)\psi^\dagger_{\alpha}({\bm k})\psi^{\ }_{\alpha}({\bm k}),
\label{Eq:2DEG}
\end{equation}
where $\psi^\dagger_{\alpha}({\bm k})$ ($\psi^{\ }_{\alpha}({\bm k})$) creates (annihilates) an electron with momentum $\bm k$ and spin index $\alpha$, $m$ is the electron's mass and  $E_F$ is the Fermi Energy. 
The spin Hamiltonian $H_{s}$ for the QSI substrate in isolation 
encodes  the exchange energy scale $J_{\rm ex}$ and the effect of geometric frustration through the emergent gauge field $\vec{A}$. Finally the dynamic degrees of freedom of each side will couple at the interface through the coupling term $H_K$. 
\footnote{Although the well-known non-Kramers doublet nature of the moments on Pr$^{3+}$\cite{Onoda10} gives rise to additional coupling between Pr quadrupole moments and conduction electron density \cite{Chen12, Lee13}, we will focus on the Kondo-type coupling in this paper for simplicity as we found the additional coupling to not affect the results in a qualitative manner (see SM section SIV).} Specifically, we consider a Kondo-like coupling\cite{Hewson}  between the conduction electron spin density $\vec{s}({\bm r},t)=\sum_{\alpha\beta} \psi^\dagger_{\alpha}({\bm r},t)\vec{\sigma}_{\alpha\beta}\psi_{\beta}({\bm r},t)$ and 
the coarse-grained spin operator $\vec{S}({\bm r},t)$ \cite{Henley05, Isakov04}:
\begin{eqnarray}
H_K&=&J_Kv_{\text{cell}}\sum_{a\alpha\beta}\int d^2{\bm r} \psi^\dagger_{\alpha}({\bm r})\vec{\sigma}_{\alpha\beta}\psi_{\beta}({\bm r}) \cdot \vec{S}({\bm r}_\perp = {\bm r},z=0)\nonumber\\
&=-&J_Kv_{\text{cell}}\sum_{a\alpha\beta}\int d^2{\bm r} \;\left(\vec{\nabla}\times \psi^\dagger_{\alpha}({\bm r})\vec{\sigma}_{\alpha\beta}\psi_{\beta}({\bm r}) \right)\cdot\vec{A}({\bm r}_\perp = {\bm r},z=0),
\label{Eq:Kondo}
\end{eqnarray}
upon integrating by parts. Here $\vec{\sigma}$ denotes the Pauli matrices, $v_{\text{cell}}$ the volume of the unit cell and $z=0$ the interface. Notice the rather obvious form of the coupling in the spin language takes a rather unusual form in the emergent gauge boson language. Usually when fermions are ``charged'' under a gauge boson $\vec{\mathcal A}$, it couples minimally 
 via ${\vec j}\cdot \vec{\mathcal A}$ coupling, with current ${\vec j}=Q\frac{\bm k}{m}\psi^\dagger_{{\bm k}\alpha}\psi_{{\bm k}\alpha}$ where $Q$ is the charge of the fermion field $\psi$ with respect to the gauge boson $\vec{\mathcal A}$.
The unusual form of coupling between the electrons and the emergent gauge boson in Eq.~\eqref{Eq:Kondo} in the form of $(\vec{\nabla}\times {\vec s})\cdot {\vec A}$ is due to the fact that the electrons are ``not charged'' under the gauge boson, i.e.  the electrons are not magnetic monopoles. This exotic coupling has striking consequences when we consider pairing possibilities.

 In the regime of interest,  the leading effect of the coupling \eqref{Eq:Kondo}
on the spin physics is to induce the RKKY interaction that can drive ordering. However, 
for a gapped spin liquid like Pr$_2$Zr$_2$O$_7$,
the QSL state would be stable as long as $J_{\rm RKKY}<J_{\rm ex}$. Hence we can ``integrate out'' the local moments and focus on the effect of the interaction induced on the metallic layer. Given a QSL substrate ( 
$\langle \vec{S}({\bf r},t)\rangle=0$ by definition)
 the leading effect of the coupling Eq.~\eqref{Eq:Kondo} is 
\begin{eqnarray}
&&H_{\rm int}(t)=-\frac{J_K^2v_{\text{cell}}^2}{2\hslash}\int dt'\int d^2{\bm r} d^2{\bm r}' \;\vec{s}({\bm r}, t) \cdot \langle \vec{S}({\bm r},0,t)\; \vec{S}({\bm r}', 0, t')\rangle \cdot\vec{s}({\bm r}', t') \label{Eq:int}\\
&&=-\frac{J_K^2v_{\text{cell}}^2}{2\hslash}\sum_{ab\alpha\beta\alpha'\beta'}\int dt'\int d^2{\bm r} d^2{\bm r}'\; \left[\left({\vec \sigma}_{\alpha\beta}\times\vec\nabla\right)_a \psi^\dagger_{{\bm r}\alpha}\psi_{{\bm r}\beta} \right] {\cal D}_{ab} \left[\left({\vec \sigma}_{\alpha'\beta'}\times\vec\nabla\right)_b \psi^\dagger_{{\bm r}'\alpha'}\psi_{{\bm r}'\beta'} \right]
\label{Eq:int-gauge}
\end{eqnarray}
 where ${\cal D}_{ab}({\bm r}-{\bm r}', t-t')\equiv\langle A_a({\bm r},0,t) A_b({\bm r}', 0, t')\rangle$ represents 
the emergent gauge field propagator, whose classical limit
in momentum space $\langle  A_a({\bm q})A_b(-{\bm q}) \rangle\sim \frac{1}{q^2}\left(\delta_{ab}-2{\hat q}_a{\hat q}_b\right)$\cite{Henley05, Isakov04} encapsulates the ice-rule. 
Note that Eq~\eqref{Eq:int}, which would apply to any QSL-based heterostructure, shows how the dynamical entanglement between spins of the QSL gets imprinted on the effective interaction between itinerant electrons. Therefore one can ``manipulate'' the  interaction between itinerant electrons through the choice of the QSL. On the other hand, Eq~\eqref{Eq:int-gauge} is specific to QSI-metal heterostructure and it reveals a critical insight into 
the leading pairing channel.
Minimally coupled gauge boson 
 prohibits pairing because the induced current-current interaction is repulsive for two electrons with opposite momenta. One way to circumvent this issue is to form finite momentum carrying pairs out of electrons with nearly parallel momenta\cite{PALee07,PALee14}. Remarkably the exotic form of fermion-gauge boson coupling (Eq.~\eqref{Eq:Kondo}) offers an alternative channel for gauge bosons to mediate pairing. Specifically, since the induced interaction in Eq.\eqref{Eq:int-gauge}  is obviously attractive between electrons of equal spin $\vec{s}(\bm r)=\vec{s}({\bm r}')$, one can anticipate triplet pairing even at the mean-field level.

Now the low energy effective theory defined by Eq.\eqref{Eq:2DEG} and Eq.\eqref{Eq:int} describes  an interacting electron problem, which is generically hard to solve. To make the problem worse, the effective interaction Eq.~\eqref{Eq:int} is highly structured as a result of the spin ice rules.  
However, we can make non-trivial progress building on the renormalization group based perspectives and the classic 
justification for the mean-field theory treatment in the 
 BCS theory. Firstly, we know from the renormalization group theory that the only weak-coupling instability of a Fermi liquid in the absence of Fermi-surface nesting is the superconducting instability\cite{Shankar94}. 
Secondly, armed with the separation of scale $\omega_{\rm sf}/E_F\ll 1$, we expect the mean-field theory treatment in the pairing channel to yield a reliable prediction for the interacting fermion problem when the  interaction is weak, i.e., $\lambda\sim N(0)V\sim J_K^2N(0)/J_{\rm ex}<1$.  Thirdly, the interaction in Eq~\eqref{Eq:int-gauge} is clearly attractive in the equal-spin pairing channel and hence we can anticipate pairing instability at mean-field level.  
All together the problem at hand promises an opportunity to predict an exotic superconductor whose pairing channel is determined by spin- and momentum-dependent interaction of Eq.~\eqref{Eq:int-gauge}, in a theoretically reliable approach.

Therefore we use  mean-field theory to look for pairing. We first note that the symmetry of the effective interaction $H_{\rm int}$ Eq.~\eqref{Eq:int} can be read off directly from the measured spin susceptibility.
For Pr$_2$Zr$_2$O$_7$, the measured spin susceptibility \cite{Kimura13} (the Fourier transform of the spin-spin correlation function $\langle S_a({\bm r}, t) S_b({\bm 0}, 0)\rangle$ in Eq.~\eqref{Eq:int}) exhibiting relaxational dynamics is well captured by the following approximate analytic form\cite{Henley05, Isakov04, Hermele04, Conlon09, Ryzhkin13,Benton12}
\begin{equation}
\chi_{ab}({\bm q}, \omega)=\frac{\chi_0}{1-i\omega\tau}\left( \delta_{ab}-\frac{q_a q_b}{q^2} +\frac{1}{1+q^2\xi^2}\frac{q_a q_b}{q^2}\right)\equiv\frac{1}{1-i\omega\tau}S_{ab}({\bm q}),
\label{Eq:chi}
\end{equation}
where $\tau=\omega_{\rm sf}^{-1}$ denotes relaxation time, $\xi$ denotes the correlation length, $\chi_0 \equiv\hslash/(J_{\rm ex} v_{\text{cell}})$, and $S_{ab}({\bm q})$ denotes the momentum dependence of the susceptibility. The first two terms are the transverse components originating from the ice-rule that reflect the propagator of the emergent gauge fields\cite{Henley05, Isakov04}, albeit with relaxational dynamics. The third term is an ice-rule breaking longitudinal component. 
Due to the entanglement between the spin direction ($a$) dependence and the momentum direction (${\bf q}$) dependence in Eq.~\eqref{Eq:chi},  the total angular momentum and the total spin are not separately conserved in $H_{\rm int}$. This situation is analogous to that of the 
the three dimensional magnetic dipole gas\cite{Li12} except
that our system is confined to two-dimensions.
 Hence the effective interaction $H_{\rm int}$ Eq.~\eqref{Eq:int} reduces the total orbital and spin rotational symmetry of $SO(3)\times SU(2)$ down to $U(1)$ leaving only the 
$z$-component of the total angular momentum $J_z=L_z+S_z$ a good quantum number for the Cooper pairs. 
In addition,
the effective interaction is even under parity and so 
parity is another good quantum number.

 Now we seek the dominant pairing channel. For this we will keep the frequency dependence implicit
 and mean-field decouple 
 $H_{\rm int}$ using pair operators formed out of 2$\times$2 matrices in the spin basis
$P_{\mu}({\bm k})\equiv\frac{1}{\sqrt{2}}\sum_{\alpha\alpha'} \left[ i\sigma_y\sigma_\mu\right]_{\alpha\alpha'}\psi_\alpha({\bm k}) \psi_{\alpha'}(-{\bm k})$, where $\sigma_\mu=\mathbb{I}$ for $\mu=0$ and Pauli matrices for $\mu=x,y,z$\cite{Sigrist91}. Further since pairing occurs near the Fermi surface, we can focus on the angular ($\varphi_{\bf k}$) dependence of the pair operators and decompose the pair operators into different partial wave components: $P_\mu(L_z)\equiv\int\frac{d\varphi_{\bf k}}{2\pi}e^{-iL_z\varphi_{\bf k}}P_\mu({\bf k})$. Guided by symmetry we then combine pair operators with different orbital angular momentum $L_z$ and total spin $S_z$ into the $J_z$ basis, and diagonalize the interaction Hamiltonian Eq~\eqref{Eq:int-gauge} into 
\begin{equation}
H_{\rm int}=\sum_{\kappa=\pm}\sum_{J_z=-\infty}^\infty V^{(\kappa)}(J_z)P^\dagger_{J_z, \kappa}P_{J_z, \kappa},
\label{Eq:pair}
\end{equation} 
where $P_{J_z, \kappa}$ denotes the pair operator with the $z$-component of total angular momentum $J_z$, and parity $\kappa$ ($+$ for even parity, $-$ for odd parity). (See SM section SI.)

Before turning to the numerical results of 
$V^{(\kappa)}(J_z)$, we can gain much insight by 
considering the limit $\xi\rightarrow \infty$ and
solving Cooper's pair-binding problem. This amounts to solving the quantum mechanics of two electrons interacting via dipole-dipole interaction
\begin{equation}
V_{\rm dd} = \frac{1}{r^3}[\vec{S}_1\cdot\vec{S}_2-3(\vec{S}_1\cdot\hat{r})(\vec{S}_2\cdot\hat{r})]\propto {\cal R}^{(2)}({\bm r}_1, {\bm r}_2)\cdot {\cal S}^{(2)}({\bm s}_1, {\bm s}_2) 
\end{equation}
 where ${\vec S}_{1, 2}$ represent the spin operators of the two electrons and 
both ${\cal R}^{(2)}$ and ${\cal S}^{(2)}$ are rank-two tensors, with ${\cal R}^{(2)}$ acting on coordinate space, and ${\cal S}^{(2)}$ (magnetic quadrupole moment) on spin space [\onlinecite{Brink}].  First we note that $\langle V_{\rm dd}\rangle=0$ in the even parity channel due to a selection rule (see SM section SIV). The celebrated selection rule that forbids optical transition between $1s$ and $2s$ state (Fig. 2A), but allows the transition between $1s$ and $2p$ state (Fig. 2B) is an example of how a tensor operator of rank $r$, $\mathcal{T}^{(r)}$ connects two angular momentum states $|l\rangle$ and $|l'\rangle$ according to the Wigner-Eckart theorem. The theorem states $\langle l'|\mathcal{T}^{(r)}|l
\rangle=0$ unless $|r-l|\leq l'\leq(r+l)$. In the case of optical transitions of a hydrogen atom, the tensor operator involved is a rank $r=1$ vector field as the photon is a spin $s=1$ boson. In our case, the scattering potential carries a quadrupolar moment which is a rank $r=2$-tensor. Hence the selection rule forbids pairing in the singlet channel, i.e., $\langle S=0|V_{\rm dd}|S=0\rangle=0$ (Fig 2C). Since the Pauli principle dictates all even parity pairing to be singlet,  the selection rule limits possible pairing to the odd-parity channel (Fig 2D). Second we note that among the odd-parity states,
\begin{equation} 
|J_z=0\rangle\sim (k_x+ik_y)|\downarrow\downarrow\rangle+(k_x-ik_y)|\uparrow\uparrow\rangle
\end{equation} and
\begin{equation}
|J_z=\pm 1\rangle\sim  (k_x\pm ik_y)\frac{|\uparrow\downarrow \rangle+|\downarrow\uparrow \rangle}{\sqrt{2}}\sim k_y \frac{|\rightrightarrows \rangle_x-|\leftleftarrows\rangle_x}{\sqrt{2}}\mp k_x \frac{|\rightrightarrows \rangle_y-|\leftleftarrows\rangle_y}{\sqrt{2}}.
\end{equation}
 states indeed yield pair-binding (see Fig 2E and SM section SIV). This is consistent with the earlier insight that Eq.~\eqref{Eq:int-gauge} is attractive in the equal-spin pairing channel. Indeed full numerical calculation of interaction strengths $V^{(\kappa)}(J_z)$ confirms the above analysis and predicts the two odd-parity pairing channels with $J_z=0$ and $J_z=\pm1$ to be overwhelmingly dominant, with predicted $T_c\sim \omega_{\rm sf}\;e^{-1/\lambda}$, with the dimensionless parameter $\lambda\sim N(0)V^{(\kappa)}{(J_z)}\sim J_K^2N(0)/J_{\rm ex}$ (see SM section SIV). 
Pairing in either of these channels on a single Fermi surface will be topological
\cite{Sato09, Qi10}.

There remains the important question of the effect of short distance physics. As we show in SM section SIV, microscopics affect the above considerations in two ways. Firstly, the Kondo coupling (Eq.\ref{Eq:Kondo}) acquires spatial dependence which gives rise to extra form factors in the pairing interaction $V_{\mu\nu}$. Secondly, the spin structure factor $S_{ab}({\bm q})$ (Eq.\ref{Eq:chi}) gains richer momentum dependence at high momentum points. Nevertheless, as long as Fermi surface stays small compared to the high momentum points, 
such short distance structures do not change the dominant pairing channels. Indeed our mean-field calculations of a microscopic model for the heterostructures (see SM section SIV) confirm that the main effect of the short distance physics to be quantitative rather than qualitative.

We now turn to a material proposal for the metallic layer that is expected to fit the above long-wave length description. 
The requirements of chemical stability at the interface between the two materials and matching of lattice constants restrict the choice of materials. In particular, we need the metallic layer to stably grow with the interface along the (111) surface in order to avoid the generation of orphan bonds and preserve the spin correlations of QSI. Electronically we require the metallic layer to be a good metal without strong Fermi surface nesting, the simplest example being an $s$-band metal. Such a metal would have several merits: (1) large band width, (2) weak correlation, which helps to avoid ordering by itself; (3) non-degenerate bands, which helps to provide odd numbers of Fermi surfaces to generate topological superconductivity. 
Unfortunately, existing metallic pyrochlores such as  
Bi$_2$Ru$_2$O$_7$ or Bi$_2$Ru$_2$O$_7$ do not satisfy these criteria since they 
show complicated Fermi surfaces due to the conduction electron of $d$-electron character. 

 A robust strategy to realize the targeted metallic system is to dope a band insulator with an empty conduction band of $s$-character.  
Simple crystal chemical rules point to compounds with Sn$^{4+}$ or Bi$^{5+}$ at the B-site of A$_2$B$_2$O$_7$, e.g., Y$_2$Sn$_2$O$_7$ or La$_2$Sn$_2$O$_7$, as prime candidates for the insulating starting material. 
 In practice, doping of the empty conduction band can be performed in a few different ways. For example, one could dope the B-site with Sb or substitute the A-site with Ce$^{4+}$
 to form, e.g., (La/Ce)$_2$Sn$_2$O$_7$. For illustration of the principle, 
we performed ab initio calculations using Density Functional Theory(DFT) on  Y$_2$Sn$_{2-x}$Sb$_x$O$_7$ /Pr$_2$Zr$_2$O$_7$ heterostructures (see Fig 3A and SM section SIII). In order to minimize the proximity effect which lowers the superconducting gap, we considered one unit-cell thick metallic layer and imposed periodic boundary condition in the direction perpendicular to the interface. As shown in Fig 3C we find a single circular Fermi surface centered at the $\Gamma$ point  for the heterostructure with Y$_2$Sn$_{2-x}$Sb$_x$O$_7$ at doping level $x=0.2$. 
Moreover, the conduction electron wavefunction penetrates into the first two layers of Pr$_2$Zr$_2$O$_7$ (Fig. 3B), generating a finite Kondo coupling across the interface. Our estimates for the parameters are $E_F = 300$ meV, $\tau^{-1}\sim 2J_{\rm ex} = 0.17$ meV, $5 \text{meV} \lesssim J_K \lesssim 14$ meV. Of these, $E_F$ was obtained directly from DFT, $J_{\rm ex}$ from experiment and $J_K$ from second order perturbation on a realistic atomic Hamiltonian (see SM section SIII). These parameters display a separation of energy scales, i.e. $\tau^{-1}/E_F\ll 1$, that ensures the reliability of the above BCS-like treatment of the superconducting instability.

Carrying out a microscopic treatment of interfacial superconductivity for the heterostructure of Fig.3A using the electronic structure of Fig.3C (see SM section SIV), we confirmed the predictions of the low energy effective model Eqs.(\ref{Eq:chi},\ref{Eq:pair}) and obtained an estimate of $T_c$ from $T_c\sim \tau^{-1} e^{-E_F J_{\rm ex}/J_K^2}$. Using our parameters, this is of order 1 K. \footnote{The precise value of $T_c$ depends sensitively on the dimensionless ratio $\lambda$ which in turn depends on the microscopic details and the actual value of $J_K$.} We note that  the predicted $T_c$  is comparable to the only other solid state candidate for topological superconductivity Sr$_2$RuO$_4$ with $T_c\simeq 1.5$ K \cite{Maeno94}. 

 The strategy we developed here for predictively achieving an exotic superconducting state we dubbed SC$|$QSL has profound theoretical and experimental consequences. At the level of theoretical principle it is an alternative approach to that of Anderson's to drive superconductivity from a quantum spin liquid.  Our proposal to 
borrow the quantum spin fluctuation of a quantum spin liquid\cite{Balents10} to form exotic superconductors through heterostructure growth bears similarities to excitonic fluctuation proposals\cite{Little64, Ginzburg70, Bardeen73, Koerting05, Hirschfeld11}. However, the latter charge fluctuation based mechanism suffers from various issues. For instance, tunneling of electrons from the metal can damage the small charge gap of the semi-conductor and kill the charge fluctuation. Further, the local charge fluctuations only drives $s$-wave pairing which needs to overcome the Coulomb repulsion. By utilizing the goodness of the quantum spin liquid, i.e., spin-fluctuation, our proposal bypasses these issues. At the level of a specific choice of quantum spin-ice, we for the first time demonstrated how emergent gauge fluctuations can mediate attractive interaction in the triplet channel through unconventional coupling between the gauge field and the fermions, which is nevertheless natural coupling in the language of spins. Our concrete material proposal can guide experimental pursuit of the proposed heterostructure. 
Clearly the experimental control over atomic interfaces and thin films have reached the state that can support superconductivity\cite{Ohtomo04, Reyren07,Gozar08, Bozovic11,WangCPL2012}.  Successful materialization of the proposed topological superconductor will not only be a  major breakthrough in superconductivity research, but would also be the first application 
emanating from the discovery of quantum spin liquids.

{\bf Acknowledgements}
The authors acknowledge useful discussions with Leon Balents, Steve Kivelson, SungBin Lee, Tyrel McQueen and Arun Paramekanti. The authors are grateful to Michel Gingras and Darrell Schlom for a careful reading of the manuscript and helpful comments and suggestions. 
J.-H.S. and E.-A.K. acknowledge support  by the U.S. Department of Energy, Office of Basic Energy Sciences, Division of Materials Science and Engineering under Award DE-SC0010313; C.H.K. and C.J.F. acknowledge support by NSF Grant No. DMR-1056441.

\newpage

\begin{figure}[t]
\begin{centering}
\includegraphics[width=0.8\linewidth]{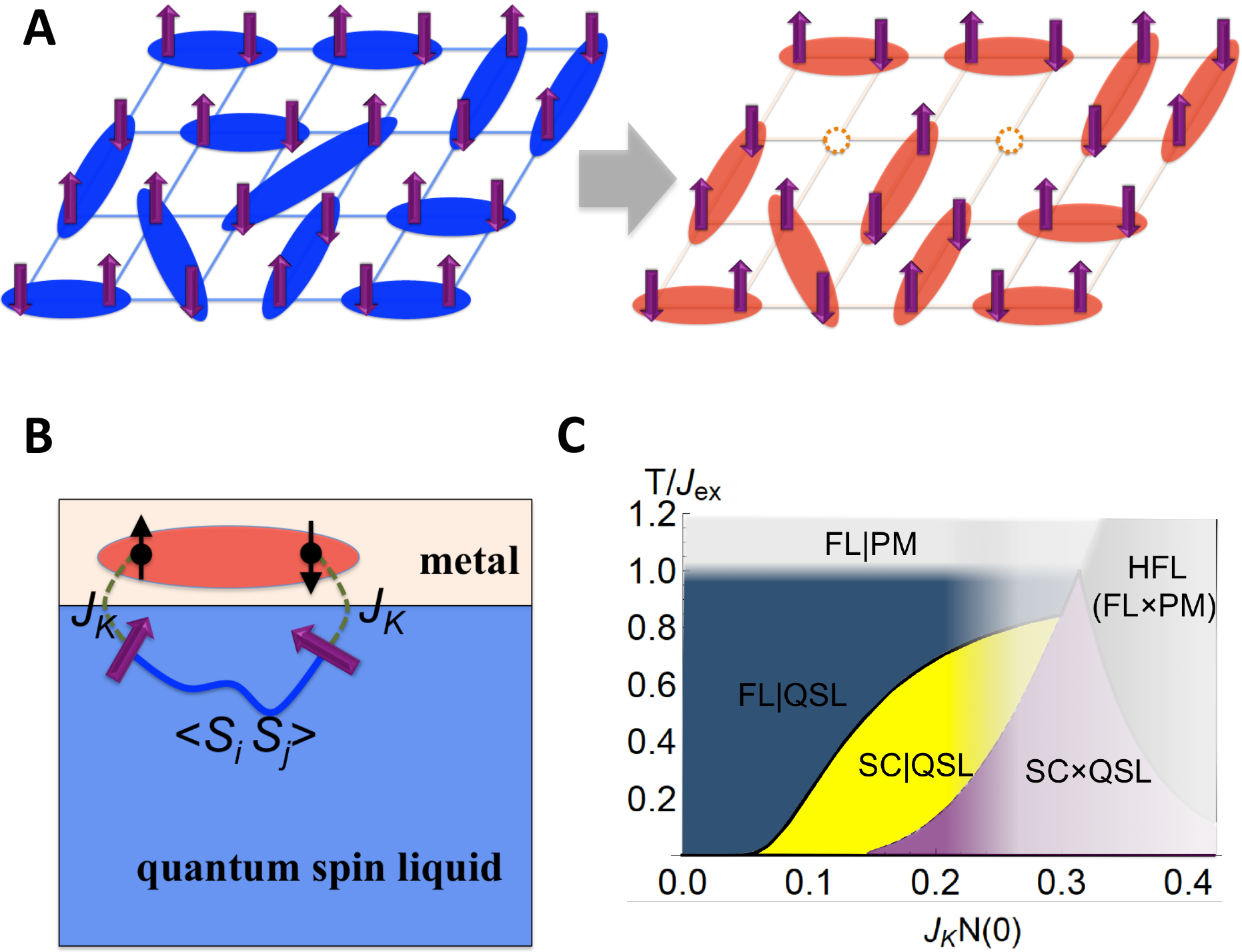} 
\end{centering}
\caption{\textbf{General considerations of spin-fluctuation-mediated-pairing in the metal/quantum-spin-liquid (QSL) heterostructure.} (${\bold A}$): The resonating valence bond (RVB) proposal of unconventional superconductivity by Anderson \cite{Anderson87}. Left represents the parent insulating system where the spins form RVB pairs (blue ellipsoid). By doping holes (dashed circle) into the system, as shown on the right, the RVB pairs become mobile (red ellipsoid), and the whole system becomes superconducting. (${\bold B}$): The proposed metal/QSL heterostructure. The metal provides the charge carriers and the QSL provides a pairing interaction via quantum paramagnetic spin-fluctuations $\langle S_i S_j\rangle$. The two systems are coupled via a Kondo type coupling $J_K$, which generates Cooper pairing among charge carriers (red ellipsoid). (${\bold C}$): Phase diagram of the metal-QSL heterostructure. In the FL$|$PM, FL$|$QSL and SC$|$QSL phases, the conduction electrons from the metal and the local moments from the QSL coexist, but are decoupled at the mean field level. The conduction electrons form a Fermi liquid (FL) or a superconductor (SC), while the local moments form an incoherent paramagnet (PM) or a coherent QSL. In the HFL and SC$\times$QSL phases, the conduction electrons and the local moments hybridize and form Kondo singlets. The aim is to design the heterostructure to be in the SC$|$QSL phase. This phase diagram applies to the parameter region $J_{\rm RKKY}<{\rm max}\left\{J_{\rm ex}, T_K\right\}$ for all coupling strength $J_KN(0)$ (see SM section SI).}
\label{Figure1}
\end{figure}

\newpage

\begin{figure}[t]
\begin{centering}
\includegraphics[width=0.8\linewidth]{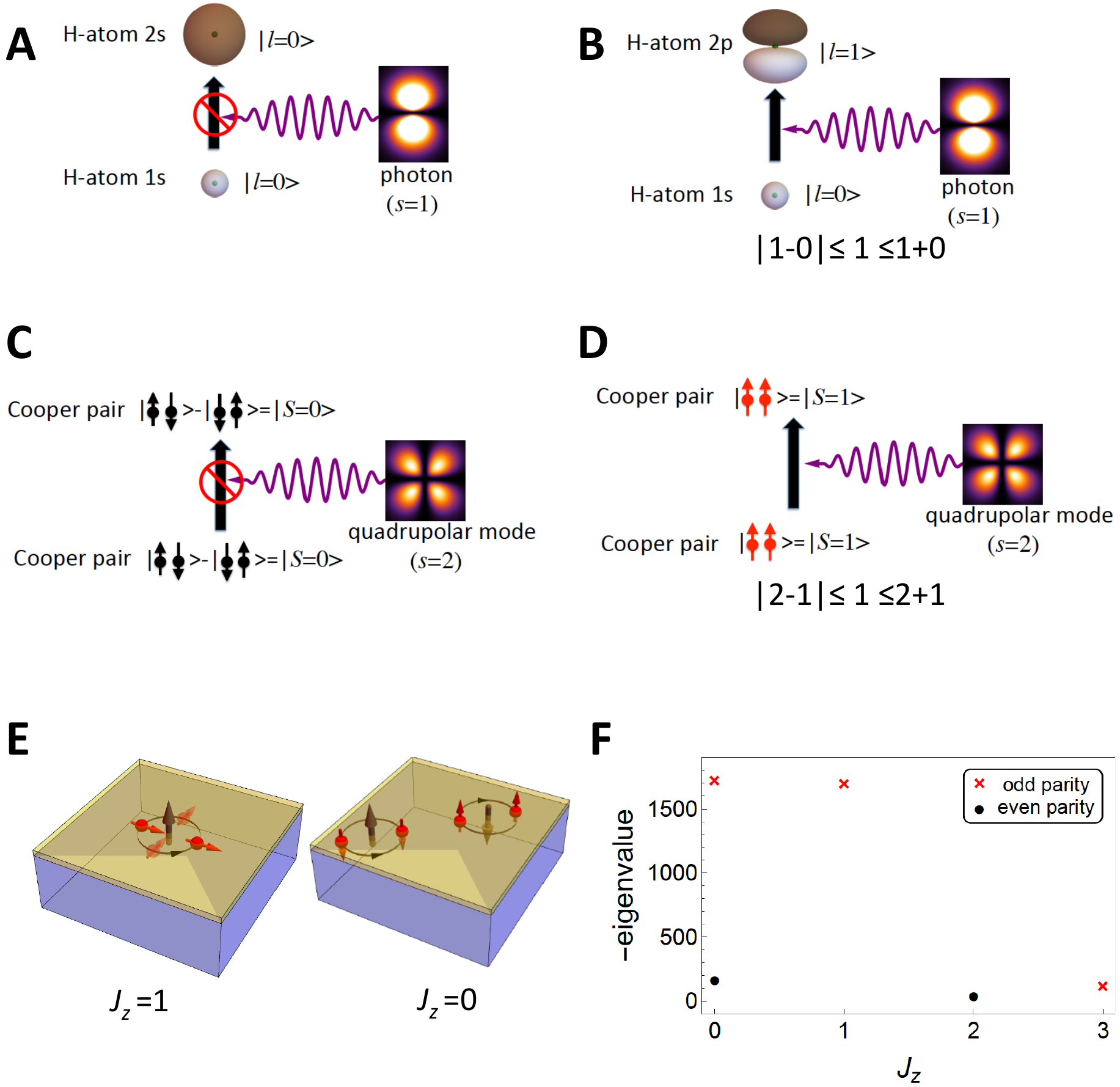} 
\end{centering}
\caption{\textbf{The dominant pairing channels in the metal/quantum-spin-ice heterostructure.} (${\bold A}, {\bold B}, {\bold C}, {\bold D}$): Understanding the emergence of parity-odd spin-triplet pairing from selection rules. ${\bold A}$ and ${\bold B}$ represent the dipole transitions for atomic hydrogen: transition from $1s$ state with angular momentum $l=0$ to $2s$ state also with $l=0$ is forbidden by the selection rule ($|1-l|\leq l'\leq 1+l$), while transition from $1s$ state to $2p$ state with $l=1$ is allowed. ${\bold C}$ and ${\bold D}$ represent the pairing problem under the rank-two magnetic dipole-dipole interaction: spin-singlet pairing with total spin $S=0$ is forbidden by the selection rule ($|2-S|\leq S\leq 2+S$), while spin-triplet pairing with total spin $S=1$ is allowed. (${\bold E}$): Illustration of spin and angular momentum configurations of the dominant pairing channels. The larger (brown) arrows represent the orbital angular momenta, and the smaller (red) arrows represent the electron spins. Spin and orbital angular momentum are coupled to yield the total angular momentum $J_z=0, 1$.  (${\bold F}$): The leading negative eigenvalues of the pairing interaction matrix for different parity and $J_z$ channels in the low energy effective model. The eigenvalues are dimensionless numbers in arbitrary units. The dominant pairing channels have odd parity with $J_z=0, \pm 1$. }
\label{Figure2}
\end{figure}

\newpage

\begin{figure}[t]
\begin{centering}
\includegraphics[width=0.6\linewidth]{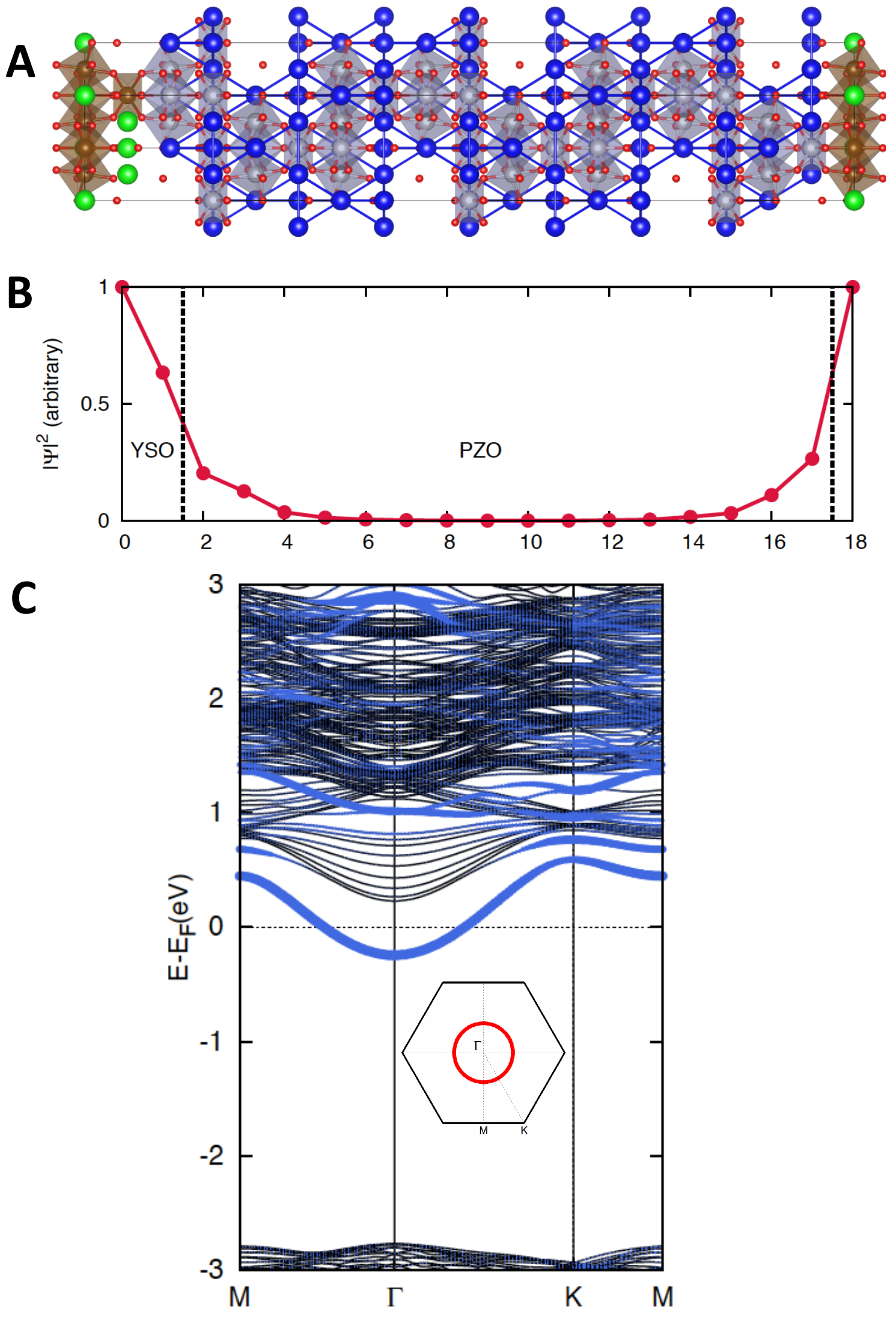} 
\end{centering}
\caption{\textbf{A concrete material realization of the metal/quantum-spin-ice heterostructure: Pr$_2$Zr$_2$O$_7$/Y$_2$Sn$_{2-x}$Sb$_x$O$_7$ (111).} (${\bold A}$): Lattice structure. Two layers of Sb-doped Y$_2$Sn$_2$O$_7$ deposited on top of 16 layers of Pr$_2$Zr$_2$O$_7$ along the [111] direction. The magnetic moments are on the Pr sites (blue), which form alternating layers of triangular and Kagome lattices. The conduction electrons are donated by the Sn atoms (brown). Red is O, Green is Y and Gray is Zr. (${\bold B}$): Amplitude of the conduction electron wavefunction in the direction perpendicular to the interface showing  penetration into the first two or three layers of Pr$_2$Zr$_2$O$_7$. (${\bold C}$): Band structure, with Fermi surface shown in the inset. There is a single band crossing the Fermi energy, and a single circular Fermi surface around the $\Gamma$ point, with Fermi energy $E_F\simeq 0.3$ eV, and Fermi momentum $k_F\simeq 0.37 (2\pi/a)$, where $a$ is the lattice constant of Pr$_2$Zr$_2$O$_7$.
 }
\label{Figure3}
\end{figure}

\clearpage
\setcounter{equation}{0}
\setcounter{figure}{0}

\centerline{\textbf{Supplemental Material}}

In the Supplemental Material, we first lay out the general framework to describe interfacial spin-fluctuation-mediated superconductivity in our heterostructure (SI). Then we include a list of possible candidate materials for the insulating substrate (SII). Finally, choosing the quantum spin ice candidate material Pr$_2$Zr$_2$O$_7$ as the insulating substrate, we study the resulting heterostructure in detail (SIII), in particular the pairing problem (SIV).

\section*{SI: Superconductivity in metal/quantum-spin-liquid heterostructure}

We include here a general description of interface superconductivity in the heterostructure of metal and quantum-spin-liquid (QSL). We consider the metal to be described by a tight-binding model with Hamiltonian $H_{\rm metal}$. The QSL is described in terms of local moments, and the moments interact via exchange interactions with Hamiltonian $H_{\rm QSL}$. The conduction electrons from the metal penetrate into the QSL, generating a Kondo type coupling with Hamiltonian $H_K$. The coupling strength is determined by the overlap of the conduction electron and the localized electron wavefunctions [\onlinecite{Jensen91}]. The heterostructure is then described by the Hamiltonian
\begin{eqnarray}
H_{\rm metal}&=&\sum_{mn\alpha}t_{mn}c^\dagger_{m\alpha}c_{n\alpha}-\mu\sum_{m\alpha} c^\dagger_{m\alpha}c_{m\alpha},\\
H_K&=&\sum_{im\alpha\beta} {\cal I}_{im}c^\dagger_{m\alpha}{\vec \sigma}_{\alpha\beta}c_{m\beta}\cdot {\vec S}_i,\\
H_{\rm QSL}&=&\sum_{ijab} J_{ij}^{ab} S_i^a S_j^b.
\end{eqnarray} 
Here we use $m, n$ to label the conduction electron sites, $i, j$ for the spin sites, ${\vec \sigma}=(\sigma^x, \sigma^y, \sigma^z)$ represent the Pauli matrices.

\subsection*{A. ~ Phase diagrams}

\begin{figure}
\begin{centering}
\subfigure[]{
\includegraphics[width=0.46\linewidth]{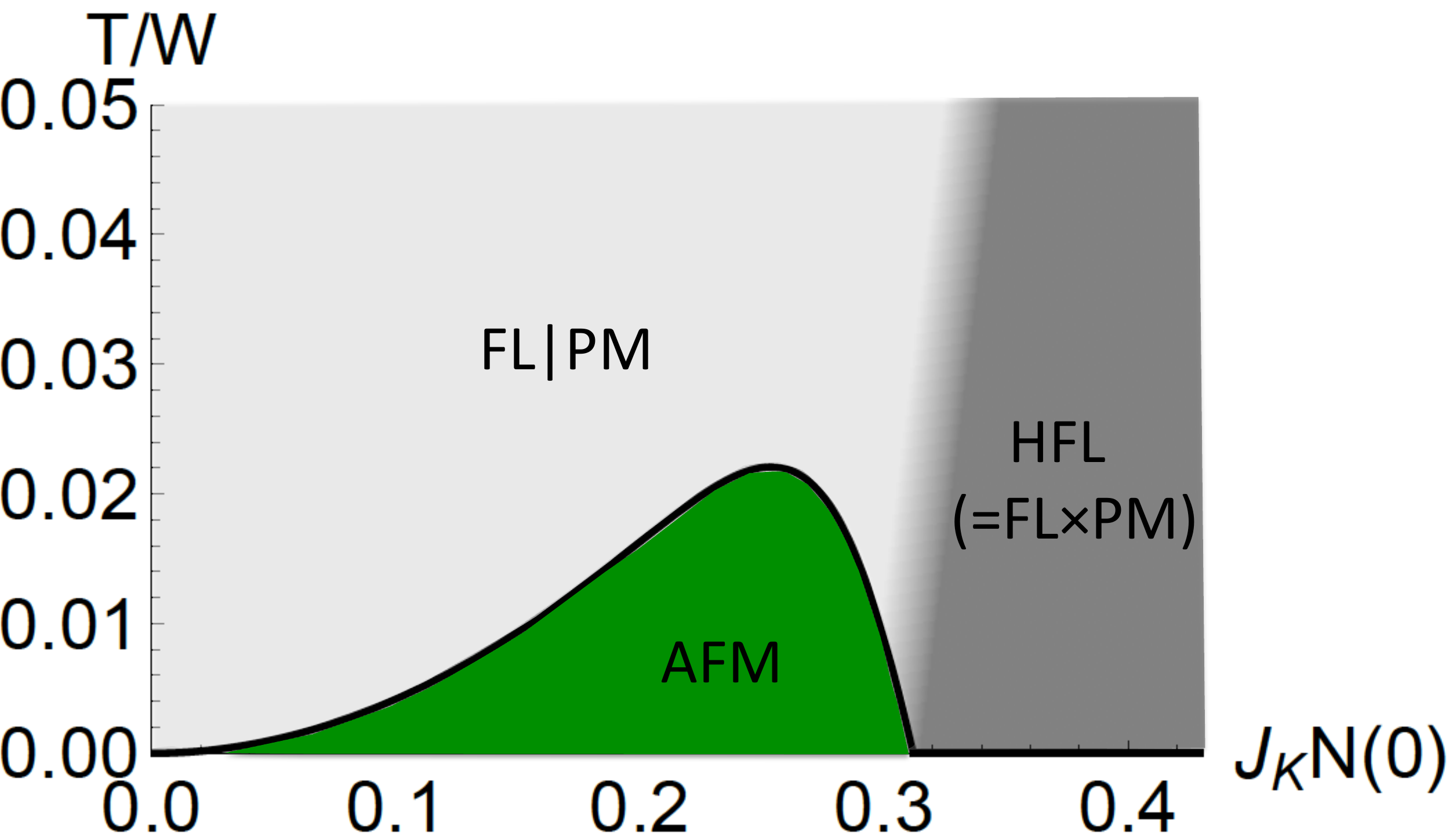} 
}
\subfigure[]{
\includegraphics[width=0.46\linewidth]{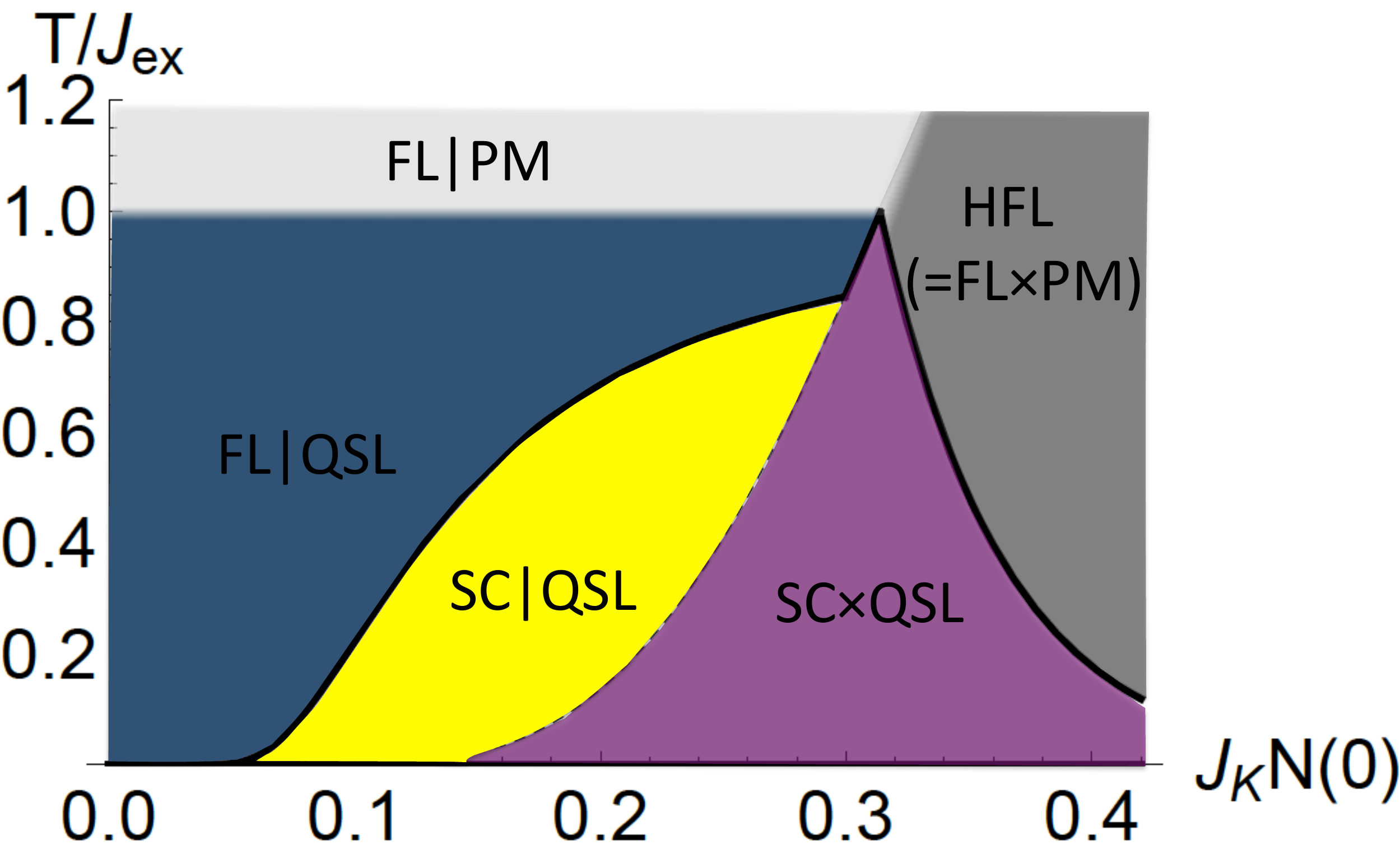} 
}
\subfigure[]{
\includegraphics[width=0.46\linewidth]{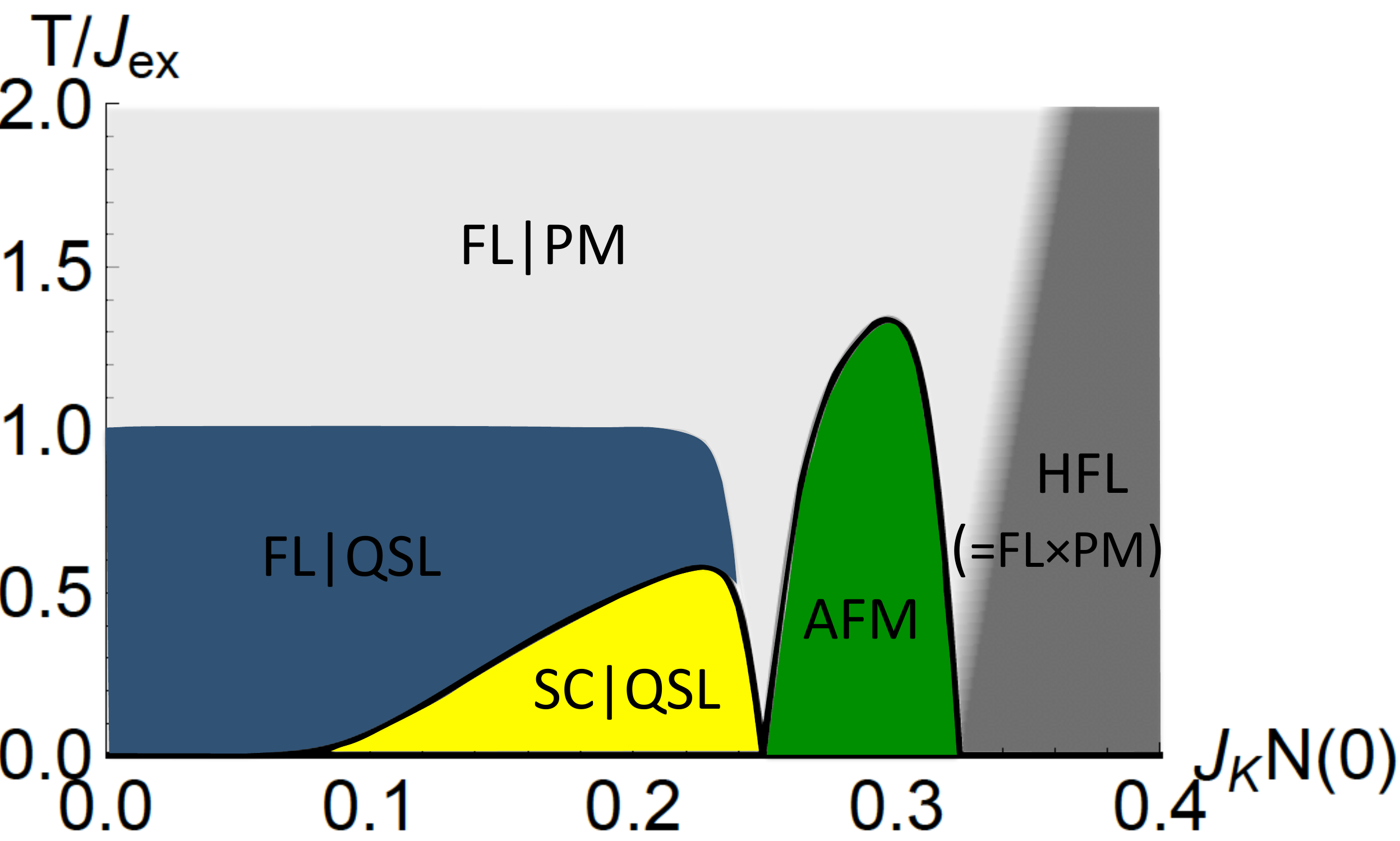} 
}
\end{centering}
\caption{(Color online) Phase diagrams for three different cases: (a) the Doniach phase diagram with $J_{\rm ex}=0$, (b) $J_{\rm RKKY}<{\rm max}\left\{J_{\rm ex}, T_K\right\}$ for all coupling strength $J_KN(0)$, and (c)  $J_{\rm ex}$, $J_{\rm RKKY}$ and $T_K$ are comparable. With $J_{\rm RKKY}=CJ_K^2/E_F$, $T_K=E_F e^{-1/J_KN(0)}$, there are two dimensionless parameters $C\sim {\cal O}(1)$ and $B\equiv E_F/J_{\rm ex}\gg 1$. When $C<(\log B)^2/B$, phase diagram (b) applies; when $C>(\log B)^2/B$, phase diagram (c) applies. The system consists of two components: conduction electrons and local moments. Here $|$ represents a phase where the two components coexist but are effectively decoupled, and $\times$ represents a phase where the two components hybridize, forming Kondo singlets. }
\label{Fig:phase}
\end{figure}

With the presence of both geometric frustration and Kondo coupling, such systems are strongly correlated, and possess a rich phase diagram. The global phase diagrams of such frustrated Kondo systems have attracted much attention recently in the context of heavy fermion systems since the pioneering work of [\onlinecite{Senthil03, Senthil04, Si06, Coleman10}]. Following the seminal work of Doniach [\onlinecite{Doniach77}], the phase diagram can be obtained by comparing the energy scales of the competing interactions in the system: (1) the spin exchange interaction $J_{\rm ex}$, (2) the Kondo temperature $T_K$, and (3) the RKKY interaction $ J_{\rm RKKY}$. The conduction electron Fermi energy $E_F$ is typically much higher than these interaction scales. 
The spin exchange interaction $J_{\rm ex}$ is a property of the insulating substrate. The Kondo temperature $T_K$ and the RKKY interaction $ J_{\rm RKKY}$ arise from coupling the metallic layer to the insulating substrate. Their forms can be found in standard textbooks (see e.g. [\onlinecite{Hewson}]). To be self-contained, we include below a short discussion of these two energy scales.

When a local moment is placed in the conduction electron sea, the conduction electron cloud Kondo-screens the local moment. This is a non-perturbative effect, and the characteristic energy scale is the Kondo temperature [\onlinecite{Hewson}]
\begin{eqnarray}
T_K\sim E_F e^{-1/J_KN(0)},
\end{eqnarray} 
where $J_K$ is the Kondo coupling strength, and $N(0)\sim 1/E_F$ is the conduction electron density of states at the Fermi level.
Above $T_K$, the local moment is essentially decoupled from the conduction electrons. Below $T_K$, the local moment forms Kondo singlets with the conduction electrons. The change around $T_K$ is not a sharp phase transition, but a crossover.

When two local moments are placed in the conduction electron sea, the conduction electrons mediate a long-range oscillating interaction among the local moments. This interaction is called the RKKY interaction, with the corresponding energy scale [\onlinecite{Hewson, Fischer75}]
\begin{eqnarray}
 J_{\rm RKKY}\sim J_K^2N(0).
\end{eqnarray} 
When the local moments form a lattice, the corresponding RKKY interactions are encoded in the Hamiltonian $H_{\rm RKKY}=\sum_{ij}J_{\rm RKKY}({\bm R}_i-{\bm R}_j) {\vec S}_i\cdot {\vec S}_j$, which generically leads to magnetic ordering of the moments.

The competition of $J_{\rm ex}$, $T_K$ and $ J_{\rm RKKY}$ gives rise to a high dimensional phase diagram. We consider below representative two dimensional cuts of such a high dimensional phase diagram in the plane expanded by the (normalized) Kondo coupling $J_K$ and temperature $T$ (see Fig.\ref{Fig:phase}). We consider three different cases here, as specified by the different choices of the dominant energy scales.

When the spin exchange interaction is small, i.e. $J_{\rm ex}\ll J_{\rm RKKY}$ and $J_{\rm ex}\ll T_K$, we recover the original Doniach phase diagram [\onlinecite{Doniach77}] (Fig.\ref{Fig:phase}a). At high temperatures, the local moments are incoherent, residing in a paramagnetic (PM) state, decoupled from the conduction electrons which form a Fermi liquid (FL). Coherent many body states develop as one lowers the temperature. In the parameter region where the Kondo coupling $J_K$ is small, one has $ J_{\rm RKKY}>T_K$, and the RKKY interaction dominates. The system develops long range magnetic order, e.g. antiferromagnetic (AFM) order. We note that since the spin lattice is frustrated, RKKY interaction can also lead to more complicated magnetic ordering patterns. In the parameter region where the Kondo coupling $J_K$ is large, one has $T_K> J_{\rm RKKY}$, and the Kondo effect dominates. The conduction electrons and the local moments form Kondo singlets, and the system is in a heavy Fermi liquid (HFL) state with a large Fermi surface, which counts both the conduction electrons and the local moments.

Of more relevance to the present paper is the case where the RKKY interaction is never the dominant energy scale, i.e. $J_{\rm RKKY}<J_{\rm ex}$ for small $J_K$ and $J_{\rm RKKY}<T_K$ for large $J_K$. The corresponding phase diagram has been studied in [\onlinecite{Senthil03, Senthil04}] (see Fig.\ref{Fig:phase}b). At low temperatures, the phase diagram is determined by the competition between $J_{\rm ex}$ and $T_K$. For large Kondo coupling $J_K$, where $T_K$ is the dominant energy scale, the system is in the HFL state as in the previous case. For $J_K$ small, where $J_{\rm ex}$ is the dominant energy scale, the local moments are in a QSL state, decoupled from the conduction electrons. Such a coexisting and decoupled FL and QSL phase (hence named SC$|$QSL here) corresponds to the FL$^*$ phase of [\onlinecite{Senthil03, Senthil04}].   

Of central importance to the present paper is the fact that at low temperatures, the FL$|$QSL phase is unstable towards pairing instability. The spin fluctuations of QSL induce pairing interactions among the conduction electrons, which then give rise to a superconducting (SC) phase at the interface, that coexists with the QSL phase in the insulating substrate. Such a SC$|$QSL phase is the main target of the present paper. We want to design our heterostructure in a proper way so that the interface lies in the SC$|$QSL phase. We note that in the slave-fermion mean field theory, for $Z_2$ spin liquids where the spinons have finite pairing amplitude, the HFL state also becomes superconducting at low temperatures [\onlinecite{Senthil03}]. Such a  phase with superconductivity entangled with spin liquid via the formation of Kondo singlets (hence the name SC$\times$QSL phase) is absent for $U(1)$ spin liquids [\onlinecite{Senthil04}].

There is also the possibility that as one varies the Kondo coupling, $J_{\rm ex}$, $J_{\rm RKKY}$ and $T_K$ dominate respectively over different parts of the 2d phase diagram (see Fig.\ref{Fig:phase}c). At small $J_K$ where $J_{\rm ex}$ dominates, the local moments are in the QSL state, decoupled from the conduction electrons. At intermediate $J_K$ where $J_{\rm RKKY}$ dominates, the local moments develop long range magnetic order, but are still decoupled from the conduction electrons. At large $J_K$ where $T_K$ dominates, the conduction electrons and local moments form Kondo singlets. In this case, we can still obtain the desired SC$|$QSL phase as  a result of the low temperature instability of the FL$|$QSL phase. Since the spin liquid correlation (here in particular spinon pairing) has been destroyed at the corresponding $J_K$, the HFL will not become superconducting at low temperatures.

\subsection*{B. ~ Spin-fluctuation mediated superconductivity}

We now study in more detail the spin-fluctuation mediated superconductivity in the SC$|$QSL phase. Coupling to the local moments induces interactions among the conduction electrons. The partition function of the metallic part becomes ${\cal Z}_{\rm metal}={\rm Tr} \exp\left( -\beta H_{\rm metal}- {\cal S}_{\rm int}\right)$, with the induced action [\onlinecite{Mahan}]
\begin{equation}
{\cal S}_{\rm int}=-\sum_{l=1}^{\infty}\frac{(-1)^l}{l} \int_0^\beta d\tau_1\cdots\int_0^\beta  \langle T_\tau H_K(\tau_1)\cdots H_K(\tau_l)\rangle.
\end{equation}
Here $\tau$ denotes imaginary time, $\beta=1/T$, and $T_\tau$ represents time ordering. We set $\hbar=k_B=1$. The expectation value is taken over the spin Hamiltonian $H_{\rm QSL}$. 
 The first order term in the action is
\begin{equation}
{\cal S}^{(1)}_{\rm int}=\sum_{ima} \int d\tau_0^\beta {\cal I}_{im}  \langle S_i^a\rangle  s_m^a(\tau),
\end{equation}
where the conduction electron spin density $s_m^a=\sum_{\alpha\beta}c^\dagger_{m\alpha} \sigma^a_{\alpha\beta}c_{m\beta}$, and the expectation value is taken over the spin Hamiltonian. One can see that when the magnetic insulator has long range order, it generates a potential that polarizes the conduction electron spins. The potential term can also be generated by other more exotic orderings, e.g. scalar spin chirality $\langle {\vec S}_i\cdot ({\vec S}_j\times{\vec S}_l) \rangle$ (see e.g. [\onlinecite{Martin08, Lee13, Flint13}]), even when $\langle S_i^a\rangle=0$. Such a potential term changes the band structure of the original metal, and is generally detrimental for superconductivity. Hence we require this term to vanish in our heterostructure. This can be achieved by choosing the proper magnetic insulator.

When the magnetic insulator does not have long range order, the leading order term of the induced interaction among the conduction electrons is of second order in the Kondo coupling,
\begin{equation}
{\cal S}^{(2)}_{\rm int}=-\frac{1}{2}\sum_{ij mn ab}\int_0^\beta d\tau \int_0^\beta d\tau'  {\cal I}_{im}{\cal I}_{jn}\langle T_\tau S_i^a(\tau) S_j^b(\tau')\rangle s_m^a(\tau) s_n^b(\tau'),
\label{Eq:H2}
\end{equation}
which represents a retarded exchange interaction among the conduction electron spin density. Rewriting the above action in the form ${\cal S}^{(2)}_{\rm int}=\int_0^\beta d t H_{\rm int}(t)$, we obtain the induced interaction Hamiltonian $H_{\rm int}(t)$ as shown in Eq.(3) of the main text.

 If the Fermi surface of the metallic layer is not too close to perfect nesting, pairing will be the only weak coupling instability [\onlinecite{Shankar94}]. We will then proceed to study the pairing problem using standard mean field theory. We first decompose the spin fluctuation induced interaction (\ref{Eq:H2}) in the pairing channel in terms of the the pair operator $P_{\alpha\alpha'}({\bm k}, {\bm q}; \omega, \Omega)\equiv c_{{\bm k}+{\bm q}/2, \omega+\Omega/2, \alpha} c_{-{\bm k}+{\bm q}/2, -\omega+\Omega/2, \alpha'}$. The resulting pairing action reads 
\begin{equation}
{\cal S}_{\rm int}=\sum_{{\bm k}{\bm k}'{\bm q}\omega\omega'\Omega}\sum_{ab\alpha\beta\alpha'\beta'}V_{ab}({\bm k}-{\bm k}', \omega-\omega')\sigma^a_{\alpha\beta}\sigma^b_{\alpha'\beta'}P^\dagger_{\alpha\alpha'}({\bm k}, {\bm q}; \omega, \Omega)P_{\beta'\beta}({\bm k}', {\bm q}; \omega', \Omega),
\end{equation}
with the interaction 
\begin{equation}
V_{ab}({\bm p}, \Omega)=-\frac{1}{2}\sum_{ijmn} \int d\tau d\tau' e^{-i{\bm p}\cdot ({\bm r}_m-{\bm r}_n)} e^{-i\Omega(\tau-\tau')} {\cal I}_{im}{\cal I}_{jn}\langle T_\tau S_i^a(\tau) S_j^b(\tau')\rangle .
\label{Eq:Vpair}
\end{equation}

However, unlike the Rashba type spin-orbit coupling ${\cal H}_R({\bm k})\sim ({\hat {\bm z}}\times{\bm k})\cdot {\vec \sigma}$, which breaks parity symmetry, the pairing interaction $V_{ab}({\bm p}, \Omega)$ induced by the insulating substrate, for which the spin dependence and momentum dependence may not decouple, is even under parity: $V_{ab}({\bm p}, \Omega)= V_{ab}(-{\bm p}, \Omega)$. This ensures that the parity odd and parity even pairing channels do not mix.
We can then organize the pair operators in different parity channels: $P_{\mu}=\frac{1}{\sqrt{2}}\sum_{\alpha\alpha'} \left[ i\sigma_y\sigma_\mu\right]_{\alpha\alpha'}P_{\alpha\alpha'}$, and $P^\dagger_{\mu}=\frac{1}{\sqrt{2}}\sum_{\alpha\alpha'} \left[ i\sigma_y\sigma_\mu\right]^*_{\alpha\alpha'}P^\dagger_{\alpha\alpha'}$, with $\mu=0$ representing parity even pairing channel, and $\mu=x, y, z$ parity odd channels. 
The pairing action written in this representation
\begin{equation}
{\cal S}_{\rm int}=\int\frac{d^2{\bm k}}{(2\pi)^2}\int \frac{d\omega}{2\pi}\int\frac{d^2{\bm k}'}{(2\pi)^2}\int \frac{d\omega'}{2\pi} V_{\mu\nu}({\bm k}-{\bm k}',\omega-\omega')P^\dagger_{\mu}({\bm k},\omega)P_{\nu}({\bm k}',\omega'),
\end{equation}
where $V_{\mu\nu}=\sum_{ab}\sum_{\alpha\alpha'\beta\beta'} V_{ab}\sigma^a_{\alpha\beta}\sigma^b_{\alpha'\beta'}[\sigma_\mu i\sigma_y]^*_{\alpha'\alpha}[\sigma_\nu i\sigma_y]_{\beta\beta'}$, is block diagonal: $V_{0x}=V_{0y}=V_{0z}=0$.

The resulting pairing order parameter is largely determined by the symmetry of the system. The presence of the interface breaks the SO(3) coordinate space rotation symmetry to $U(1)$ rotation around the $z$-axis (the interface is at $z=0$). The spin-orbit entanglement in the pairing interaction $V_{ab}({\bm q}, \Omega)$ then leaves only the $z$-component of the total angular momentum $J_z=L_z+S_z$ conserved, i.e. the full $SO(3)_{\bm L}\times SU(2)_{\bm S}$ symmetry is reduced down to $U(1)_{J_z}$. This imposes severe constraints on the possible pairing channels.

With both parity and $J_z$ being good quantum numbers, the pairing interaction can be organized in the parity and $J_z$ basis. We first decompose the pairing interaction in different partial wave channels:
\begin{eqnarray}
V_{\mu\nu}({\bm k}-{\bm k}')=\sum_{L_z, L'_z} V^{\mu\nu}_{L_zL'_z}(k, k')e^{iL_z\varphi_{\bm k}}e^{-iL'_z\varphi_{{\bm k}'}}.
\end{eqnarray}
Here $\varphi_{\bm k}$ represents the angle of the momentum, i.e. ${\bm k}=k(\cos\varphi_{\bm k}, \sin\varphi_{\bm k} )$.
 To simplify the notation, we will keep the frequency dependence implicit. The different partial wave components read
\begin{equation}
V^{\mu\nu}_{L_zL'_z}(k, k')=\int\frac{d\varphi_{\bm k}}{2\pi} \int\frac{d\varphi_{{\bm k}'}}{2\pi}  V_{\mu\nu}({\bm k}-{\bm k}') e^{-iL_z\varphi_{\bm k}}e^{iL'_z\varphi_{{\bm k}'}}.
\end{equation}
Since pairing happens near the Fermi surface, we can approximate $V^{\mu\nu}_{L_z, L'_z}(k, k')\simeq V^{\mu\nu}_{L_z, L'_z}(k_F, k_F)\equiv V_{\mu\nu}(L_z, L'_z)$. Correspondingly the partial wave components of the pair operator can be written as 
\begin{eqnarray}
P_\mu(L_z)=\int\frac{d\varphi_{\bm k}}{2\pi} e^{-iL_z\varphi_{\bm k}}P_\mu({\bm k}).
\end{eqnarray}
The pairing Hamiltonian then becomes of the form
\begin{eqnarray}
H_{\rm int}=\sum_{L_z, L'_z} V_{\mu\nu}(L_z, L'_z) P^\dagger_\mu(L_z)P_\nu(L'_z).
\end{eqnarray}

With parity a good quantum number, the pairty-even and parity-odd parts of the pairing interactio decouple, and we can write the above Hamiltonain as $H_{\rm int}=H^{(+)}_{\rm int}+H^{(-)}_{\rm int}$, with $+$ for even parity and $-$ for odd parity. The even parity part involves only even orbital angular momentum (e.g. $s$- and $d$-waves), and spin singlet channels:
 \begin{eqnarray}
H^{(+)}_{\rm int}=\sum_{L_z, L'_z {\rm even}} V^{(+)}_{00}(L_z, L'_z) P^\dagger_0(L_z)P_0(L'_z).
\end{eqnarray}
The odd parity part involves only odd orbital angular momentum (e.g. $p$- and $f$-waves), and spin triplet channels:
 \begin{eqnarray}
H^{(-)}_{\rm int}=\sum_{L_z, L'_z {\rm odd}}\sum_{\mu,\nu=x,y,z} V^{(-)}_{\mu\nu}(L_z, L'_z) P^\dagger_\mu(L_z)P_\nu(L'_z).
\end{eqnarray}
In the even parity channels, the Cooper pair has $S_z=0$, and hence $J_z=L_z$. In the odd parity channels, the orbital and spin parts combine to form different $J_z$ channels. We reorganize the pair operator in $S_z$-basis as
\begin{eqnarray}
P(L_z, S_z=0)&=&-P_z(L_z),\\
P(L_z, S_z=1)&=&\frac{1}{\sqrt{2}}\left[ P_x(L_z)-iP_y(L_z) \right],\\
P(L_z, S_z=-1)&=&-\frac{1}{\sqrt{2}}\left[ P_x(L_z)+iP_y(L_z) \right].
\end{eqnarray}
The odd parity pairing interaction thus reads
\begin{eqnarray}
H^{(-)}_{\rm int}=\sum_{L_z, L'_z {\rm odd}}\sum_{S_z=0, \pm 1} V^{(-)}(L_z, S_z; L'_z, S'_z) P^\dagger(L_z, S_z)P(L'_z, S'_z).
\end{eqnarray}
Then combining different $(L_z, S_z)$ pair operators with the same $J_z=L_z+S_z$, one arrives at 
the pairing Hamiltonian in the parity and $J_z$ basis as
\begin{eqnarray}
H_{\rm int}=\sum_{\kappa=\pm}\sum_{J_z=-\infty}^{\infty} V^{(\kappa)}(J_z)P^\dagger_{\kappa}(J_z)P_\kappa(J_z),
\end{eqnarray}
which is diagonal in both parity $\kappa$ and $J_z$. 
The gap function can be defined as 
\begin{eqnarray}
\Delta^{(\kappa)}_{J_z}= V^{(\kappa)}(J_z) \langle P_\kappa(J_z)\rangle.
\end{eqnarray}



\subsubsection*{Estimation of superconducting $T_c$}

While the frequency dependence of the pairing interaction does not affect the dominant pairing channel, it does determine the superconducting $T_c$. As in phonon mediated superconductivity, where the phonon frequency sets the scale of $T_c$, here $T_c$ is determined by the frequency scale of the spin correlations in the spin-fluctuating insulator. To give an estimation of $T_c$, we then separate the frequency part of the pairing interaction from the momentum and spin dependent part. The frequency part of the pairing interaction is determined by the uniform spin susceptibility is of the spin-fluctuating insulator $\chi(\omega)$. We can write the pairing interaction as $V_{\mu\nu}({\hat{\bm k}},{\hat{\bm k}}'; \omega_n)=V_0 {\hat V}_{\mu\nu}({\hat{\bm k}},{\hat{\bm k}}')\chi(\omega_n)$. Correspondingly the gap can be factorized as $d_{\mu}({\hat {\bm k}}, \omega_n)={\hat d}_{\mu}({\hat{\bm k}})\Delta(\omega_n)$, with the gap magnitude encoded in $\Delta(\omega_n)$.
The frequency dependent part of the gap equation
\begin{equation}
\Delta(\omega_n)=N_0V_0T\sum_{n'}\chi(\omega_n-\omega_{n'})  \frac{\Delta(\omega_{n'})}{|\omega_{n'}|},
\end{equation}
then determines $T_c$. 

Consider a general Debye type relaxation for the magnetic substrate, one has
\begin{equation}
 \chi(\omega)\sim \frac{1}{1-i\omega\tau},
\end{equation}
 with the relaxation rate $\tau^{-1}$. 
Then we carry out the analytic continuation to the Matsubara frequency domain,
\begin{equation}
 \chi(\omega_n)= -\frac{1}{\pi}\int d\nu\frac{\chi''(\nu)}{i\omega_n-\nu},
\end{equation}
where $\chi''$ denotes the imaginary part of $\chi$. The result is
\begin{equation}
 \chi(\omega_n)\sim \frac{1}{1+|\omega_n|\tau},
\end{equation}
which is an even function of $\omega_n$.
We can then substitute $\chi(\omega_n)$ into the frequency part of the gap equation
\begin{equation}
\Delta(\omega_n)=T\sum_{n'}\frac{N_0V_0}{1+|\omega_n-\omega_{n'}|\tau} \frac{\Delta(\omega_{n'})}{|\omega_{n'}|}.
\label{Eq:Delta}
\end{equation}
This can be compared with the case of phonon mediated pairing. The dominant scattering processes for Cooper pairing occur around momentum transfer $q\simeq 2k_F$. In this region, the phonon spectrun can be approximated by the Einstein form, and the phonon propagator is  $D({\bm q},\omega_n)=\omega_0/(\omega_n^2+\omega_0^2)$, with the typical phonon frequency $\omega_0$, which is of the order of the Debye frequency. One can see that the relaxation rate $\tau^{-1}$ plays the role of phonon frequency $\omega_0$.

The superconducting $T_c$ can be obtained by solving the integral equation (\ref{Eq:Delta}). To estimate the energy scale of $T_c$, one can use a BCS type approximation [\onlinecite{BCS57}]. Since the pairing interaction is largely suppressed for frequencies larger than the relaxation rate $\tau^{-1}$, we can truncate frequency at $\omega_c$, with $\omega_c$ of order $\tau^{-1}$, and only sum over frequences $|\omega_n|, |\omega_{n'}|<\omega_c$. The resulting $T_c$ equation is of the well-known BCS form $T_c\sim \omega_c e^{-1/\lambda}$, with the effective coupling $\lambda=N_0V$. The strength of pairing interaction can be determined from $V\sim J_K^2\langle S_iS_j\rangle$. With the exchange interaction $J_{\rm ex} S_iS_j$, one has $\langle S_iS_j\rangle\sim 1/J_{\rm ex}$, and hence $V\sim J_K^2/J_{\rm ex}$, and 
\begin{equation}
T_c\sim \tau^{-1}\exp( -E_F J_{\rm ex}/J_K^2).
\end{equation}
Therefore $T_c$ critically depends on having appreciable scales of relaxation rate and interfacial Kondo coupling.

\section*{SII: Candidate materials for the spin-fluctuating insulator}

\begin{table}[t]
    \begin{tabular}{ | c | c | c | c |c | c | c |c|}
    \hline
Material &    magnetic ion  & moment  & magnetic lattice & exchange interaction & $T^*$ bound & reference  \\ \hline
 Pr$_2$Zr$_2$O$_7$,  Pr$_2$Sn$_2$O$_7$ & Pr$^{3+}$ & $j=4$ & pyrochlore & $J\sim 1.4$K &20 mK & [\onlinecite{Kimura13}] \\ \hline
 Yb$_2$Ti$_2$O$_7$& Yb$^{3+}$ & $j=7/2$ & pyrochlore & $J\sim -0.65$K  & 30 mK & [\onlinecite{Thompson11, Ross11}] \\ \hline
Tb$_2$Ti$_2$O$_7$& Tb$^{3+}$ & $j=6$ & pyrochlore & $J\sim 14$K & 50 mK & [\onlinecite{Gardner99}] \\ \hline
ZnCu$_3$(OH)$_6$Cl$_2$ & Cu$^{2+}$ & $s=1/2$ & kagome & $J\sim 200$K  & 20 mK &  [\onlinecite{Helton07, Han12}] \\ \hline
[NH$_4$]$_2$[C$_7$H$_{14}$N][V$_7$O$_6$F$_{18}$] & V$^{4+}$ & $s=1/2$ & kagome & $J\sim 60$K  & 40 mK &  [\onlinecite{Clark13}] \\ \hline
AA$'$VO(PO$_4$)$_2$ & V$^{4+}$ & $s=1/2$ & square & $-J_1\sim J_2\sim 5$K & 0.4 K & [\onlinecite{Nath08, Tsirlin09}]   \\ \hline
Na$_4$Ir$_3$O$_8$ & Ir$^{4+}$ & $s=1/2$ & hyperkagome & $J\sim 300$K  & 7 K& [\onlinecite{Okamoto07, Shockley15, Norman10, Micklitz10}] \\ \hline
NiGa$_2$S$_4$ & Ni$^{2+}$ & $s=1$ & triangular & $J\sim 80$K  & 10 K& [\onlinecite{Nakatsuji05, MacLaughlin10, Stock10}] \\ \hline
$\kappa$-(ET)$_2$Cu$_2$(CN)$_3$ & ET dimer & $s=1/2$ & triangular & $J\sim 250$K & 32 mK& [\onlinecite{Shimizu03}]\\ \hline
    \end{tabular}\par
\caption{List of candidate materials for spin-fluctuating insulator. Magnetic lattice represents the lattice of magnetic ions. $T^*$ bound denotes the lowest temperature at which neither long-range magnetic ordering nor spin freezing has been detected experimentally.}
\end{table}

As discussed in the main text, the criteria for the material choice for the spin-fluctuating insulator are (1) no long range order, and (2) having strong dynamic spin fluctuations. Practically one can only expect the candidate material to have such properties in certain temperature window $T^*<T<J_{\rm ex}$, where the exchange interaction $J_{\rm ex}$ is on the order of the Curie-Weiss temperature, and $T^*$ represents the temperature at which other effects set in to destroy the paramagnetic spin correlations. Below $T^*$, the material can develop magnetic ordering, or the spins may just freeze out. What is crucial is that the superconducting $T_c$, which is a function of $J_{\rm ex}$, should be much higher than $T^*$, i.e. $T_c(J_{\rm ex})\gg T^*$, so that there is a finite temperature range $T^*<T\leq T_c$ where the material superconducts.

A list of candidate materials that have such desired properties is included in Table I. All these materials are magnetic insulators with strong frustration, and are usually termed quantum spin liquids. For these materials, $T^*$ is usually at least two orders of magnitude smaller than the exchange interaction, providing sufficiently large temperature window that can accommodate superconductivity.
The first class of these materials are the quantum spin ice materials: Pr$_2$Zr$_2$O$_7$,  Pr$_2$Sn$_2$O$_7$, Yb$_2$Ti$_2$O$_7$, Tb$_2$Ti$_2$O$_7$ (see [\onlinecite{Gingras14}] for a comprehensive review).  They involve strong spin-orbit coupling and relatively large magnetic moment. Here we use the term quantum spin ice in a broader sense as quantum mechanical generalization of classical spin ice, which does not necessarily imply the existence of photon like excitations. Classical spin ice materials will not generate superconductivity, since their relaxation rate is extremely low, and hence the resulting superconducting $T_c$ is too low to be of any relevance.  One extra merit of quantum spin ice materials is that since their moment is relatively large, the conduction electrons from the metallic layer can not easily Kondo screen them. Hence their magnetic properties are more stable. Another class of materials are quantum spin liquids with small magnetic moments, mostly $s=1/2$ (see [\onlinecite{Balents10}] for a comprehensive review). Among them, herbertsmithite ZnCu$_3$(OH)$_6$Cl$_2$ is a spin-$1/2$ kagome-lattice antiferromagnet, and a prominent candidate for quantum spin liquid [\onlinecite{Helton07, Han12}]. High quality single crystals are available, and neutron scattering has been performed, showing a continuum of spin excitations at low temperatures [\onlinecite{Han12}]. The large exchange energy scale of herbertsmithite can potentially lead to a high superconducting transition temperature in the heterostructure. A third class of materials are the organic spin liquids. We expect it is technically more challenging to fabric superconducting heterostructures using these materials.

\section*{SIII: The metal/quantum-spin-ice heterostructure}

In this section, we include a detailed study of a concrete example of metal/QSL heterostructure. The insulator part is taken to be a candidate material for quantum spin ice, namely Pr$_2$Zr$_2$O$_7$, for which thermodynamic and neutron scattering data are available [\onlinecite{Kimura13}]. We first present the modeling of spin correlations for this material (SIII A). Then we search for the proper metallic layer that provides a good match for Pr$_2$Zr$_2$O$_7$, and density functional theory (DFT) is employed to calculate the band structure of the resulting heterostructure (SIII B). Having the two parts of the heterostructure ready, we then proceed to study the coupling between the two parts (SIII C). Finally we study the pairing problem in this heterostructure, and determine the dominant pairing channel (SIII D).

\subsection*{A. ~ The insulator: spin correlations in quantum spin ice Pr$_2$Zr$_2$O$_7$}

One can see from Eq.(\ref{Eq:Vpair}) that the pairing interaction in the metal/spin-fluctuating-insulator heterostructure depends crucially on the spin-spin correlation function $\langle S_i^a(\tau) S_j^b(\tau')\rangle$ of the spin fluctuating insulator. Hence to study the pairing problem in a microscopic setup, we need to obtain the spin-spin correlation function in the lattice spin system, which in principle can be obtained from the neutron scattering measurements. However for the purpose of theoretical calculations, it is valuable to obtain the correlation functions in closed analytic forms. Having in mind the experimental results for the spin-spin correlation function in quantum spin ice (QSI) material Pr$_2$Zr$_2$O$_7$ [\onlinecite{Kimura13}], here we will take a semi-phenomenological approach to model the spin-spin correlation function in QSI. 

The elastic neutron spectrum of QSI [\onlinecite{Kimura13}] is essentially the same as that of classical spin ice [\onlinecite{Bramwell01}], showing clear pinch point structures. Hence we start by modeling the spatial/momentum part of the spin-spin correlation function in QSI by that of classical spin ice, the analytic form of which is known [\onlinecite{Garanin99, Garanin01, Isakov04}]. As is manifest from the smearing of the pinch points in the inelastic neutron spectrum [\onlinecite{Kimura13}], quantum fluctuations play important roles in QSI, the effect of which is captured by the inclusion of the correlation length and relaxation rate. The relaxation rate determines the frequency dependence of the resulting pairing interaction, and will determine the superconducting $T_c$. The correlation length controls the momentum dependence of the pairing interaction, which enters the calculation of the pairing symmetry.

The analytic form of the spin-spin correlation function for classical spin ice can be obtained in the large-$N$ limit, where $N$ is the number of spin components. For pyrochlore magnets, this method was first developed in [\onlinecite{Garanin99, Garanin01}]. Generalizing the Ising spins $\sigma_i$ to $O(N)$ spins $\phi^\alpha_i$, with $\alpha=1,\cdots, N$, the spin Hamiltonian can be written as [\onlinecite{Garanin99, Garanin01}]
\begin{equation}
H_S=\frac{J}{2}\sum_{\langle ij\rangle}\sum_{\alpha=1}^{N}\phi^\alpha_i\phi^\alpha_j,
\end{equation}
with the constraint $\sum_{\alpha=1}^N \phi_i^\alpha\phi_i^\alpha=N$. Explicit forms of the spin correlators can be found in [\onlinecite{Isakov04}]
\begin{eqnarray}
\langle \phi_1^\alpha({\bm q})\phi_1^\beta(-{\bm q}) \rangle&=&2\delta_{\alpha\beta}\frac{3-{\bar c}^2_{xy}-{\bar c}^2_{xz}-{\bar c}^2_{yz}}{3-Q},\nonumber\\
\langle \phi_1^\alpha({\bm q})\phi_2^\beta(-{\bm q}) \rangle&=&2\delta_{\alpha\beta}\frac{\cos\left( \frac{q_y}{2}\right){\bar c}_{xz}-c_{xz}}{3-Q},
\label{Eq:correlator}
\end{eqnarray}
where $c_{ab}=\cos\frac{q_a+q_b}{4}$, ${\bar c}_{ab}=\cos\frac{q_a-q_b}{4}$, and $Q=c^2_{xy}+c^2_{yz}+c^2_{xz}+{\bar c}^2_{xy}+{\bar c}^2_{yz}+{\bar c}^2_{xz}-3$. The long wavelength limit of the spin structure factor obtained from Eq.(\ref{Eq:correlator}) is of the form [\onlinecite{Henley05, Isakov04}]
\begin{equation}
S_{ab}({\bm q})\sim  \delta_{ab}-\frac{q_a q_b}{q^2}.
\label{Eq:SSr}
\end{equation}
which is long-ranged and strongly anisotropic, and is the $\xi\to\infty$ limit of Eq.(5) of the main text.

\begin{figure}
\begin{centering}
\includegraphics[width=0.5\linewidth]{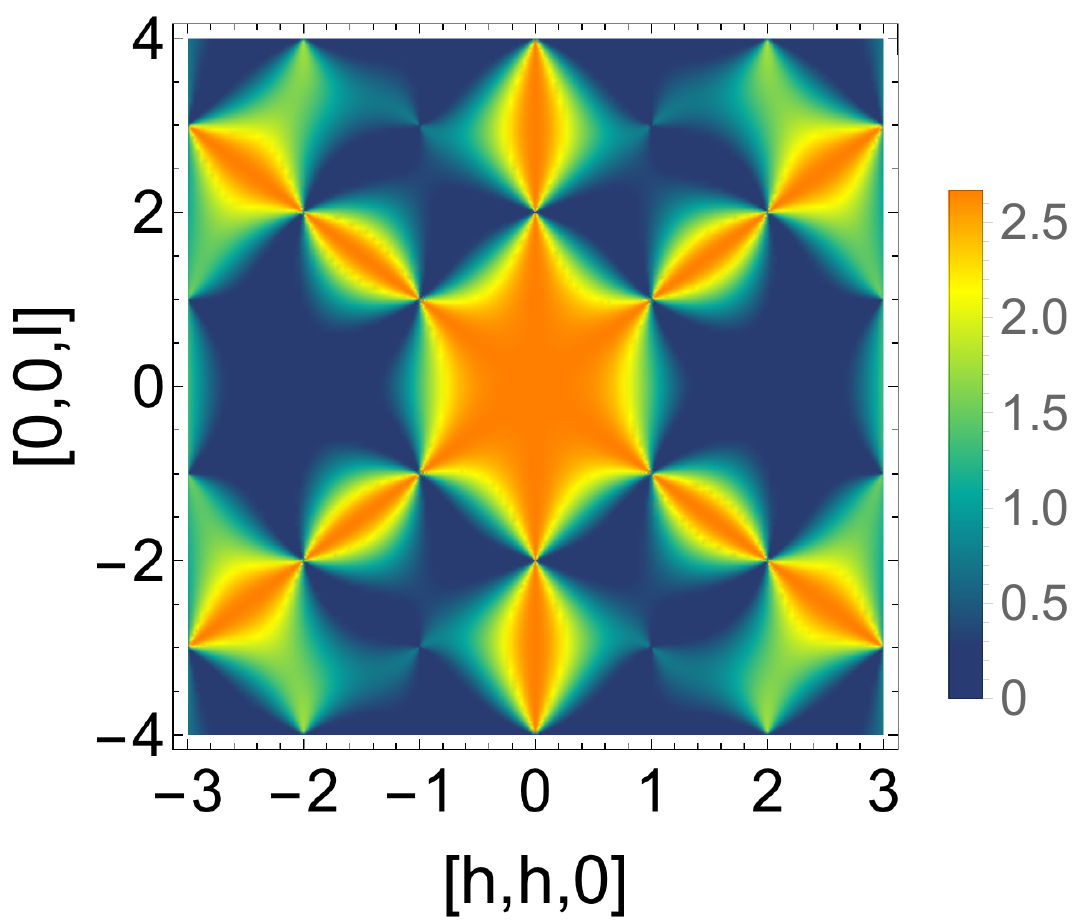} 
\end{centering}
\caption{(Color online) Momentum depenence of the static spin structure factor in the plane where the wavevector is of the form [h, h, l], as obtained from the large-$N$ calculation. One can see clearly the pinch-point structure. We also note that there is substantial spectral weight near ${\bm Q}=0$.}
\label{Fig:neutron}
\end{figure}

In the neutron scattering experiments, the incident neutron polarization is parallel to the crystalline $[1{\bar 1}0]$ axis, which determines the direction of $S_z$. $S_x$ direction is chosen to be parallel to the wavevector $(h, h, l)$. Hence we have the unit vectors ${\hat e}_z=\frac{1}{\sqrt{2}}(1, -1, 0)$, ${\hat e}_x=\frac{1}{\sqrt{2h^2+l^2}}(h, h, l)$, ${\hat e}_y=\frac{1}{\sqrt{2l^2+4h^2}}(l, l, -2h)$. The Ising spin directions ${\hat z}_{1-4}$ now need to be written in the basis of ${\hat e}_{x,y,z}$. With ${\vec S}_A\sim \phi_A{\hat z}_A$, we have 
\begin{eqnarray}
\langle S_A^a({\bm q}) S_B^b(-{\bm q}) \rangle\sim {\hat z}_A^a {\hat z}_B^b \langle \phi_A({\bm q})\phi_B(-{\bm q}) \rangle.
\end{eqnarray}
To compare with the experimental results, let us consider the spin flip spin structure factor in the $(h, h, l)$ plane, which can be shown to be of the form
\begin{eqnarray}
S_{yy}({\bm q})&=&\sum_{AB}\langle S_A^y({\bm q}) S_B^y(-{\bm q}) \rangle \sim \frac{64 \left[q_x (\cos\frac{q_x}{4} + \cos\frac{q_z}{4}) \sin\frac{q_x}{4} + q_z \cos\frac{q_x}{4} \sin\frac{q_z}{4}\right]^2}{3 (2 q_x^2 + q_z^2) (5 - \cos q_x - 4 \cos\frac{q_x}{2} \cos\frac{q_z}{2})}.
\label{neutron}
\end{eqnarray}
The result is plotted in Fig.~\ref{Fig:neutron}.
This reproduces the elastic neutron scattering results for classical and quantum spin ice [\onlinecite{Fennell09, Kimura13}] in the full momentum space.

For the metal/QSI heterostructure, we use the coordinate system with the $[1,{\bar 1},1]$ plane as the $xy$ plane. The basis vectors are ${\hat\kappa}_x=\frac{1}{\sqrt{2}}(1, 1, 0)$, ${\hat\kappa}_y=\frac{1}{\sqrt{6}}(-1, 1, 2)$, ${\hat\kappa}_z=\frac{1}{\sqrt{3}}(1, -1, 1)$. In the $\hat\kappa$ coordinate system, the spins are in the directions ${\hat z}_1=\frac{1}{3}(-\sqrt{6}, -\sqrt{2}, -1)$, ${\hat z}_2=(0, 0, 1)$, ${\hat z}_3=\frac{1}{3}(\sqrt{6}, -\sqrt{2}, -1)$, ${\hat z}_4=\frac{1}{3}(0, 2\sqrt{2}, -1)$. The spin structure factors can be obtained in this coordinate system by substituting $q_x\to \frac{q_x}{\sqrt{2}}- \frac{q_y}{\sqrt{6}}+ \frac{q_z}{\sqrt{3}}$, $q_y\to \frac{q_x}{\sqrt{2}}+ \frac{q_y}{\sqrt{6}}- \frac{q_z}{\sqrt{3}}$, $q_z\to \frac{2q_y}{\sqrt{6}}+ \frac{q_z}{\sqrt{3}}$.

\subsection*{B. ~ The metal: material choices and DFT calculations}

With the insulator part chosen to be Pr$_2$Zr$_2$O$_7$, we now search for proper metallic materials that can be deposited on top of it.  We try to keep the connection between the real materials and the minimal theoretical model that we constructed as tight and realistic as possible for definiteness and for making progress of the basic idea. We look for metals in which the conduction electrons are from $s$-orbitals, since $s$-orbitals have several merits: (1) large bandwidth, and hence large Fermi energy, which is good for superconductivity; (2) weak correlation, which helps to avoid ordering by itself; (3) non-degenerate bands, which helps to provide odd numbers of Fermi surfaces to generate topological superconductivity. We first searched for elemental metals with fcc structure, such as alkaline earth metals and some of the transition metals, that can match with the cubic pyrochlore lattice structure of Pr$_2$Zr$_2$O$_7$. Unfortunately, they have complicated band structure, and the Fermi surfaces do not have the desired peroperties. We then searched for stoichiometric metallic pyrochlores with chemical formula A$_2$B$_2$O$_7$, e.g. Bi$_2$Ru$_2$O$_7$, Bi$_2$Ir$_2$O$_7$. However, the conduction electrons in these materials have strong $d$-electron component, and hence correlation effects play important roles, which can give rise to local moments or large mass renormalization. Actually to the best of our knowledge, there is no stoichiometric metallic pyrochlore with $s$-electrons forming the conduction band. 

\begin{figure}
\includegraphics[width=0.9\textwidth]{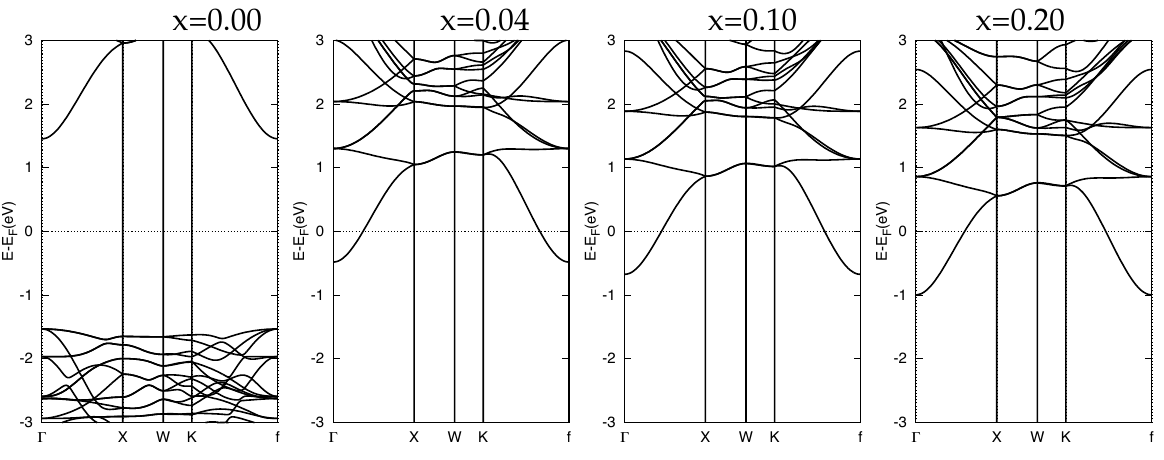}
\caption{Evolution of the band structure of Y$_2$Sn$_{2-x}$Sb$_x$O$_7$ with Sb doping from DFT calculation. }
\label{Fig:YSO}
\end{figure}

Then we turn to the case of doping an insulating pyrochlore to make it metallic. We choose Sn as the B-site in the parent compound A$_2$B$_2$O$_7$ since it can provide $s$-electrons. 
 Both La and Y, which form $3+$ cations, can be the A-site. Hence two possible parent compounds are La$_2$Sn$_2$O$_7$ and Y$_2$Sn$_2$O$_7$. We can dope either the A-site or the B-site. Ce with robust $4+$ cation can be used to dope the A-site, and Sb can be used to dope the B-site. Actually DFT calculations show that Ce-doped La$_2$Sn$_2$O$_7$, Sb-doped La$_2$Sn$_2$O$_7$ and Sb-doped Y$_2$Sn$_2$O$_7$ have very similar band structures near the Fermi level. At doping levels around $x=0.2$, they all have the desired electronic properties for our heterostructure. The evolution of the band structure of Y$_2$Sn$_{2-x}$Sb$_x$O$_7$ with Sb doping is shown in Fig.~\ref{Fig:YSO}.

We have performed DFT calculations on the heterostructure Pr$_2$Zr$_2$O$_7$/ Y$_2$Sn$_{2-x}$Sb$_x$O$_7$ (see Fig.3A in main text).
For the structural relaxations, we used PBEsol approximation using projector augmented-wave potentials, as implemented in the Vienna ab initio simulation package (VASP) \cite{Kresse1996,Kresse1999}. 
Structural optimization was performed with a force criterion of 0.001eV/\AA.
To find lattice parameters of Y$_2$Sn$_2$O$_7$ films which is grown on top of Pr$_2$Zr$_2$O$_7$, we performed `strained-bulk' calculations.
With the theoretical in-plane lattice constant of Pr$_2$Zr$_2$O$_7$ (spacegroup $Fd\bar{3}m$), 10.686\AA,
we optimized $c$-axis lattice constant of Y$_2$Sn$_2$O$_7$ bulk in (111) oriented geometry.
Once we have lattice parameters of Y$_2$Sn$_2$O$_7$, we then construct (111)-oriented (Pr$_2$Zr$_2$O$_7$)$_{16}$/(Y$_2$Sn$_{2-x}$Sb$_x$O$_7$)$_2$ superlattice.
For the calculation of the electronic properties of the superlattice,
we employed the linear-combination-of-pseudo-atomic-orbitals (LCPAO) method as implemented in OpenMX \cite{openmx}.
To describe the Sb-doping, a virtual atom is treated with fractional nuclear charge by using a pseudopotential with the corresponding fractional nuclear charge.
And the $4f$ state of Pr is treated as core state by considering occupation of two electrons, and thereby not involed in calculations explicitly. The resulting electronic properties of the superlattice are shown in Fig.3B and Fig.3C of the main text.

\subsection*{C. ~ The interfacial coupling: estimation from microscopic model}

We consider next the coupling between the metallic layer and the insulating substrate. There is extra complication due to the fact that the ground state of Pr$^{3+}$ is a non-Kramers doublet. The Ising component of Pr moment along the local $[111]$ direction forms magnetic dipole moment $\tau^z\sim J^z$, and the planar components form quadrupole moments $\tau^{\pm}\sim \left\{ J^z, J^{\pm}\right\}$. Under time reversal, one has $\tau^z\to -\tau^z$, and $\tau^{\pm}\to \tau^{\pm}$. Therefore the Ising component $\tau^z$ couples to the spin density of the conduction electrons $s_m^a=\sum_{\alpha\beta}c^\dagger_{m\alpha} \sigma^a_{\alpha\beta}c_{m\beta}$, i.e. a Kondo type coupling, and the planar components $\tau^{\pm}$ couple to the charge density of the conduction electrons $\rho_m=\sum_{\alpha}c^\dagger_{m\alpha} c_{m\alpha}$ [\onlinecite{Chen12, Lee13}]. The couplings depend crucially on the symmetry properties of both the ground state atomic configuration and the lowest excited state atomic configuration at Pr Site, which give rise to several selection rules [\onlinecite{Cox98}]. Similarly neutron spin only couples to the Ising component of Pr moment, and hence what is measured in neutron scattering [\onlinecite{Kimura13}] is the spin susceptibility of the Ising component of Pr moment [\onlinecite{Onoda11}]. Therefore we have a good knowledge about the Ising spin correlations. The quadrupole-quadrupole correlations are currently not available experimentally. Here we will first focus on the Kondo coupling between the Ising moment and the conduction electron spin, and then show that the inclusion of planar quadrupole fluctuations with reasonable strength will not change the dominant pairing channel.

We now estimate the strength of the Kondo coupling from second-order perturbation theory. To benchmark this estimaiton, let us also estimate the Kondo coupling in the closely related material Pr$_2$Ir$_2$O$_7$ for which the Kondo coupling can be compared with experiments. The local environments on Pr$^{3+}$ sites are essentially the same for both systems. There are two $4f$ electrons at Pr$^{3+}$. If we neglect the nonsphericity of the Coulomb and exchange interaction (which is exact in the case of the fully occupied or empty orbital), the local atomic Hamiltonian has the form:
\begin{align}
  \mathcal{H}_{f} = \sum_{imm'} \epsilon^f_{mm'} f^\dag_{im}f_{im'} + \lambda_{\rm SO} \sum_{i}  \mathbf{L}_i \cdot \mathbf{S}_i
  + \frac12 \sum_{imm',\alpha\neq\beta} U n_{im\alpha} n_{im'\beta} + \frac12  \sum_{im\neq m', \alpha} (U - J_H) n_{m\alpha}n_{m'\alpha},
\end{align}
where $i$ is the site index; $m, m'$ are the orbital quantum number; $\alpha=\uparrow,\downarrow$ is the spin index; $n_{im\alpha}=f^\dagger_{im\alpha}f_{im\alpha}$ is the $f$-electron density; $\epsilon_f$ is the effective one-electron energies; $\lambda_{\rm SO}$ is the spin-orbit coupling energy; $U$ is the Hubbard interaction; $J_H$ is the Hund's rule coupling. The dominant energy scales are the Hund's coupling $J_H\sim$ 1 eV [\onlinecite{Norman1995}] and the Hubbard interaction $U\sim$ 4 eV [\onlinecite{Norman1995}] (crystal field splitting $\sim$ 40 meV [\onlinecite{Zhou08}], spin-orbit coupling $\lambda_{\rm SO}\sim$ 0.25 eV [\onlinecite{handbook}]).

\begin{figure}
\includegraphics[width=0.9\textwidth]{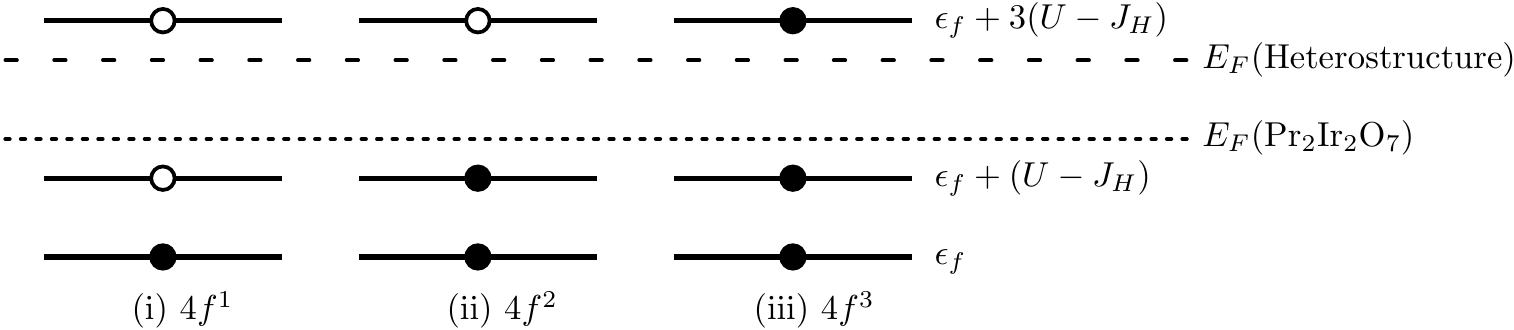}
\caption{Energy level diagram of metal/QSI heterostructure and Pr$_2$Ir$_2$O$_7$ for Pr site in 4$f^1$, 4$f^2$ and 4$f^3$ states. The filled circles represent filled levels, and the empty circles  represent empty levels. The dashed lines denote the corresponding Fermi energy. }
\label{diagram}
\end{figure}

The Kondo coupling arises from the overlap of the conduction electron and the local $4f$ electron wavefunctions. The hybridization Hamiltonian reads
\begin{eqnarray}
H_{cf}=\sum_{ij\alpha}\left(V^{cf}_{ij}f^\dagger_{i\alpha}c_{j\alpha}+h.c.\right),
\end{eqnarray} 
with the transfer integral $V_{cf}$. The leading order contribution to the Kondo coupling arises from the following 
two processes: (1) an electron hops from the conduction electron site (5s orbitals of Sn, 5d orbitals of Ir) to the $4f$ orbitals of Pr and then hops back, (2) an electron hops from $4f$ orbitals of Pr to the conduction electron site and then hops back. From the energy level diagram Fig.\ref{diagram}, one can see that these two processes give rise to the coupling 
\begin{equation}
J_K\sim V^2_{cf}\left[ \frac{1}{\epsilon_f+3(U-J_H)-E_F}+\frac{1}{E_F-\epsilon_f} \right] \sim V^2_{cf}\frac{3(U-J_H)}{\Delta_{cf}[3(U-J_H)-\Delta_{cf}]},
\end{equation}
with the charge transfer energy $\Delta_{cf}\equiv E_F-\epsilon_f$. $\Delta_{cf}$ can be estimated from the band structure, with $\Delta_{cf}({\rm Hetero})\sim$ 8.5 eV, and $\Delta_{cf}$(Pr$_2$Ir$_2$O$_7$)$\sim$ 3.5 eV.

To estimate $V_{cf}$ we have performed DFT calculations of (1) one-dimensional chains with alternating Pr and Sn sites and (2) one-dimensional chains with alternating Pr and Ir sites. The lattice constants in these chains are chosen to be the same as in the corresponding pyrochlore structures. We constructed Wannier function to fit DFT band structure. With this constructed Wannier function, we can obtain tight-binding matrix element of $4f$(Pr)-$5d$(Ir) and $4f$(Pr)-$5s$(Sn) orbitals. As a result, we confirmed that both transfer integrals are about 0.05-0.08eV. This is one order of magnitude smaller than the estimates of hybridization parameter $V$ for Anderson impurity model of transition metal impurity which are on the order of 0.1-0.7 eV [\onlinecite{handbook}].



Using the above parameters, the Kondo coupling for Pr$_2$Ir$_2$O$_7$ can be estimated to be $J_K$(Pr$_2$Ir$_2$O$_7$)$\sim$ 12-30 K. This is of the same order as the estimated $J_K\sim 45 K$ using the relation $J_{\rm RKKY}\sim J_K^2/E_F$,
with the RKKY interaction $J_{\rm RKKY}\sim$ 20 K determined from the Curie-Weiss temperature [\onlinecite{Nakatsuji06}], and the Fermi energy $E_F\sim$ 10 meV with mass renormalization $m^*\sim 20 m_e$.  For the heterostructure, one obtains $J_K({\rm Hetero})\sim$ 53-136 K, and $\lambda=J_K^2/(E_FJ_{\rm ex})\sim$ 0.9-6, which is of order one. We emphasize again that this is just a rough estimate of the order of magnitude, and the real materials are much more complicated.

\section*{SIV: Pairing in the metal/quantum-spin-ice heterostructure}

We include here the detailed study of the pairing problem in the metal/QSI heterostructure. The pairing problem can be treated using the standard BCS type approach, by diagonalizing the pairing interaction matrix to find the most negative eigenvalue. However the essential physics can be most clearly demonstrated by considering a simplified model with magnetic dipole-dipole interaction. We will first study such a simplified model analytically to gain insight into the pairing mechanism, and then proceed to study the effects of microscopic details using the BCS approach.

\subsection*{A. ~ Magnetic dipole-dipole interaction}

The spin-spin correlation function of QSI gives rise to a long-ranged, anisotropic spin exchange interaction among the conduction electrons. In the limit of large correlation length, the induced interaction is asymptotically of the form of the magnetic dipole-dipole interaction at large separations. As we show below, such an interaction vanishes identically in the even-parity spin-singlet pairing channels. Thus pairing only occurs in the odd-parity spin-triplet channels.

\subsubsection*{1. Selection rules}

To better understand the pairing problem, we consider first the quantum mechanical problem of two electrons interacting via the magnetic dipole-dipole interaction: 
\begin{equation}
V_{\rm dd}({\bm r})\sim\frac{1}{r^3}\left[ {\vec S}_1\cdot{\vec S}_2-3({\vec S}_1\cdot{\hat{\bm r}})({\vec S}_2\cdot{\hat{\bm r}}) \right],
\end{equation} 
 where ${\vec S}_{1, 2}$ represent the spin operators of the two electrons. A generic two body interaction between spinful particles can be decomposed in terms of the irreducible components of a general rank two tensor, which consist of a rank-zero scalar interaction ($\propto {\vec S}_1\cdot{\vec S}_2$), a rank-one vector interaction ($\propto {\vec L}\cdot {\vec S}$, with the relative orbital angular momentum ${\vec L}$ and the total spin ${\vec S}={\vec S}_1+{\vec S}_2$) and a traceless and symmetric rank-two tensor interaction [\onlinecite{Brink}]. The magnetic dipole-dipole interaction has the unique property that it is a rank-two tensor interaction, and it carries magnetic quadrupole moment. We can then write $V_{\rm dd}({\bm r})$ in the form  [\onlinecite{Brink}]
\begin{equation}
V_{\rm dd}({\bm r})\sim{\cal R}^{(2)}({\bm r}_1, {\bm r}_2)\cdot {\cal S}^{(2)}({\bm s}_1, {\bm s}_2),
\label{Eq:Vrank2}
\end{equation} 
where both ${\cal R}^{(2)}$ and ${\cal S}^{(2)}$ are rank-two tensors, with ${\cal R}^{(2)}$ acting on coordinate space, and ${\cal S}^{(2)}$ on spin space. In particular, ${\cal S}^{(2)}$ represents the magnetic quadrupole moment. Here $\cdot$ denotes the scalar product of two tensors.
The rank-two character of $V_{\rm dd}$ gives rise to selection rules that dictate the possible ground states of the two-body system. In particular, the spin tensor ${\cal S}^{(2)}$ determines the spin part of the ground state wavefunction. 
 
Let us consider the spin part of the two electron Hilbert space. The two electrons can form a spin singlet state with total spin $S=0$: $\frac{|\uparrow\downarrow\rangle-|\downarrow\uparrow\rangle}{\sqrt{2}}\equiv|S=0 \rangle$. And they can form spin triplet states with $S=1$: $|\uparrow\uparrow\rangle\equiv|S=1, S_z=1\rangle$, $|\downarrow\downarrow\rangle\equiv|S=1, S_z=-1\rangle$, $\frac{|\uparrow\downarrow\rangle+|\downarrow\uparrow\rangle}{\sqrt{2}}\equiv|S=1, S_z=0\rangle$. As parity is a good quantum number, the spin singlet states, which have even parity, and the spin triplet states, which have odd parity, can not be transformed into each other by the interaction term, i.e. $\langle S=0|V_{\rm dd}|S=1, S_z\rangle=0$. The question is then which states can have nonzero expectation value under the interaction $V_{\rm dd}$.

This question is answered by a powerful theorem of quantum mechanics, namely the Wigner-Eckart theorem [\onlinecite{Sakurai}]. To set the stage, let us first review the well-known example of forbidden transition in atomic hydrogen [\onlinecite{Sakurai}]. Consider a single  atomic electron interacting with electromagnetic field via the coupling $\frac{e}{m}{\vec A}\cdot {\vec p}$. Within the electric dipole approximation, the transition amplitude from an initial state with quantum numbers $l$ and $m$ to a final state with quantum numbers $l'$ and $m'$ is $\langle l'm' |e{\hat{\bm \varepsilon}}\cdot {\vec p}|lm \rangle$, where ${\hat{\bm \varepsilon}}$ is the polarization direction of the vector potential ${\vec A}$. Since ${\bm p}\sim [H_0, {\bm r}]$, where $H_0=p^2/2m$ is the free electron Hamiltonian, the transition amplitude is proportional to
\begin{equation}
 \langle l'm' |e{\hat{\bm \varepsilon}}\cdot {\bm r}|lm \rangle,
\end{equation}   
from which one recognizes the electric dipole operator $e{\bm r}$. Since the dipole operator is a vector, i.e. rank-one tensor, Wigner-Eckart theorem leads to the selection rule (``triangular relation'')
\begin{equation}
|1-l|\leq l' \leq 1+l.
\end{equation}   
Hence for example the $2s\to 1s$ transition with $l=l'=0$ is forbidden, while the $2p\to 1s$ transition with $l=1$, $l'=0$ is allowed.

The above result of single particle transition under the electric dipole operator can be easily generalized to our case of two electron transition under the magnetic quadrupole operator. According to the Wigner-Eckart theorem, the matrix elements of the rank-two tensor interaction $\langle S'S'_z|{\cal S}^{(2)}|SS_z\rangle$ are nonzero only if the following selection rule is satisfied: 
\begin{equation}
|2-S|\leq S' \leq 2+S.
\end{equation}   
It follows immediately that even parity pairing states with $S=0$ have zero expectation value under the magnetic dipole-dipole interaction [\onlinecite{Brink}]
\begin{equation}
\langle S=0|V_{\rm dd}|S=0\rangle =0. 
\end{equation}   
The vanishing of magnetic dipole-dipole interaction in the even parity pairing channels has been pointed out in [\onlinecite{Li12}]. Odd parity pairing states which have total spin $S=1$ satisfy the above selection rule and hence can have nonzero matrix elements
\begin{equation}
\langle S=1, S'_z|V_{\rm dd}|S=1, S_z\rangle \neq 0. 
\end{equation}

\subsubsection*{2. Binding energy}

We then proceed to consider in more detail which odd parity pairing states are favored by $V_{\rm dd}$, i.e. have negative binding energy $E_{\rm dd}\equiv\langle V_{\rm dd}\rangle$.
We consider first the case of 3d magnetic dipole-dipole interaction [\onlinecite{Li12}], which has higher symmetry $SO(3)_{\bm J}$. The two electron states can be classified according to their $J$ value. Since the interaction does not contain appreciable higher harmonic components, we consider orbital angular momentum channel $L=1$. With $L=1$ and $S=1$, the total angular momentum can take values $J=0, 1, 2$. The corresponding pairing states can be easily obtained from the Clebsch-Gordan coefficients $\langle JJ_z|LL_zSS_z\rangle$. In particular, the $J=0$  state reads
\begin{eqnarray}
|J=0\rangle &=&\frac{1}{\sqrt{3}} \left(|L_z=1, S_z=-1\rangle +|L_z=-1, S_z=1\rangle -|L_z=0, S_z=0\rangle \right),
\end{eqnarray}
the $J=1$ states
\begin{eqnarray}
|J=1, J_z=0\rangle &=&\frac{1}{\sqrt{2}} \left(|L_z=1, S_z=-1\rangle -|L_z=-1, S_z=1\rangle \right) ,\\
|J=1, J_z=\pm 1\rangle &=&\frac{1}{\sqrt{2}} \left(|L_z=\pm 1, S_z=0\rangle -|L_z=0, S_z=\pm 1\rangle \right),
\end{eqnarray}
and similarly for the $J=2$ states. More explicitly, in terms of the spherical harmonic $Y_{lm}({\hat{\bm r}})$, one has  
\begin{eqnarray}
|J=0\rangle&=&\frac{1}{\sqrt{3}}\left(Y_{1, 1}|\downarrow\downarrow\rangle +Y_{1, -1}|\uparrow\uparrow\rangle-Y_{1,0}\frac{|\uparrow\downarrow\rangle+|\downarrow\uparrow\rangle}{\sqrt{2}}\right),
\end{eqnarray}
and 
\begin{eqnarray}
|J=1, J_z=0\rangle&=&\frac{1}{\sqrt{2}}\left(Y_{1, 1}|\downarrow\downarrow\rangle -Y_{1, -1}|\uparrow\uparrow\rangle\right),\\
|J=1, J_z=1\rangle&=&\frac{1}{\sqrt{2}}\left(Y_{1, 1}\frac{|\uparrow\downarrow\rangle+|\downarrow\uparrow\rangle}{\sqrt{2}}-Y_{1,0}|\uparrow\uparrow\rangle\right),\\
|J=1, J_z=-1\rangle&=&\frac{1}{\sqrt{2}}\left(Y_{1, -1}\frac{|\uparrow\downarrow\rangle+|\downarrow\uparrow\rangle}{\sqrt{2}}-Y_{1,0}|\downarrow\downarrow\rangle\right).
\end{eqnarray}

To calculate the binding energy, we write Eq.(\ref{Eq:Vrank2}) explicitly in terms of the spherical harmonic $Y_{lm}({\hat{\bm r}})$ and the rank-two spin tensors $\Sigma_{2,m}$ [\onlinecite{Brink, Pethick}]
\begin{equation}
V_{\rm dd}({\bm r})=V_0 \sum_{m=-2}^2 Y^*_{2m}({\hat{\bm r}})\Sigma_{2,m}.
\end{equation}   
The spin tensors read [\onlinecite{Pethick}]
\begin{eqnarray}
\Sigma_{2,0}&=&-\sqrt{\frac{3}{2}}\left(S_{1z}S_{2z}-\frac{1}{3}{\vec S}_1\cdot{\vec S}_2\right),\nonumber\\
\Sigma_{2, \pm 1}&=&\pm\frac{1}{2}\left(S_{1z}S_{2\pm}+ S_{1\pm}S_{2z}\right),\nonumber\\
\Sigma_{2,\pm 2}&=&-\frac{1}{2}S_{1\pm}S_{2\pm}.
\end{eqnarray}
One can see that $\Sigma_{2,m}$ raises the $z$-component of the electron spin by $m$, while $Y^*_{2m}$ raises the $z$-component of the orbital angular momentum by $-m$, and hence their product $Y^*_{2m}\Sigma_{2,m}$ conserves $J_z$.

One can then check that the binding energy is negative in the $J=1$ triplet channel, and positive in the $J=0$ and $J=2$ channels [\onlinecite{Li12}]. In particular, in the $J=1$ channel the binding energy reads
 \begin{eqnarray}
E_{\rm dd}(|J=1, J_z=0, \pm 1\rangle)=-\frac{1}{4}\sqrt{\frac{15}{2\pi}}\frac{1}{3}V_0,
\end{eqnarray}
where the three $J_z$ channels are degenerate as required by symmetry. In fact, in the corresponding many-body pairing problem, the magnetic dipole-dipole energy of a spin triplet pairing state can be determined from its ${\vec d}$-vector as [\onlinecite{Leggett75}]
\begin{equation}
E_{\rm dd}\sim \int \frac{d\Omega_{\bm k}}{4\pi}\left( |{\vec d}({\hat{\bm k}})\cdot{\hat{\bm k}}|^2-|{\vec d}({\hat{\bm k}})|^2 \right),
\end{equation}
with the solid angle $\Omega_{\bm k}$ in momentum space. Here one assumes that pairing involves only a single orbital angular momentum (say $L=1$). The second term in the above equation is fixed by normalization $\int \frac{d\Omega_{\bm k}}{4\pi}|{\vec d}({\hat{\bm k}})|^2 =1$. The first term depends on the orientation of the ${\vec d}$-vector, and is minimized when
\begin{eqnarray}
 {\vec d}({\hat{\bm k}})\cdot{\hat{\bm k}}=0.
\end{eqnarray}
The ${\vec d}$-vector has 5 degrees of freedom (three complex variables modulo an overall factor). The $ {\vec d}({\hat{\bm k}})\cdot{\hat{\bm k}}=0$ condition imposes two constraints (real and imaginary parts). Hence we are left with $5-2=3$ independent degrees of freedom. It is thus a triplet state and hence $J=1$. One can check explicitly that the $J=1$ pairing states, with
the correspondingly ${\vec d}$-vectors 
\begin{eqnarray}
 {\vec d}(J=1, J_z=0)&=&\frac{1}{2} ({\hat k}_y, -{\hat k}_x, 0),\\
{\vec d}(J=1, J_z=\pm 1)&=&\frac{1}{2\sqrt{2}} (-{\hat k}_z, \mp i{\hat k}_z, {\hat k}_x\pm i{\hat k}_y),
\end{eqnarray}
 satisfy the above transverse condition, and hence minimizes the dipole energy.

Going from the above 3d case to the 2d interface, the $SO(3)_{\bm J}$ symmetry is reduced to $U(1)_{{\bm J}_z}$ symmetry. The binding energy can be similarly calculated, noting that at the interface $z=0$, and hence $Y_{1,0}\to 0$, $Y_{2,0}\to 0$. The resulting negative binding energy states are 
\begin{eqnarray}
|J_z=0\rangle&=&\frac{1}{\sqrt{2}}\left(Y_{1, 1}|\downarrow\downarrow\rangle -Y_{1, -1}|\uparrow\uparrow\rangle\right),\\
|J_z=\pm1\rangle &=&Y_{1,\pm 1}\frac{|\uparrow\downarrow\rangle+|\downarrow\uparrow\rangle}{\sqrt{2}}.
\end{eqnarray}
The states at the 2d interface can be regarded as descendants of the 3d case. In particular, the  negative binding energy states $|J_z=0, \pm 1\rangle$ are descendants of the 3d $J=1$ triplet states. These three states are still degenerate, with the same binding energy as in 3d:
 \begin{eqnarray}
E_{\rm dd}(|J_z=0, \pm 1\rangle)=-\frac{1}{4}\sqrt{\frac{15}{2\pi}}\frac{1}{3}V_0.
\end{eqnarray}
But while the degeneracy in 3d is protected by $SO(3)_{\bm J}$ symmetry, the degeneracy at the 2d interface is not protected by any symmetry, and hence will be lifted by perturbations.

\subsubsection*{3. Emergent gauge structure}

The appearence of novel superconductivity at the metal/QSI interface has its origin from the underlying emergent gauge structure of spin ice materials [\onlinecite{Henley05, Isakov04, Gingras14}]. The classical spin ice Hamiltonian possesses a largely degenerate ground state manifold, consisting of all Ising spin configurations that satisfy the ice rule condition, namely for the four spins on each tetrahedron, two point in towards the center of the tetrahedron and two point out. Such an ice rule condition is the lattice version of the zero divergence condition. The coarse-grained spin field ${\vec S}({\bm r})$ then satisfies the constraint \onlinecite{Henley05, Isakov04}
 \begin{eqnarray}
{\vec\nabla}\cdot{\vec S}({\bm r})=0.
\end{eqnarray}
One can then define an emergent gauge field ${\vec A}({\bm r})$ so that 
\begin{eqnarray}
{\vec S}({\bm r})={\vec\nabla}\times{\vec A}({\bm r}).
\end{eqnarray}
The gauge field propagator reads
\begin{eqnarray}
 \langle  A_a({\bm q})A_b(-{\bm q}) \rangle\sim \frac{1}{q^2}\left(\delta_{ab}-2{\hat q}_a{\hat q}_b\right),
\label{Eq:gauge}
\end{eqnarray}
from which one can obtian the spin-spin correlation function 
\begin{eqnarray}
 S_{ab}({\bm q})\equiv \langle  S_a({\bm q})S_b(-{\bm q}) \rangle\sim \delta_{ab}-{\hat q}_a{\hat q}_b.
\end{eqnarray}
Quantum fluctuations can lead to different types of dynamical spin correlations. A theoretically appealing possibility is the emergence of the full photon-like excitations [\onlinecite{Hermele04, Benton12}]. However in the particular QSI candidate material that we consider, namely Pr$_2$Zr$_2$O$_7$, the dynamics is actually simply relaxational as observed from neutron scattering [\onlinecite{Kimura13}]. Hence we will consider the full dynamical spin susceptibility to be of the form $\chi_{ab}({\bm q}, \omega)\sim \frac{1}{1-i\omega\tau}S_{ab}({\bm q})$.

The Kondo coupling can be written in the form 
\begin{eqnarray}
H_K=J_K\sum_{{\bm r}\alpha\beta}\psi^\dagger_{{\bm r}\alpha}{\vec \sigma}_{\alpha\beta}\psi_{{\bm r}\beta}\cdot\left[{\vec\nabla}\times{\vec A}({\bm r})\right].
\end{eqnarray}
For simplicity, let us first consider the coupling in the whole 3d space. A partial integration can be performed to yield 
\begin{eqnarray}
H_K=J_K\sum_{{\bm r}\alpha\beta}  {\vec A}({\bm r})\cdot \left({\vec \sigma}_{\alpha\beta}\times {\vec\nabla}\right) \psi^\dagger_{{\bm r}\alpha}\psi_{{\bm r}\beta},
\end{eqnarray}
i.e. the gauge field couples to a magnetic current. We can rewrite the coupling in momentum space as
\begin{eqnarray}
H_K=J_K\sum_{{\bm k}{\bm q}\alpha\beta}  {\vec A}({\bm q})\cdot \left({\vec \sigma}_{\alpha\beta}\times {\bm q}\right) \psi^\dagger_{{\bm k}+\frac{\bm q}{2},\alpha}\psi_{{\bm k}-\frac{\bm q}{2},\beta}.
\end{eqnarray}

The coupling here should be contrasted to the usual minimal coupling between fermions and gauge fields obtained from the momentum substitution: 
\begin{eqnarray}
H_J=\sum_{\bm q} {\vec j}({\bm q})\cdot {\vec A}({\bm q})=e\sum_{{\bm k}{\bm q}\alpha}  {\vec A}({\bm q})\cdot  \frac{\bm k}{m} \psi^\dagger_{{\bm k}+\frac{\bm q}{2},\alpha}\psi_{{\bm k}-\frac{\bm q}{2},\alpha}.
\end{eqnarray}
Choosing the Coulomb gauge where ${\vec\nabla}\cdot {\vec A}=0$, the gauge field mediated interaction among the fermions is [\onlinecite{PALee07,PALee14}]
\begin{eqnarray}
H_{\rm int}\sim -\sum_{{\bm p}_1{\bm p}_2{\bm q}\alpha} D(q)\frac{\left({\bm p}_1\times{\hat {\bm q}}\right)\cdot \left({\bm p}_2\times{\hat {\bm q}}\right) }{m^2} \psi^\dagger_{{\bm p}_1+{\bm q},\alpha}\psi_{{\bm p}_1,\alpha} \psi^\dagger_{{\bm p}_2-{\bm q},\beta}\psi_{{\bm p}_2,\beta},
\end{eqnarray}
with the propagator of the transverse gauge field $D(q)>0$. Such an interaction can not generate the usual BCS type pairing where two electrons in a Cooper pair have opposite momenta. Taking ${\bm p}_1\simeq -{\bm p}_2$, one can see that the above interaction is repulsive. This is indeed the well-known result of Ampere's force law: two antiparallel currents repel each other. To generate pairing using the current-current interaction mediated by gauge field, one thus needs to have parallel currents which attract each other, that is, to pair electrons with nearly equal momenta, i.e. ${\bm p}_1\simeq {\bm p}_2$. This is the idea of Ampere pairing [\onlinecite{PALee07,PALee14}]. It has been shown that when the gauge field propagator is singular enough, the gauge field mediated interaction can indeed generate Ampere type finite momentum Cooper pairing [\onlinecite{PALee07,PALee14}]. 
 
A similar analysis can be carried out for our problem. The Kondo coupling induced interaction  reads
\begin{eqnarray}
 H_{\rm int}\sim -J_K^2\sum_{{\bm p}_1{\bm p}_2{\bm q}} \sum_{ab}\sum_{\alpha\beta\alpha'\beta'}\langle  A_a({\bm q})A_b(-{\bm q}) \rangle\left({\vec \sigma}_{\alpha\beta}\times{\bm q} \right)_a \left({\vec \sigma}_{\alpha'\beta'}\times{\bm q} \right)_b   \psi^\dagger_{{\bm p}_1+{\bm q},\alpha}\psi_{{\bm p}_1,\beta}\psi^\dagger_{{\bm p}_2-{\bm q},\alpha'}\psi_{{\bm p}_2,\beta'},
\end{eqnarray}
The second term in the gauge field propagator (Eq.\ref{Eq:gauge}) does not contribute, since $\left({\vec \sigma}\times{\bm q}\right)\cdot{\bm q}=0$. The interaction can then be simplified to
\begin{eqnarray}
 H_{\rm int}\sim -J_K^2\sum_{{\bm p}_1{\bm p}_2{\bm q}}\sum_{\alpha\beta\alpha'\beta'}\left({\vec \sigma}_{\alpha\beta}\times{\hat{\bm q}} \right)\cdot \left({\vec \sigma}_{\alpha'\beta'}\times{\hat{\bm q}} \right)   \psi^\dagger_{{\bm p}_1+{\bm q},\alpha}\psi^\dagger_{{\bm p}_2-{\bm q},\alpha'}\psi_{{\bm p}_2,\beta'}\psi_{{\bm p}_1,\beta}.
\end{eqnarray}

The pairing states can be understood from the above interaction. The $|J=1, J_z=0\rangle$ pairing state for the 3d case, and correspondingly its descendant 2d pairing state: 
\begin{eqnarray}
|J_z=0\rangle\sim (k_x+ik_y)|\downarrow\downarrow \rangle+ (k_x-ik_y)|\uparrow\uparrow \rangle, 
\end{eqnarray}
correspond to equal spin pairing. They make use of the $\alpha=\beta=\alpha'=\beta'=\uparrow$ term in the interaction 
\begin{eqnarray}
 H_{\rm int}\sim -J_K^2\sum_{{\bm p}_1{\bm p}_2{\bm q}}|{\vec \sigma}_{\uparrow\uparrow}\times{\hat{\bm q}} |^2   \psi^\dagger_{{\bm p}_1+{\bm q},\uparrow}\psi^\dagger_{{\bm p}_2-{\bm q},\uparrow}\psi_{{\bm p}_2,\uparrow}\psi_{{\bm p}_1,\uparrow},
\label{Eq:Hz}
\end{eqnarray}
and its spin down counterpart with $\alpha=\beta=\alpha'=\beta'=\downarrow$. These terms are attractive when ${\sigma}_{\alpha\alpha}\times{\hat{\bm q}}\neq 0$. Since ${\sigma}_{\alpha\alpha}=(0, 0, \pm 1)$, this condition is satisfied when the momentum transfer ${\bm q}$ is in the $xy$ plane, or the Cooper pair momenta in the $xy$ plane.

The $|J=1, J_z=\pm 1\rangle$ pairing states in 3d and the $|J_z=\pm 1\rangle$ pairing state in 2d can be similarly undersood. The degeneracy of the $|J=1, J_z=\pm 1\rangle$ states with the $|J=1, J_z=0\rangle$ state in the 3d case is guaranteed by $SO(3)_{\bm J}$ symmetry. For the 2d case, we can rewrite the states as 
\begin{eqnarray}
|J_z=\pm 1\rangle\sim  (k_x\pm ik_y)\frac{|\uparrow\downarrow \rangle+|\downarrow\uparrow \rangle}{\sqrt{2}}\sim k_y \frac{|\rightrightarrows \rangle_x-|\leftleftarrows\rangle_x}{\sqrt{2}}\mp k_x \frac{|\rightrightarrows \rangle_y-|\leftleftarrows\rangle_y}{\sqrt{2}}.
\end{eqnarray}
The resulting four terms correspond to spin-orbit configurations with attractive interactions. For example the term $k_y|\rightrightarrows \rangle_x$ represents equal spin pairing with spins pointing in the $x$ direction, i.e. $\alpha=\beta=\alpha'=\beta'=\rightarrow$. The corresponding interaction term reads
\begin{eqnarray}
 H_{\rm int}\sim -J_K^2\sum_{{\bm p}_1{\bm p}_2{\bm q}}|{\vec \sigma}_{\rightrightarrows}\times{\hat{\bm q}} |^2   \psi^\dagger_{{\bm p}_1+{\bm q},\rightarrow}\psi^\dagger_{{\bm p}_2-{\bm q},\rightarrow}\psi_{{\bm p}_2,\rightarrow}\psi_{{\bm p}_1,\rightarrow},
\end{eqnarray}
which is attractive as its $z$ direction counterpart (Eq.\ref{Eq:Hz}). Here the momentum is along the $y$ direction and hence ${\vec \sigma}\times{\hat {\bm q}}\neq 0$.

\subsection*{B.~ BCS approach including microscopic details}

We have given above an analytic understanding of the pairing states at the metal/QSI interface using a much simplified model, namely continuum Fermi gas with magnetic dipole-dipole interaction. In this subsection, we study the effect of longitudinal spin fluctuations, the effect of the lattice, and the effect of quadrupole fluctuations. These effects can no longer be treated analytically. We will proceed using the BCS type approach by diagonalizing the pairing interaction matrix. The most negative eigenvalue corresponds to the dominant pairing channel.

\subsubsection*{1. Low energy effective model}
We start with the simpler case, namely the low energy effective model, and then proceed to study the effect of the lattice. In the low energy effective model, the momentum part of the spin structure factor is 
\begin{equation}
S_{ab}({\bm q})\sim  \delta_{ab}-\left( 1-\frac{1}{1+q^2\xi^2} \right)\frac{q_a q_b}{q^2}.
\end{equation}
When the correlation length $\xi\to\infty$, one has 
\begin{equation}
S_{ab}({\bm q})\sim  \delta_{ab}-\frac{q_a q_b}{q^2},
\end{equation}
which gives rise to the magnetic dipole-dipole interaction as considered above.

Following the procedure outlined in Section SI, we can obtain the interaction matrix ${\hat V}$. We first project $S_{ab}$ to different pairing channels to obtain 
\begin{equation}
S_{\mu\nu}=\sum_{ab}\sum_{\alpha\alpha'\beta\beta'} S_{ab}\sigma^a_{\alpha\beta}\sigma^b_{\alpha'\beta'}[\sigma_\mu i\sigma_y]^*_{\alpha'\alpha}[\sigma_\nu i\sigma_y]_{\beta\beta'}. 
\end{equation}
Then we integrate over momentum in the direction perpendicular to the interface to obtain the pairing interaction at the interface:
\begin{eqnarray}
V_{\mu\nu}({\bm p})=-J_K^2\int_{-\infty}^{\infty}\frac{dq_z}{2\pi}S_{\mu\nu}(p_x, p_y, q_z),
\end{eqnarray}

\begin{figure}
\begin{centering}
\subfigure[]{
\includegraphics[width=0.4\linewidth]{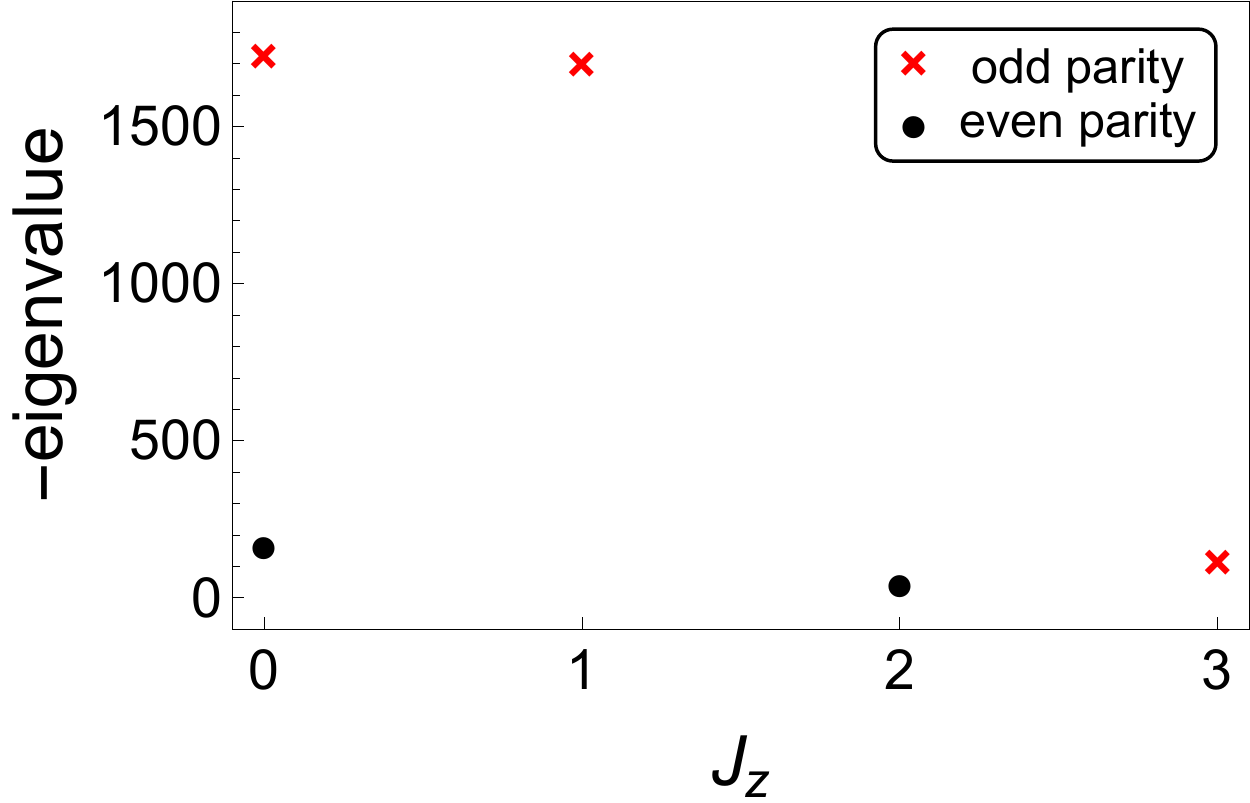} 
}
\subfigure[]{
\includegraphics[width=0.4\linewidth]{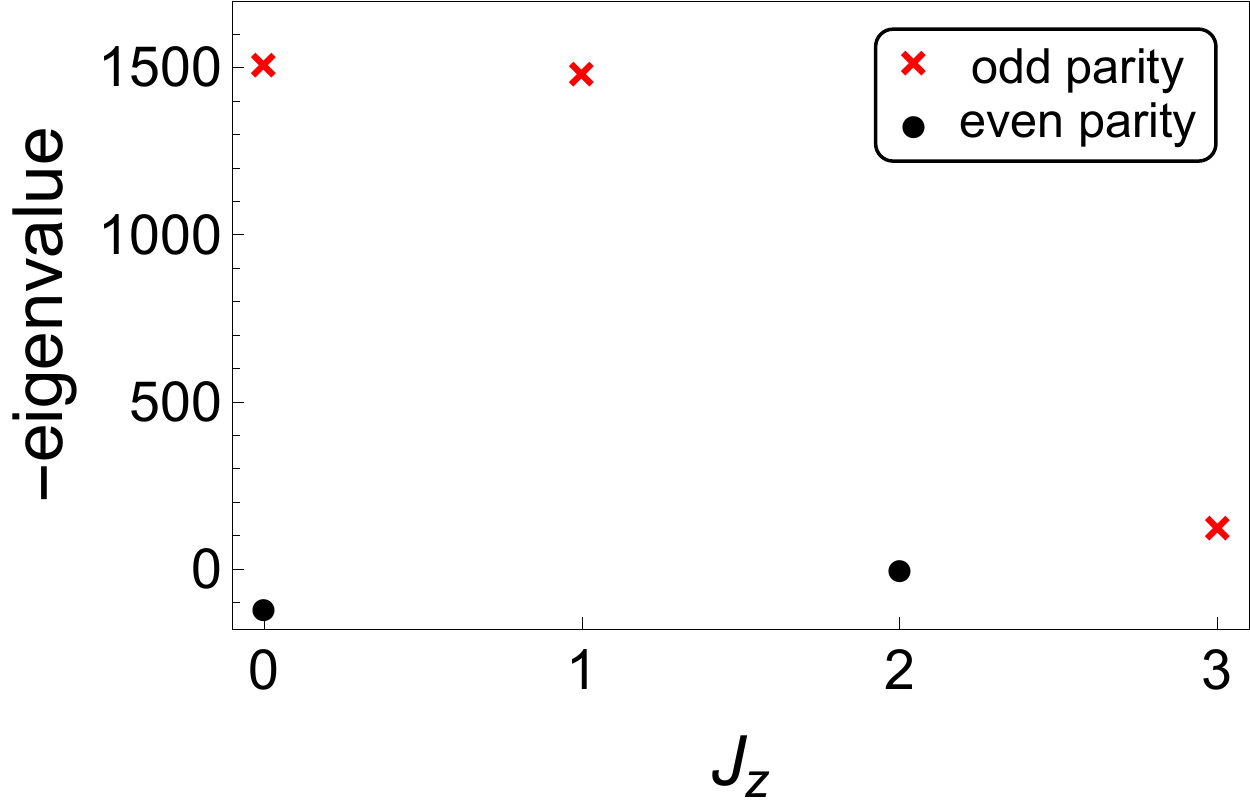} 
}
\end{centering}
\caption{(Color online) The leading negative eigenvalues of the pairing interaction matrix for different parity and $J_z$ channels in the low energy effective model (a), and the lattice model (b). The eigenvalues are dimensionless numbers in arbitrary units. The dominant pairing channels have odd parity with $L=1$ ($p$-wave), $S=1$, $J_z=0, \pm 1$. Also shown other pairing channels: $s$-wave (even parity, $J_z=0$), $d$-wave (even parity, $J_z=2$), $f$-wave (odd parity, $J_z=3$). }
\label{Fig:eigen}
\end{figure}

\begin{figure}
\begin{centering}
\subfigure[]{
\includegraphics[width=0.4\linewidth]{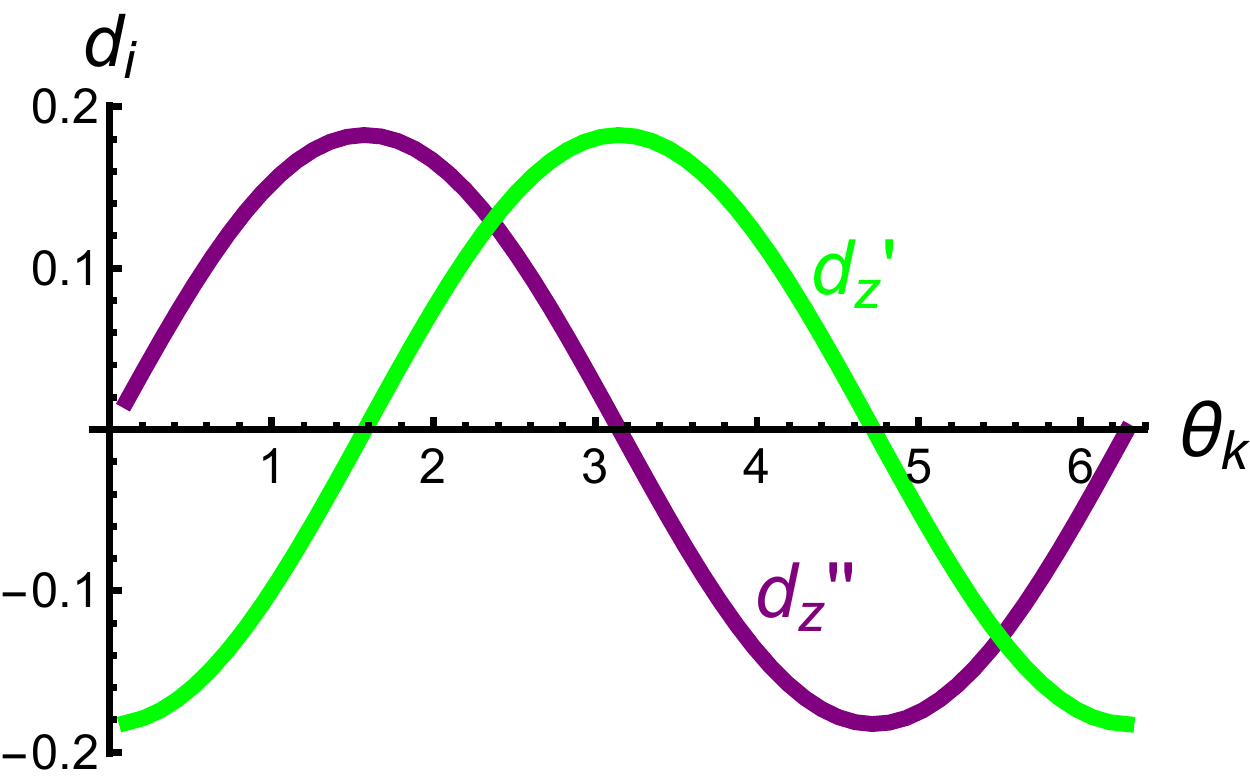} 
}
\subfigure[]{
\includegraphics[width=0.4\linewidth]{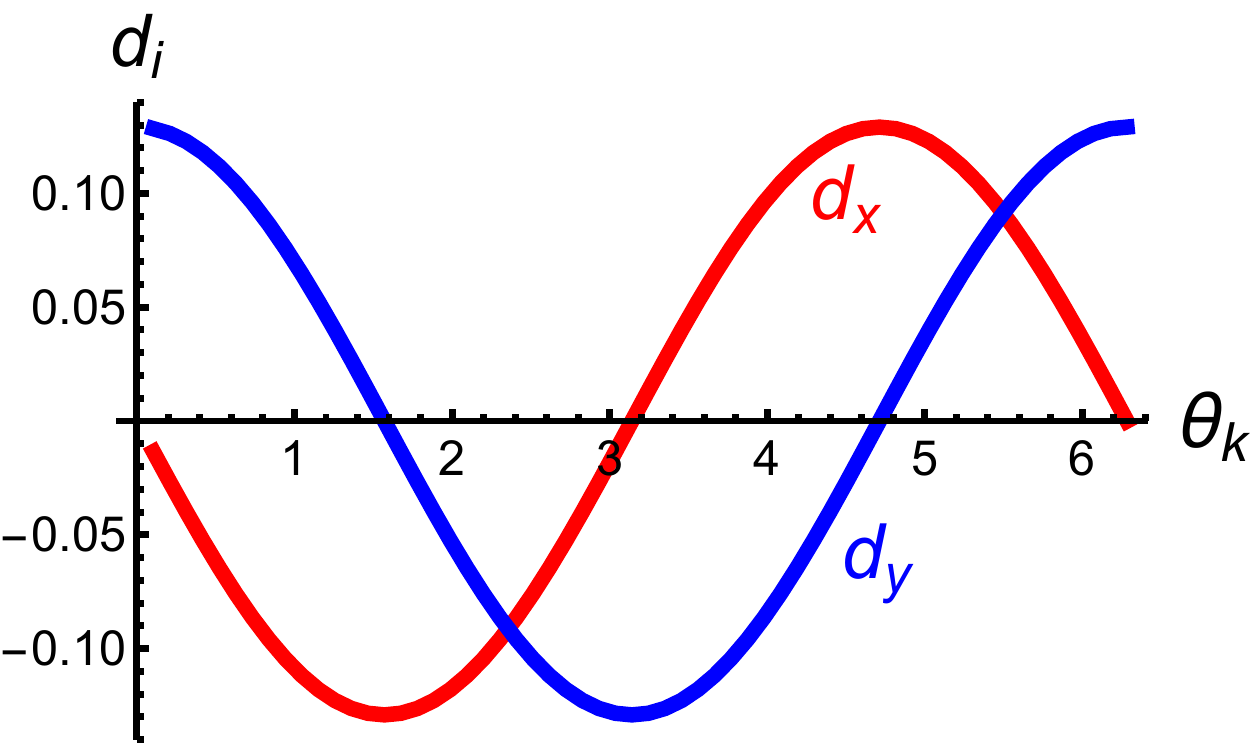} 
}
\end{centering}
\caption{(Color online) The superconducting gap function in the dominant pairing channels obtained from both the lattice model, and the low energy effective model. (a) represents odd parity pairing with $J_z=\pm 1$, where $d_x=d_y=0$, $d_z'$ and $d_z''$ denote real and imaginary parts of $d_z$. (b) represents odd parity pairing with $J_z=0$, where $d_z=0$. }
\label{Fig:gap}
\end{figure}

From the ab initio calculations, one obtains a circular Fermi surface for the heterostructure Pr$_2$Zr$_2$O$_7$/Y$_2$Sn$_{2-x}$Sb$_x$O$_7$, with Fermi momemtum $k_F=0.37(2\pi/a)$ (lattice constant of Pr$_2$Zr$_2$O$_7$ is $a\simeq 10.7{\rm\AA}$). Hence we can parameterize momentum on the Fermi surface as ${\bm k}=k_F(\cos\theta_{\bm k}, \sin\theta_{\bm k})$. To solve the gap equation numerically, we then discretize the angle $\theta_{\bm k}=\frac{2\pi}{N}n$, with $n=0,1,2,\cdots, N-1$. 
Due to the presence of parity symmetry, the coupling between the even and odd parity pairing channels vanishes. Hence the interaction matrix ${\hat V}$ is block diagonal, and the even parity part and odd parity part can be separately diagonalized. Due to the presence of the $U(1)_{J_z}$ symmetry, the eigenvectors can be further organized in the $J_z$ basis. The leading eigenvalues  and the corresponding parity, $J_z$, and orbital angular momentum are listed in Table II, and plotted in Fig.\ref{Fig:eigen}a. The dominant pairing channels have odd parity, orbital angular momentum $L=1$ ($p$-wave), with $J_z=0, \pm 1$. Up to a $U(1)$ phase factor, the eigenvectors can be written in matrix form as (see Fig.\ref{Fig:gap})
\begin{equation}
{\hat \Delta}({\bm k})=i {\vec d}({\bm k})\cdot {\vec \sigma}\sigma_y\sim \begin{pmatrix} k_x-ik_y & 0 \\ 0 & k_x+ik_y \end{pmatrix}, ~~\begin{pmatrix} 0& k_x\pm ik_y  \\   k_x\pm ik_y & 0\end{pmatrix}.
\label{d-vector}
\end{equation}
These are exactly the negative binding energy states $|J_z=0, \pm 1\rangle$ as obtained in the two-body problem with magnetic dipole-dipole interaction. We also find that with decreasing correlation length, the eigenvalue decreases, while the dominant pairing channels do not change. Hence the effect of longitudinal spin fluctuations is to reduce the superconducting $T_c$.

\begin{table}[t]
    \begin{tabular}{ | c | c | c | c |c | c |}
    \hline
$J_z$ & 0 & 0 & 1 & 2 & 3 \\ \hline
parity & even & odd & odd & even & odd \\ \hline
orbital & $s$& $p$&$p$ & $d$ & $f$ \\ \hline
 eigenvalue (LEEM) &    -163.994 & -1743.41 & -1716.57 & -41.78 & -133.978 \\ \hline
 eigenvalue (LM) &    118.746 & -1526.9 & -1499.41 & -0.188 & -139.588 \\ \hline
    \end{tabular}\par
\caption{The leading negative eigenvalues of the pairing interaction matrix in the low energy effective model (LEEM) and the lattice model (LM). Here the Fermi momentum $k_F=0.37(2\pi/a)$, where $a$ is the lattice constant, the correlation length $\xi=10{\rm \AA}$ [\onlinecite{Kimura13}]. The results are plotted in Fig.\ref{Fig:eigen} below.}
\end{table}

\subsubsection*{2. Lattice model}

We then consider the pairing problem on the lattice.
From the ab initio calculations, one can see that the conduction electron wavefunction mostly only penetrates into the first layer of tetrahedra in QSI. Hence only spins on these tetrahedra have appreciable coupling with the conduction electrons. These tetrahedra form a triangular lattice in the $[1,{\bar 1},1]$ plane, with lattice constant $a'=1/\sqrt{2}$ (the lattice constant of the Pr$_2$Zr$_2$O$_7$ is set to $1$). To proceed, we make the approximation that the Kondo couplings between the conduction electrons and the four spins in a tetrahedron are of equal strength. The induced exchange coupling in the Hamiltonian $H^{(2)}_{\rm int}=\sum_{ab{\bm r}{\bm r}'} J_{ab}({\bm r}, {\bm r}') s_a({\bm r}) s_b({\bm r}')$ is thus of the form
\begin{equation}
 J_{ab}({\bm r}, {\bm r}')=-\frac{1}{2}\sum_{{\bm R}{\bm R}'}{\cal I}({\bm r}, {\bm R}) {\cal I}({\bm r}', {\bm R}') \sum_{\bm q}e^{i{\bm q}\cdot ({\bm R}-{\bm R}')}S_{ab}({\bm q}),
\label{Eq:Jab}
\end{equation}
with ${\bm r}$ labeling the position of the conduction electron sites, and ${\bm R}$ the center of the tetrahedra in QSI. Under such an approximation, all the information about the local moment part needed for the pairing problem is encoded in the spin structure factor 
\begin{equation}
 S_{ab}({\bm q})=\sum_{AB} {\hat z}^a_A  {\hat z}^b_B \langle \phi_A({\bm q}) \phi_B(-{\bm q})\rangle,
\end{equation}
which can be extracted from neutron scattering data (neutron scattering averages over the four sites in a tetrahedron).

In the metal part of the heterostructure, the conduction electrons are predominantly on the B sites of A$_2$B$_2$O$_7$. We consider the first layer of conduction electrons near the interface, which also form a triangular lattice, with its sites right above the center of the spin triangular lattice. Since the Kondo coupling decays very fast in space, we will only include the coupling between the conduction electrons with the nearest tetrahedra of spins. For such a bilayer system, we include the coupling with the three nearest tetrahedra, and the exchange interaction is
\begin{equation}
 J_{ab}({\bm r}_i, {\bm r}_j)=-\frac{J_K^2}{9}\sum_{m,n=1,2,3}\int \frac{d^3{\bm q}}{(2\pi)^3}  S_{ab}({\bm q})e^{i{\bm q}\cdot ({\bm r}_i-{\bm r}_j+{\bm b}_m-{\bm b}_n)},
\end{equation}
with ${\bm b}_{1,2}=(\pm\frac{a'}{2}, \frac{a'}{2\sqrt{3}})$, and ${\bm b} _3=(0,\frac{a'}{\sqrt{3}})$, $a'=1/\sqrt{2}$. The resulting pairing interaction is then
\begin{eqnarray}
V_{\mu\nu}({\bm p})=-J_K^2f({\bm p})\int\frac{dq_z}{2\pi}S_{\mu\nu}(p_x, p_y, q_z),
\end{eqnarray}
with the form factor $f({\bm p})=\frac{1}{3}+\frac{2}{9}\left[ \cos\left(p_x a'\right)+\cos\left(\frac{p_x a'}{2}+\frac{\sqrt{3}p_ya'}{2} \right)+\cos\left(\frac{p_x a'}{2}-\frac{\sqrt{3}p_ya'}{2} \right) \right]$, and the projected spin structure factor $S_{\mu\nu}=\sum_{ab}\sum_{\alpha\alpha'\beta\beta'} S_{ab}\sigma^a_{\alpha\beta}\sigma^b_{\alpha'\beta'}[\sigma_\mu i\sigma_y]^*_{\alpha'\alpha}[\sigma_\nu i\sigma_y]_{\beta\beta'}$. The $q_z$ integral is over the Brillouin zone
\begin{eqnarray}
|\frac{p_x}{\sqrt{2}}- \frac{p_y}{\sqrt{6}}+ \frac{q_z}{\sqrt{3}}|&<&2\pi,\nonumber\\
|\frac{p_x}{\sqrt{2}}+ \frac{p_y}{\sqrt{6}}- \frac{q_z}{\sqrt{3}}|&<&2\pi,\nonumber\\
|\frac{2p_y}{\sqrt{6}}+ \frac{q_z}{\sqrt{3}}|&<&2\pi,\nonumber\\
|\frac{p_x}{\sqrt{2}}- \frac{p_y}{\sqrt{6}}+ \frac{q_z}{\sqrt{3}}|+
|\frac{p_x}{\sqrt{2}}+ \frac{p_y}{\sqrt{6}}- \frac{q_z}{\sqrt{3}}|+
|\frac{2p_y}{\sqrt{6}}+ \frac{q_z}{\sqrt{3}}|&<&3\pi.
\end{eqnarray}
 
With the knowledge of the momentum and spin dependence of the pairing interaction, the gap equation can be solved following the same procedure as in the low energy effective model. We discretize the Fermi surface to obtain the interaction matrix ${\hat V}$, which is also block diagonal. Organizing the eigenvectors in the $J_z$ basis, the leading eigenvalues and the corresponding quantum numbers are listed in Table II, and plotted in Fig.\ref{Fig:eigen}b. The dominant pairing channels are the same as in the low energy effective model. We emphasize again that the degeneracy of the $J_z=0$ and $J_z=\pm 1$ pairing channels are accidental, i.e. not protected by any symmetry. Hence such degeneracy can be lifted in real materials due to microscopic details.

\subsubsection*{3. Quadrupole fluctuations}

We have been focusing on the spin fluctuations associated with the Ising component of Pr moments in QSI, and we consider next the effect of quadrupole fluctuations associated with the planar components. The planar components $\tau^{\pm}$ couple to the charge density of conduction electrons $\rho_m=\sum_{\alpha}c^\dagger_{m\alpha} c_{m\alpha}$ at site $m$ in the form [\onlinecite{Chen12, Lee13}]
 \begin{equation}
H_\rho=\sum_{im}\left(M_{im}\tau^+_i\rho_m+h.c.\right),
\end{equation}
with the coupling $M_{im}$.
Such a coupling induces density-density interactions among the conduction electrons, 
\begin{equation}
{\cal S}_{\rm int}=\sum_{mn}\int dt \int dt' V_{\rho}({\bm r}_{mn}, t-t')\rho_m(t)\rho_n(t').
\end{equation}
Such an interaction can generate pairing in the parity even channels that competes with odd-parity pairing mediated by the Ising spin fluctuations. We can mean-field decompose this interaction into the even-parity pairing channel
\begin{equation}
{\cal S}_{\rm int}=\int\frac{d^2{\bm k}}{(2\pi)^2}\int \frac{d\omega}{2\pi}\int\frac{d^2{\bm k}'}{(2\pi)^2}\int \frac{d\omega'}{2\pi} V_{\rho}({\bm k}-{\bm k}',\omega-\omega')P^\dagger_0({\bm k},\omega)P_0({\bm k}',\omega'),
\end{equation}
where the pairing interaction $V_\rho({\bm p})\sim |M|^2\int dq_z {\cal Q}_{+-}(p_x, p_y, q_z)$ is determined by the quadrupole structure factor ${\cal Q}_{+-}$ of QSI. The momentum dependence of ${\cal Q}_{+-}({\bm q})$ will determine the pairing symmetry.
Since the planar components and the Ising components are strongly entangled, one expects the momentum dependence of the quadrupole correlation functions to be correlated with that of the spin correlation functions. Since the spin correlation functions do not contain appreciable harmonic components higher than $p$-wave, it is reasonable to expect that quadrupole correlation functions also do not contain appreciable high harmonic components. One thus expects that quadrupole correlations induce paring in the $s$-wave channel. However, Coulomb repulsion suppresses $s$-wave pairing, while pairings in higher angular momentum channels are much less affected. Hence we expect that including quadrupole fluctuations will not change the dominant pairing channel in our heterostructure.

\bibliographystyle{apsrev}
\bibliography{strings,refs}

\begin{thebibliography}{85}
\expandafter\ifx\csname natexlab\endcsname\relax\def\natexlab#1{#1}\fi
\expandafter\ifx\csname bibnamefont\endcsname\relax
  \def\bibnamefont#1{#1}\fi
\expandafter\ifx\csname bibfnamefont\endcsname\relax
  \def\bibfnamefont#1{#1}\fi
\expandafter\ifx\csname citenamefont\endcsname\relax
  \def\citenamefont#1{#1}\fi
\expandafter\ifx\csname url\endcsname\relax
  \def\url#1{\texttt{#1}}\fi
\expandafter\ifx\csname urlprefix\endcsname\relax\def\urlprefix{URL }\fi
\providecommand{\bibinfo}[2]{#2}
\providecommand{\eprint}[2][]{\url{#2}}

\bibitem[{\citenamefont{Anderson}(1987)}]{Anderson87}
\bibinfo{author}{\bibfnamefont{P.~W.} \bibnamefont{Anderson}},
  \bibinfo{journal}{Science} \textbf{\bibinfo{volume}{235}},
  \bibinfo{pages}{1196} (\bibinfo{year}{1987}),
  \urlprefix\url{http://www.sciencemag.org/content/235/4793/1196.abstract}.

\bibitem[{\citenamefont{Anderson}(1973)}]{Anderson73}
\bibinfo{author}{\bibfnamefont{P.~W.} \bibnamefont{Anderson}},
  \bibinfo{journal}{Mater. Res. Bull.} \textbf{\bibinfo{volume}{8}},
  \bibinfo{pages}{153} (\bibinfo{year}{1973}).

\bibitem[{\citenamefont{Balents}(2010)}]{Balents10}
\bibinfo{author}{\bibfnamefont{L.}~\bibnamefont{Balents}},
  \bibinfo{journal}{Nature} \textbf{\bibinfo{volume}{464}},
  \bibinfo{pages}{199} (\bibinfo{year}{2010}).

\bibitem[{\citenamefont{Lee}(2008)}]{Lee08}
\bibinfo{author}{\bibfnamefont{P.~A.} \bibnamefont{Lee}},
  \bibinfo{journal}{Science} \textbf{\bibinfo{volume}{321}},
  \bibinfo{pages}{1306} (\bibinfo{year}{2008}),
  \eprint{http://www.sciencemag.org/content/321/5894/1306.full.pdf},
  \urlprefix\url{http://www.sciencemag.org/content/321/5894/1306.short}.

\bibitem[{\citenamefont{Gingras and McClarty}(2014)}]{Gingras14}
\bibinfo{author}{\bibfnamefont{M.~J.~P.} \bibnamefont{Gingras}}
  \bibnamefont{and} \bibinfo{author}{\bibfnamefont{P.~A.}
  \bibnamefont{McClarty}}, \bibinfo{journal}{Rep. Prog. Phys.}
  \textbf{\bibinfo{volume}{77}}, \bibinfo{pages}{056501}
  (\bibinfo{year}{2014}).

\bibitem[{\citenamefont{Migdal}(1958)}]{Migdal58}
\bibinfo{author}{\bibfnamefont{A.~B.} \bibnamefont{Migdal}},
  \bibinfo{journal}{Sov. Phys. JETP} \textbf{\bibinfo{volume}{7}},
  \bibinfo{pages}{996} (\bibinfo{year}{1958}).

\bibitem[{\citenamefont{Hewson}(1993)}]{Hewson}
\bibinfo{author}{\bibfnamefont{A.~C.} \bibnamefont{Hewson}},
  \emph{\bibinfo{title}{The Kondo Problem to Heavy Fermions}}
  (\bibinfo{publisher}{Cambridge University Press}, \bibinfo{year}{1993}).

\bibitem[{\citenamefont{Ohtomo and Hwang}(2004)}]{Ohtomo04}
\bibinfo{author}{\bibfnamefont{A.}~\bibnamefont{Ohtomo}} \bibnamefont{and}
  \bibinfo{author}{\bibfnamefont{H.~Y.} \bibnamefont{Hwang}},
  \bibinfo{journal}{Nature} \textbf{\bibinfo{volume}{427}},
  \bibinfo{pages}{423} (\bibinfo{year}{2004}),
  \urlprefix\url{http://dx.doi.org/10.1038/nature02308}.

\bibitem[{\citenamefont{Reyren et~al.}(2007)\citenamefont{Reyren, Thiel,
  Caviglia, Kourkoutis, Hammerl, Richter, Schneider, Kopp, R{\"u}etschi,
  Jaccard et~al.}}]{Reyren07}
\bibinfo{author}{\bibfnamefont{N.}~\bibnamefont{Reyren}},
  \bibinfo{author}{\bibfnamefont{S.}~\bibnamefont{Thiel}},
  \bibinfo{author}{\bibfnamefont{A.~D.} \bibnamefont{Caviglia}},
  \bibinfo{author}{\bibfnamefont{L.~F.} \bibnamefont{Kourkoutis}},
  \bibinfo{author}{\bibfnamefont{G.}~\bibnamefont{Hammerl}},
  \bibinfo{author}{\bibfnamefont{C.}~\bibnamefont{Richter}},
  \bibinfo{author}{\bibfnamefont{C.~W.} \bibnamefont{Schneider}},
  \bibinfo{author}{\bibfnamefont{T.}~\bibnamefont{Kopp}},
  \bibinfo{author}{\bibfnamefont{A.-S.} \bibnamefont{R{\"u}etschi}},
  \bibinfo{author}{\bibfnamefont{D.}~\bibnamefont{Jaccard}},
  \bibnamefont{et~al.}, \bibinfo{journal}{Science}
  \textbf{\bibinfo{volume}{317}}, \bibinfo{pages}{1196} (\bibinfo{year}{2007}),
  \eprint{http://www.sciencemag.org/content/317/5842/1196.full.pdf},
  \urlprefix\url{http://www.sciencemag.org/content/317/5842/1196.abstract}.

\bibitem[{\citenamefont{Mannhart and Schlom}(2010)}]{Schlom10}
\bibinfo{author}{\bibfnamefont{J.}~\bibnamefont{Mannhart}} \bibnamefont{and}
  \bibinfo{author}{\bibfnamefont{D.~G.} \bibnamefont{Schlom}},
  \bibinfo{journal}{Science} \textbf{\bibinfo{volume}{327}},
  \bibinfo{pages}{1607} (\bibinfo{year}{2010}),
  \eprint{http://www.sciencemag.org/content/327/5973/1607.full.pdf},
  \urlprefix\url{http://www.sciencemag.org/content/327/5973/1607.abstract}.

\bibitem[{\citenamefont{Coleman}(2015)}]{Coleman}
\bibinfo{author}{\bibfnamefont{P.}~\bibnamefont{Coleman}},
  \emph{\bibinfo{title}{Introduction to Many Body Physics}}
  (\bibinfo{publisher}{Cambridge University Press}, \bibinfo{year}{2015}).

\bibitem[{\citenamefont{Doniach}(1977)}]{Doniach77}
\bibinfo{author}{\bibfnamefont{S.}~\bibnamefont{Doniach}},
  \bibinfo{journal}{Physica B+C} \textbf{\bibinfo{volume}{91}},
  \bibinfo{pages}{231 } (\bibinfo{year}{1977}), ISSN \bibinfo{issn}{0378-4363},
  \urlprefix\url{http://www.sciencedirect.com/science/article/pii/0378436377901905}.

\bibitem[{\citenamefont{Coleman and Andrei}(1989)}]{Coleman89}
\bibinfo{author}{\bibfnamefont{P.}~\bibnamefont{Coleman}} \bibnamefont{and}
  \bibinfo{author}{\bibfnamefont{N.}~\bibnamefont{Andrei}},
  \bibinfo{journal}{Journal of Physics: Condensed Matter}
  \textbf{\bibinfo{volume}{1}}, \bibinfo{pages}{4057} (\bibinfo{year}{1989}),
  \urlprefix\url{http://stacks.iop.org/0953-8984/1/i=26/a=003}.

\bibitem[{\citenamefont{Senthil et~al.}(2003)\citenamefont{Senthil, Sachdev,
  and Vojta}}]{Senthil03}
\bibinfo{author}{\bibfnamefont{T.}~\bibnamefont{Senthil}},
  \bibinfo{author}{\bibfnamefont{S.}~\bibnamefont{Sachdev}}, \bibnamefont{and}
  \bibinfo{author}{\bibfnamefont{M.}~\bibnamefont{Vojta}},
  \bibinfo{journal}{Phys. Rev. Lett.} \textbf{\bibinfo{volume}{90}},
  \bibinfo{pages}{216403} (\bibinfo{year}{2003}),
  \urlprefix\url{http://link.aps.org/doi/10.1103/PhysRevLett.90.216403}.

\bibitem[{\citenamefont{Lee and Zimanyi}(1989)}]{DHLee89}
\bibinfo{author}{\bibfnamefont{D.~H.} \bibnamefont{Lee}} \bibnamefont{and}
  \bibinfo{author}{\bibfnamefont{G.~T.} \bibnamefont{Zimanyi}},
  \bibinfo{journal}{Phys. Rev. B} \textbf{\bibinfo{volume}{40}},
  \bibinfo{pages}{9404} (\bibinfo{year}{1989}),
  \urlprefix\url{http://link.aps.org/doi/10.1103/PhysRevB.40.9404}.

\bibitem[{\citenamefont{Shankar}(1994)}]{Shankar94}
\bibinfo{author}{\bibfnamefont{R.}~\bibnamefont{Shankar}},
  \bibinfo{journal}{Rev. Mod. Phys.} \textbf{\bibinfo{volume}{66}},
  \bibinfo{pages}{129} (\bibinfo{year}{1994}),
  \urlprefix\url{http://link.aps.org/doi/10.1103/RevModPhys.66.129}.

\bibitem[{\citenamefont{Hermele et~al.}(2004)\citenamefont{Hermele, Fisher, and
  Balents}}]{Hermele04}
\bibinfo{author}{\bibfnamefont{M.}~\bibnamefont{Hermele}},
  \bibinfo{author}{\bibfnamefont{M.~P.~A.} \bibnamefont{Fisher}},
  \bibnamefont{and} \bibinfo{author}{\bibfnamefont{L.}~\bibnamefont{Balents}},
  \bibinfo{journal}{Phys. Rev. B} \textbf{\bibinfo{volume}{69}},
  \bibinfo{pages}{064404} (\bibinfo{year}{2004}),
  \urlprefix\url{http://link.aps.org/doi/10.1103/PhysRevB.69.064404}.

\bibitem[{\citenamefont{Onoda and Tanaka}(2010)}]{Onoda10}
\bibinfo{author}{\bibfnamefont{S.}~\bibnamefont{Onoda}} \bibnamefont{and}
  \bibinfo{author}{\bibfnamefont{Y.}~\bibnamefont{Tanaka}},
  \bibinfo{journal}{Phys. Rev. Lett.} \textbf{\bibinfo{volume}{105}},
  \bibinfo{pages}{047201} (\bibinfo{year}{2010}),
  \urlprefix\url{http://link.aps.org/doi/10.1103/PhysRevLett.105.047201}.

\bibitem[{\citenamefont{Ross et~al.}(2011{\natexlab{a}})\citenamefont{Ross,
  Savary, Gaulin, and Balents}}]{Balents11}
\bibinfo{author}{\bibfnamefont{K.~A.} \bibnamefont{Ross}},
  \bibinfo{author}{\bibfnamefont{L.}~\bibnamefont{Savary}},
  \bibinfo{author}{\bibfnamefont{B.~D.} \bibnamefont{Gaulin}},
  \bibnamefont{and} \bibinfo{author}{\bibfnamefont{L.}~\bibnamefont{Balents}},
  \bibinfo{journal}{Phys. Rev. X} \textbf{\bibinfo{volume}{1}},
  \bibinfo{pages}{021002} (\bibinfo{year}{2011}{\natexlab{a}}),
  \urlprefix\url{http://link.aps.org/doi/10.1103/PhysRevX.1.021002}.

\bibitem[{\citenamefont{Lee et~al.}(2012)\citenamefont{Lee, Onoda, and
  Balents}}]{Balents12}
\bibinfo{author}{\bibfnamefont{S.}~\bibnamefont{Lee}},
  \bibinfo{author}{\bibfnamefont{S.}~\bibnamefont{Onoda}}, \bibnamefont{and}
  \bibinfo{author}{\bibfnamefont{L.}~\bibnamefont{Balents}},
  \bibinfo{journal}{Phys. Rev. B} \textbf{\bibinfo{volume}{86}},
  \bibinfo{pages}{104412} (\bibinfo{year}{2012}),
  \urlprefix\url{http://link.aps.org/doi/10.1103/PhysRevB.86.104412}.

\bibitem[{\citenamefont{Gardner et~al.}(1999)\citenamefont{Gardner, Dunsiger,
  Gaulin, Gingras, Greedan, Kiefl, Lumsden, MacFarlane, Raju, Sonier
  et~al.}}]{Gardner99}
\bibinfo{author}{\bibfnamefont{J.~S.} \bibnamefont{Gardner}},
  \bibinfo{author}{\bibfnamefont{S.~R.} \bibnamefont{Dunsiger}},
  \bibinfo{author}{\bibfnamefont{B.~D.} \bibnamefont{Gaulin}},
  \bibinfo{author}{\bibfnamefont{M.~J.~P.} \bibnamefont{Gingras}},
  \bibinfo{author}{\bibfnamefont{J.~E.} \bibnamefont{Greedan}},
  \bibinfo{author}{\bibfnamefont{R.~F.} \bibnamefont{Kiefl}},
  \bibinfo{author}{\bibfnamefont{M.~D.} \bibnamefont{Lumsden}},
  \bibinfo{author}{\bibfnamefont{W.~A.} \bibnamefont{MacFarlane}},
  \bibinfo{author}{\bibfnamefont{N.~P.} \bibnamefont{Raju}},
  \bibinfo{author}{\bibfnamefont{J.~E.} \bibnamefont{Sonier}},
  \bibnamefont{et~al.}, \bibinfo{journal}{Phys. Rev. Lett.}
  \textbf{\bibinfo{volume}{82}}, \bibinfo{pages}{1012} (\bibinfo{year}{1999}),
  \urlprefix\url{http://link.aps.org/doi/10.1103/PhysRevLett.82.1012}.

\bibitem[{\citenamefont{Thompson et~al.}(2011)\citenamefont{Thompson, McClarty,
  R\o{}nnow, Regnault, Sorge, and Gingras}}]{Thompson11}
\bibinfo{author}{\bibfnamefont{J.~D.} \bibnamefont{Thompson}},
  \bibinfo{author}{\bibfnamefont{P.~A.} \bibnamefont{McClarty}},
  \bibinfo{author}{\bibfnamefont{H.~M.} \bibnamefont{R\o{}nnow}},
  \bibinfo{author}{\bibfnamefont{L.~P.} \bibnamefont{Regnault}},
  \bibinfo{author}{\bibfnamefont{A.}~\bibnamefont{Sorge}}, \bibnamefont{and}
  \bibinfo{author}{\bibfnamefont{M.~J.~P.} \bibnamefont{Gingras}},
  \bibinfo{journal}{Phys. Rev. Lett.} \textbf{\bibinfo{volume}{106}},
  \bibinfo{pages}{187202} (\bibinfo{year}{2011}),
  \urlprefix\url{http://link.aps.org/doi/10.1103/PhysRevLett.106.187202}.

\bibitem[{\citenamefont{Ross et~al.}(2011{\natexlab{b}})\citenamefont{Ross,
  Yaraskavitch, Laver, Gardner, Quilliam, Meng, Kycia, Singh, Proffen,
  Dabkowska et~al.}}]{Ross11}
\bibinfo{author}{\bibfnamefont{K.~A.} \bibnamefont{Ross}},
  \bibinfo{author}{\bibfnamefont{L.~R.} \bibnamefont{Yaraskavitch}},
  \bibinfo{author}{\bibfnamefont{M.}~\bibnamefont{Laver}},
  \bibinfo{author}{\bibfnamefont{J.~S.} \bibnamefont{Gardner}},
  \bibinfo{author}{\bibfnamefont{J.~A.} \bibnamefont{Quilliam}},
  \bibinfo{author}{\bibfnamefont{S.}~\bibnamefont{Meng}},
  \bibinfo{author}{\bibfnamefont{J.~B.} \bibnamefont{Kycia}},
  \bibinfo{author}{\bibfnamefont{D.~K.} \bibnamefont{Singh}},
  \bibinfo{author}{\bibfnamefont{T.}~\bibnamefont{Proffen}},
  \bibinfo{author}{\bibfnamefont{H.~A.} \bibnamefont{Dabkowska}},
  \bibnamefont{et~al.}, \bibinfo{journal}{Phys. Rev. B}
  \textbf{\bibinfo{volume}{84}}, \bibinfo{pages}{174442}
  (\bibinfo{year}{2011}{\natexlab{b}}),
  \urlprefix\url{http://link.aps.org/doi/10.1103/PhysRevB.84.174442}.

\bibitem[{\citenamefont{Kimura et~al.}(2013)\citenamefont{Kimura, Nakatsuji,
  Wen, Broholm, Stone, Nishibori, and Sawa}}]{Kimura13}
\bibinfo{author}{\bibfnamefont{K.}~\bibnamefont{Kimura}},
  \bibinfo{author}{\bibfnamefont{S.}~\bibnamefont{Nakatsuji}},
  \bibinfo{author}{\bibfnamefont{J.-J.} \bibnamefont{Wen}},
  \bibinfo{author}{\bibfnamefont{C.}~\bibnamefont{Broholm}},
  \bibinfo{author}{\bibfnamefont{M.~B.} \bibnamefont{Stone}},
  \bibinfo{author}{\bibfnamefont{E.}~\bibnamefont{Nishibori}},
  \bibnamefont{and} \bibinfo{author}{\bibfnamefont{H.}~\bibnamefont{Sawa}},
  \bibinfo{journal}{Nat. Commun.} \textbf{\bibinfo{volume}{4}},
  \bibinfo{pages}{1934} (\bibinfo{year}{2013}),
  \urlprefix\url{http://dx.doi.org/10.1038/ncomms2914}.

\bibitem[{\citenamefont{Henley}(2005)}]{Henley05}
\bibinfo{author}{\bibfnamefont{C.~L.} \bibnamefont{Henley}},
  \bibinfo{journal}{Phys. Rev. B} \textbf{\bibinfo{volume}{71}},
  \bibinfo{pages}{014424} (\bibinfo{year}{2005}),
  \urlprefix\url{http://link.aps.org/doi/10.1103/PhysRevB.71.014424}.

\bibitem[{\citenamefont{Isakov et~al.}(2004)\citenamefont{Isakov, Gregor,
  Moessner, and Sondhi}}]{Isakov04}
\bibinfo{author}{\bibfnamefont{S.~V.} \bibnamefont{Isakov}},
  \bibinfo{author}{\bibfnamefont{K.}~\bibnamefont{Gregor}},
  \bibinfo{author}{\bibfnamefont{R.}~\bibnamefont{Moessner}}, \bibnamefont{and}
  \bibinfo{author}{\bibfnamefont{S.~L.} \bibnamefont{Sondhi}},
  \bibinfo{journal}{Phys. Rev. Lett.} \textbf{\bibinfo{volume}{93}},
  \bibinfo{pages}{167204} (\bibinfo{year}{2004}),
  \urlprefix\url{http://link.aps.org/doi/10.1103/PhysRevLett.93.167204}.

\bibitem[{\citenamefont{Lee et~al.}(2007)\citenamefont{Lee, Lee, and
  Senthil}}]{PALee07}
\bibinfo{author}{\bibfnamefont{S.-S.} \bibnamefont{Lee}},
  \bibinfo{author}{\bibfnamefont{P.~A.} \bibnamefont{Lee}}, \bibnamefont{and}
  \bibinfo{author}{\bibfnamefont{T.}~\bibnamefont{Senthil}},
  \bibinfo{journal}{Phys. Rev. Lett.} \textbf{\bibinfo{volume}{98}},
  \bibinfo{pages}{067006} (\bibinfo{year}{2007}),
  \urlprefix\url{http://link.aps.org/doi/10.1103/PhysRevLett.98.067006}.

\bibitem[{\citenamefont{Lee}(2014)}]{PALee14}
\bibinfo{author}{\bibfnamefont{P.~A.} \bibnamefont{Lee}},
  \bibinfo{journal}{Phys. Rev. X} \textbf{\bibinfo{volume}{4}},
  \bibinfo{pages}{031017} (\bibinfo{year}{2014}),
  \urlprefix\url{http://link.aps.org/doi/10.1103/PhysRevX.4.031017}.

\bibitem[{\citenamefont{Conlon and Chalker}(2009)}]{Conlon09}
\bibinfo{author}{\bibfnamefont{P.~H.} \bibnamefont{Conlon}} \bibnamefont{and}
  \bibinfo{author}{\bibfnamefont{J.~T.} \bibnamefont{Chalker}},
  \bibinfo{journal}{Phys. Rev. Lett.} \textbf{\bibinfo{volume}{102}},
  \bibinfo{pages}{237206} (\bibinfo{year}{2009}),
  \urlprefix\url{http://link.aps.org/doi/10.1103/PhysRevLett.102.237206}.

\bibitem[{\citenamefont{{Ryzhkin} et~al.}(2013)\citenamefont{{Ryzhkin},
  {Ryzhkin}, and {Bramwell}}}]{Ryzhkin13}
\bibinfo{author}{\bibfnamefont{M.~I.} \bibnamefont{{Ryzhkin}}},
  \bibinfo{author}{\bibfnamefont{I.~A.} \bibnamefont{{Ryzhkin}}},
  \bibnamefont{and} \bibinfo{author}{\bibfnamefont{S.~T.}
  \bibnamefont{{Bramwell}}}, \bibinfo{journal}{EPL (Europhysics Letters)}
  \textbf{\bibinfo{volume}{104}}, \bibinfo{eid}{37005} (\bibinfo{year}{2013}),
  \eprint{1306.0653}.

\bibitem[{\citenamefont{Benton et~al.}(2012)\citenamefont{Benton, Sikora, and
  Shannon}}]{Benton12}
\bibinfo{author}{\bibfnamefont{O.}~\bibnamefont{Benton}},
  \bibinfo{author}{\bibfnamefont{O.}~\bibnamefont{Sikora}}, \bibnamefont{and}
  \bibinfo{author}{\bibfnamefont{N.}~\bibnamefont{Shannon}},
  \bibinfo{journal}{Phys. Rev. B} \textbf{\bibinfo{volume}{86}},
  \bibinfo{pages}{075154} (\bibinfo{year}{2012}),
  \urlprefix\url{http://link.aps.org/doi/10.1103/PhysRevB.86.075154}.

\bibitem[{\citenamefont{Li and Wu}(2012)}]{Li12}
\bibinfo{author}{\bibfnamefont{Y.}~\bibnamefont{Li}} \bibnamefont{and}
  \bibinfo{author}{\bibfnamefont{C.}~\bibnamefont{Wu}}, \bibinfo{journal}{Sci.
  Rep.} \textbf{\bibinfo{volume}{2}} (\bibinfo{year}{2012}),
  \urlprefix\url{http://dx.doi.org/10.1038/srep00392}.

\bibitem[{\citenamefont{Sigrist and Ueda}(1991)}]{Sigrist91}
\bibinfo{author}{\bibfnamefont{M.}~\bibnamefont{Sigrist}} \bibnamefont{and}
  \bibinfo{author}{\bibfnamefont{K.}~\bibnamefont{Ueda}},
  \bibinfo{journal}{Rev. Mod. Phys.} \textbf{\bibinfo{volume}{63}},
  \bibinfo{pages}{239} (\bibinfo{year}{1991}),
  \urlprefix\url{http://link.aps.org/doi/10.1103/RevModPhys.63.239}.

\bibitem[{\citenamefont{Brink and Satchler}(1962)}]{Brink}
\bibinfo{author}{\bibfnamefont{D.~M.} \bibnamefont{Brink}} \bibnamefont{and}
  \bibinfo{author}{\bibfnamefont{G.~R.} \bibnamefont{Satchler}},
  \emph{\bibinfo{title}{Angular Momentum}} (\bibinfo{publisher}{Oxford
  University Press}, \bibinfo{year}{1962}).

\bibitem[{\citenamefont{Sato}(2009)}]{Sato09}
\bibinfo{author}{\bibfnamefont{M.}~\bibnamefont{Sato}}, \bibinfo{journal}{Phys.
  Rev. B} \textbf{\bibinfo{volume}{79}}, \bibinfo{pages}{214526}
  (\bibinfo{year}{2009}),
  \urlprefix\url{http://link.aps.org/doi/10.1103/PhysRevB.79.214526}.

\bibitem[{\citenamefont{Qi et~al.}(2010)\citenamefont{Qi, Hughes, and
  Zhang}}]{Qi10}
\bibinfo{author}{\bibfnamefont{X.-L.} \bibnamefont{Qi}},
  \bibinfo{author}{\bibfnamefont{T.~L.} \bibnamefont{Hughes}},
  \bibnamefont{and} \bibinfo{author}{\bibfnamefont{S.-C.} \bibnamefont{Zhang}},
  \bibinfo{journal}{Phys. Rev. B} \textbf{\bibinfo{volume}{81}},
  \bibinfo{pages}{134508} (\bibinfo{year}{2010}),
  \urlprefix\url{http://link.aps.org/doi/10.1103/PhysRevB.81.134508}.

\bibitem[{\citenamefont{Maeno et~al.}(1994)\citenamefont{Maeno, Hashimoto,
  Yoshida, Nishizaki, Fujita, Bednorz, and Lichtenberg}}]{Maeno94}
\bibinfo{author}{\bibfnamefont{Y.}~\bibnamefont{Maeno}},
  \bibinfo{author}{\bibfnamefont{H.}~\bibnamefont{Hashimoto}},
  \bibinfo{author}{\bibfnamefont{K.}~\bibnamefont{Yoshida}},
  \bibinfo{author}{\bibfnamefont{S.}~\bibnamefont{Nishizaki}},
  \bibinfo{author}{\bibfnamefont{T.}~\bibnamefont{Fujita}},
  \bibinfo{author}{\bibfnamefont{J.~G.} \bibnamefont{Bednorz}},
  \bibnamefont{and}
  \bibinfo{author}{\bibfnamefont{F.}~\bibnamefont{Lichtenberg}},
  \bibinfo{journal}{Nature} \textbf{\bibinfo{volume}{372}},
  \bibinfo{pages}{532} (\bibinfo{year}{1994}).

\bibitem[{\citenamefont{Little}(1964)}]{Little64}
\bibinfo{author}{\bibfnamefont{W.~A.} \bibnamefont{Little}},
  \bibinfo{journal}{Phys. Rev.} \textbf{\bibinfo{volume}{134}},
  \bibinfo{pages}{A1416} (\bibinfo{year}{1964}),
  \urlprefix\url{http://link.aps.org/doi/10.1103/PhysRev.134.A1416}.

\bibitem[{\citenamefont{Ginzburg}(1970)}]{Ginzburg70}
\bibinfo{author}{\bibfnamefont{V.~L.} \bibnamefont{Ginzburg}},
  \bibinfo{journal}{Usp. Fiz. Nauk} \textbf{\bibinfo{volume}{101}},
  \bibinfo{pages}{185} (\bibinfo{year}{1970}), \bibinfo{note}{[Sov. Phys.-Usp.
  13, 335 (1970)]}.

\bibitem[{\citenamefont{Allender et~al.}(1973)\citenamefont{Allender, Bray, and
  Bardeen}}]{Bardeen73}
\bibinfo{author}{\bibfnamefont{D.}~\bibnamefont{Allender}},
  \bibinfo{author}{\bibfnamefont{J.}~\bibnamefont{Bray}}, \bibnamefont{and}
  \bibinfo{author}{\bibfnamefont{J.}~\bibnamefont{Bardeen}},
  \bibinfo{journal}{Phys. Rev. B} \textbf{\bibinfo{volume}{7}},
  \bibinfo{pages}{1020} (\bibinfo{year}{1973}),
  \urlprefix\url{http://link.aps.org/doi/10.1103/PhysRevB.7.1020}.

\bibitem[{\citenamefont{Koerting et~al.}(2005)\citenamefont{Koerting, Yuan,
  Hirschfeld, Kopp, and Mannhart}}]{Koerting05}
\bibinfo{author}{\bibfnamefont{V.}~\bibnamefont{Koerting}},
  \bibinfo{author}{\bibfnamefont{Q.}~\bibnamefont{Yuan}},
  \bibinfo{author}{\bibfnamefont{P.~J.} \bibnamefont{Hirschfeld}},
  \bibinfo{author}{\bibfnamefont{T.}~\bibnamefont{Kopp}}, \bibnamefont{and}
  \bibinfo{author}{\bibfnamefont{J.}~\bibnamefont{Mannhart}},
  \bibinfo{journal}{Phys. Rev. B} \textbf{\bibinfo{volume}{71}},
  \bibinfo{pages}{104510} (\bibinfo{year}{2005}),
  \urlprefix\url{http://link.aps.org/doi/10.1103/PhysRevB.71.104510}.

\bibitem[{\citenamefont{Stephanos et~al.}(2011)\citenamefont{Stephanos, Kopp,
  Mannhart, and Hirschfeld}}]{Hirschfeld11}
\bibinfo{author}{\bibfnamefont{C.}~\bibnamefont{Stephanos}},
  \bibinfo{author}{\bibfnamefont{T.}~\bibnamefont{Kopp}},
  \bibinfo{author}{\bibfnamefont{J.}~\bibnamefont{Mannhart}}, \bibnamefont{and}
  \bibinfo{author}{\bibfnamefont{P.~J.} \bibnamefont{Hirschfeld}},
  \bibinfo{journal}{Phys. Rev. B} \textbf{\bibinfo{volume}{84}},
  \bibinfo{pages}{100510} (\bibinfo{year}{2011}),
  \urlprefix\url{http://link.aps.org/doi/10.1103/PhysRevB.84.100510}.

\bibitem[{\citenamefont{Gozar et~al.}(2008)\citenamefont{Gozar, Logvenov,
  Kourkoutis, Bollinger, Giannuzzi, Muller, and Bozovic}}]{Gozar08}
\bibinfo{author}{\bibfnamefont{A.}~\bibnamefont{Gozar}},
  \bibinfo{author}{\bibfnamefont{G.}~\bibnamefont{Logvenov}},
  \bibinfo{author}{\bibfnamefont{L.~F.} \bibnamefont{Kourkoutis}},
  \bibinfo{author}{\bibfnamefont{A.~T.} \bibnamefont{Bollinger}},
  \bibinfo{author}{\bibfnamefont{L.~A.} \bibnamefont{Giannuzzi}},
  \bibinfo{author}{\bibfnamefont{D.~A.} \bibnamefont{Muller}},
  \bibnamefont{and} \bibinfo{author}{\bibfnamefont{I.}~\bibnamefont{Bozovic}},
  \bibinfo{journal}{Nature} \textbf{\bibinfo{volume}{455}},
  \bibinfo{pages}{782} (\bibinfo{year}{2008}),
  \urlprefix\url{http://dx.doi.org/10.1038/nature07293}.

\bibitem[{\citenamefont{{Pereiro} et~al.}(2011)\citenamefont{{Pereiro},
  {Petrovic}, {Panagopoulos}, and {Bo{\v z}ovi{\'c}}}}]{Bozovic11}
\bibinfo{author}{\bibfnamefont{J.}~\bibnamefont{{Pereiro}}},
  \bibinfo{author}{\bibfnamefont{A.}~\bibnamefont{{Petrovic}}},
  \bibinfo{author}{\bibfnamefont{C.}~\bibnamefont{{Panagopoulos}}},
  \bibnamefont{and} \bibinfo{author}{\bibfnamefont{I.}~\bibnamefont{{Bo{\v
  z}ovi{\'c}}}}, \bibinfo{journal}{Phys. Express} \textbf{\bibinfo{volume}{1}},
  \bibinfo{pages}{208} (\bibinfo{year}{2011}).

\bibitem[{\citenamefont{Wang et~al.}(2012)\citenamefont{Wang, Li, Zhang, Zhang,
  Zhang, Li, Ding, Ou, Deng, Chang et~al.}}]{WangCPL2012}
\bibinfo{author}{\bibfnamefont{Q.-Y.} \bibnamefont{Wang}},
  \bibinfo{author}{\bibfnamefont{Z.}~\bibnamefont{Li}},
  \bibinfo{author}{\bibfnamefont{W.-H.} \bibnamefont{Zhang}},
  \bibinfo{author}{\bibfnamefont{Z.-C.} \bibnamefont{Zhang}},
  \bibinfo{author}{\bibfnamefont{J.-S.} \bibnamefont{Zhang}},
  \bibinfo{author}{\bibfnamefont{W.}~\bibnamefont{Li}},
  \bibinfo{author}{\bibfnamefont{H.}~\bibnamefont{Ding}},
  \bibinfo{author}{\bibfnamefont{Y.-B.} \bibnamefont{Ou}},
  \bibinfo{author}{\bibfnamefont{P.}~\bibnamefont{Deng}},
  \bibinfo{author}{\bibfnamefont{K.}~\bibnamefont{Chang}},
  \bibnamefont{et~al.}, \bibinfo{journal}{Chinese Physics Letters}
  \textbf{\bibinfo{volume}{29}}, \bibinfo{pages}{037402}
  (\bibinfo{year}{2012}), \eprint{arXiv:1201.5694}.

\bibitem[{\citenamefont{Jensen and Mackintosh}(1991)}]{Jensen91}
\bibinfo{author}{\bibfnamefont{J.}~\bibnamefont{Jensen}} \bibnamefont{and}
  \bibinfo{author}{\bibfnamefont{A.~R.} \bibnamefont{Mackintosh}},
  \emph{\bibinfo{title}{Rare earth magnetism: structures and excitations}}
  (\bibinfo{publisher}{Oxford University Press}, \bibinfo{address}{New York},
  \bibinfo{year}{1991}).

\bibitem[{\citenamefont{Senthil et~al.}(2004)\citenamefont{Senthil, Vojta, and
  Sachdev}}]{Senthil04}
\bibinfo{author}{\bibfnamefont{T.}~\bibnamefont{Senthil}},
  \bibinfo{author}{\bibfnamefont{M.}~\bibnamefont{Vojta}}, \bibnamefont{and}
  \bibinfo{author}{\bibfnamefont{S.}~\bibnamefont{Sachdev}},
  \bibinfo{journal}{Phys. Rev. B} \textbf{\bibinfo{volume}{69}},
  \bibinfo{pages}{035111} (\bibinfo{year}{2004}),
  \urlprefix\url{http://link.aps.org/doi/10.1103/PhysRevB.69.035111}.

\bibitem[{\citenamefont{{Si}}(2006)}]{Si06}
\bibinfo{author}{\bibfnamefont{Q.}~\bibnamefont{{Si}}},
  \bibinfo{journal}{Physica B Condensed Matter} \textbf{\bibinfo{volume}{378}},
  \bibinfo{pages}{23} (\bibinfo{year}{2006}), \eprint{cond-mat/0601001}.

\bibitem[{\citenamefont{{Coleman} and {Nevidomskyy}}(2010)}]{Coleman10}
\bibinfo{author}{\bibfnamefont{P.}~\bibnamefont{{Coleman}}} \bibnamefont{and}
  \bibinfo{author}{\bibfnamefont{A.~H.} \bibnamefont{{Nevidomskyy}}},
  \bibinfo{journal}{Journal of Low Temperature Physics}
  \textbf{\bibinfo{volume}{161}}, \bibinfo{pages}{182} (\bibinfo{year}{2010}),
  \eprint{1007.1723}.

\bibitem[{\citenamefont{Fischer and Klein}(1975)}]{Fischer75}
\bibinfo{author}{\bibfnamefont{B.}~\bibnamefont{Fischer}} \bibnamefont{and}
  \bibinfo{author}{\bibfnamefont{M.~W.} \bibnamefont{Klein}},
  \bibinfo{journal}{Phys. Rev. B} \textbf{\bibinfo{volume}{11}},
  \bibinfo{pages}{2025} (\bibinfo{year}{1975}),
  \urlprefix\url{http://link.aps.org/doi/10.1103/PhysRevB.11.2025}.

\bibitem[{\citenamefont{Mahan}(2000)}]{Mahan}
\bibinfo{author}{\bibfnamefont{G.~D.} \bibnamefont{Mahan}},
  \emph{\bibinfo{title}{Many-particle physics}} (\bibinfo{publisher}{Springer},
  \bibinfo{year}{2000}).

\bibitem[{\citenamefont{Martin and Batista}(2008)}]{Martin08}
\bibinfo{author}{\bibfnamefont{I.}~\bibnamefont{Martin}} \bibnamefont{and}
  \bibinfo{author}{\bibfnamefont{C.~D.} \bibnamefont{Batista}},
  \bibinfo{journal}{Phys. Rev. Lett.} \textbf{\bibinfo{volume}{101}},
  \bibinfo{pages}{156402} (\bibinfo{year}{2008}),
  \urlprefix\url{http://link.aps.org/doi/10.1103/PhysRevLett.101.156402}.

\bibitem[{\citenamefont{Lee et~al.}(2013)\citenamefont{Lee, Paramekanti, and
  Kim}}]{Lee13}
\bibinfo{author}{\bibfnamefont{S.}~\bibnamefont{Lee}},
  \bibinfo{author}{\bibfnamefont{A.}~\bibnamefont{Paramekanti}},
  \bibnamefont{and} \bibinfo{author}{\bibfnamefont{Y.~B.} \bibnamefont{Kim}},
  \bibinfo{journal}{Phys. Rev. Lett.} \textbf{\bibinfo{volume}{111}},
  \bibinfo{pages}{196601} (\bibinfo{year}{2013}),
  \urlprefix\url{http://link.aps.org/doi/10.1103/PhysRevLett.111.196601}.

\bibitem[{\citenamefont{Flint and Senthil}(2013)}]{Flint13}
\bibinfo{author}{\bibfnamefont{R.}~\bibnamefont{Flint}} \bibnamefont{and}
  \bibinfo{author}{\bibfnamefont{T.}~\bibnamefont{Senthil}},
  \bibinfo{journal}{Phys. Rev. B} \textbf{\bibinfo{volume}{87}},
  \bibinfo{pages}{125147} (\bibinfo{year}{2013}),
  \urlprefix\url{http://link.aps.org/doi/10.1103/PhysRevB.87.125147}.

\bibitem[{\citenamefont{Bardeen et~al.}(1957)\citenamefont{Bardeen, Cooper, and
  Schrieffer}}]{BCS57}
\bibinfo{author}{\bibfnamefont{J.}~\bibnamefont{Bardeen}},
  \bibinfo{author}{\bibfnamefont{L.~N.} \bibnamefont{Cooper}},
  \bibnamefont{and} \bibinfo{author}{\bibfnamefont{J.~R.}
  \bibnamefont{Schrieffer}}, \bibinfo{journal}{Phys. Rev.}
  \textbf{\bibinfo{volume}{108}}, \bibinfo{pages}{1175} (\bibinfo{year}{1957}),
  \urlprefix\url{http://link.aps.org/doi/10.1103/PhysRev.108.1175}.

\bibitem[{\citenamefont{Helton et~al.}(2007)\citenamefont{Helton, Matan,
  Shores, Nytko, Bartlett, Yoshida, Takano, Suslov, Qiu, Chung
  et~al.}}]{Helton07}
\bibinfo{author}{\bibfnamefont{J.~S.} \bibnamefont{Helton}},
  \bibinfo{author}{\bibfnamefont{K.}~\bibnamefont{Matan}},
  \bibinfo{author}{\bibfnamefont{M.~P.} \bibnamefont{Shores}},
  \bibinfo{author}{\bibfnamefont{E.~A.} \bibnamefont{Nytko}},
  \bibinfo{author}{\bibfnamefont{B.~M.} \bibnamefont{Bartlett}},
  \bibinfo{author}{\bibfnamefont{Y.}~\bibnamefont{Yoshida}},
  \bibinfo{author}{\bibfnamefont{Y.}~\bibnamefont{Takano}},
  \bibinfo{author}{\bibfnamefont{A.}~\bibnamefont{Suslov}},
  \bibinfo{author}{\bibfnamefont{Y.}~\bibnamefont{Qiu}},
  \bibinfo{author}{\bibfnamefont{J.-H.} \bibnamefont{Chung}},
  \bibnamefont{et~al.}, \bibinfo{journal}{Phys. Rev. Lett.}
  \textbf{\bibinfo{volume}{98}}, \bibinfo{pages}{107204}
  (\bibinfo{year}{2007}),
  \urlprefix\url{http://link.aps.org/doi/10.1103/PhysRevLett.98.107204}.

\bibitem[{\citenamefont{Han et~al.}(2012)\citenamefont{Han, Helton, andDaniel
  G.~Nocera, Rodriguez-Rivera, Broholm, and Lee}}]{Han12}
\bibinfo{author}{\bibfnamefont{T.-H.} \bibnamefont{Han}},
  \bibinfo{author}{\bibfnamefont{J.~S.} \bibnamefont{Helton}},
  \bibinfo{author}{\bibfnamefont{S.~C.} \bibnamefont{andDaniel G.~Nocera}},
  \bibinfo{author}{\bibfnamefont{J.~A.} \bibnamefont{Rodriguez-Rivera}},
  \bibinfo{author}{\bibfnamefont{C.}~\bibnamefont{Broholm}}, \bibnamefont{and}
  \bibinfo{author}{\bibfnamefont{Y.~S.} \bibnamefont{Lee}},
  \bibinfo{journal}{Nature} \textbf{\bibinfo{volume}{492}},
  \bibinfo{pages}{406} (\bibinfo{year}{2012}).

\bibitem[{\citenamefont{Clark et~al.}(2013)\citenamefont{Clark, Orain, Bert,
  De~Vries, Aidoudi, Morris, Lightfoot, Lord, Telling, Bonville
  et~al.}}]{Clark13}
\bibinfo{author}{\bibfnamefont{L.}~\bibnamefont{Clark}},
  \bibinfo{author}{\bibfnamefont{J.~C.} \bibnamefont{Orain}},
  \bibinfo{author}{\bibfnamefont{F.}~\bibnamefont{Bert}},
  \bibinfo{author}{\bibfnamefont{M.~A.} \bibnamefont{De~Vries}},
  \bibinfo{author}{\bibfnamefont{F.~H.} \bibnamefont{Aidoudi}},
  \bibinfo{author}{\bibfnamefont{R.~E.} \bibnamefont{Morris}},
  \bibinfo{author}{\bibfnamefont{P.}~\bibnamefont{Lightfoot}},
  \bibinfo{author}{\bibfnamefont{J.~S.} \bibnamefont{Lord}},
  \bibinfo{author}{\bibfnamefont{M.~T.~F.} \bibnamefont{Telling}},
  \bibinfo{author}{\bibfnamefont{P.}~\bibnamefont{Bonville}},
  \bibnamefont{et~al.}, \bibinfo{journal}{Phys. Rev. Lett.}
  \textbf{\bibinfo{volume}{110}}, \bibinfo{pages}{207208}
  (\bibinfo{year}{2013}),
  \urlprefix\url{http://link.aps.org/doi/10.1103/PhysRevLett.110.207208}.

\bibitem[{\citenamefont{Nath et~al.}(2008)\citenamefont{Nath, Tsirlin, Kaul,
  Baenitz, B\"uttgen, Geibel, and Rosner}}]{Nath08}
\bibinfo{author}{\bibfnamefont{R.}~\bibnamefont{Nath}},
  \bibinfo{author}{\bibfnamefont{A.~A.} \bibnamefont{Tsirlin}},
  \bibinfo{author}{\bibfnamefont{E.~E.} \bibnamefont{Kaul}},
  \bibinfo{author}{\bibfnamefont{M.}~\bibnamefont{Baenitz}},
  \bibinfo{author}{\bibfnamefont{N.}~\bibnamefont{B\"uttgen}},
  \bibinfo{author}{\bibfnamefont{C.}~\bibnamefont{Geibel}}, \bibnamefont{and}
  \bibinfo{author}{\bibfnamefont{H.}~\bibnamefont{Rosner}},
  \bibinfo{journal}{Phys. Rev. B} \textbf{\bibinfo{volume}{78}},
  \bibinfo{pages}{024418} (\bibinfo{year}{2008}),
  \urlprefix\url{http://link.aps.org/doi/10.1103/PhysRevB.78.024418}.

\bibitem[{\citenamefont{Tsirlin and Rosner}(2009)}]{Tsirlin09}
\bibinfo{author}{\bibfnamefont{A.~A.} \bibnamefont{Tsirlin}} \bibnamefont{and}
  \bibinfo{author}{\bibfnamefont{H.}~\bibnamefont{Rosner}},
  \bibinfo{journal}{Phys. Rev. B} \textbf{\bibinfo{volume}{79}},
  \bibinfo{pages}{214417} (\bibinfo{year}{2009}),
  \urlprefix\url{http://link.aps.org/doi/10.1103/PhysRevB.79.214417}.

\bibitem[{\citenamefont{Okamoto et~al.}(2007)\citenamefont{Okamoto, Nohara,
  Aruga-Katori, and Takagi}}]{Okamoto07}
\bibinfo{author}{\bibfnamefont{Y.}~\bibnamefont{Okamoto}},
  \bibinfo{author}{\bibfnamefont{M.}~\bibnamefont{Nohara}},
  \bibinfo{author}{\bibfnamefont{H.}~\bibnamefont{Aruga-Katori}},
  \bibnamefont{and} \bibinfo{author}{\bibfnamefont{H.}~\bibnamefont{Takagi}},
  \bibinfo{journal}{Phys. Rev. Lett.} \textbf{\bibinfo{volume}{99}},
  \bibinfo{pages}{137207} (\bibinfo{year}{2007}),
  \urlprefix\url{http://link.aps.org/doi/10.1103/PhysRevLett.99.137207}.

\bibitem[{\citenamefont{Shockley et~al.}(2015)\citenamefont{Shockley, Bert,
  Orain, Okamoto, and Mendels}}]{Shockley15}
\bibinfo{author}{\bibfnamefont{A.~C.} \bibnamefont{Shockley}},
  \bibinfo{author}{\bibfnamefont{F.}~\bibnamefont{Bert}},
  \bibinfo{author}{\bibfnamefont{J.-C.} \bibnamefont{Orain}},
  \bibinfo{author}{\bibfnamefont{Y.}~\bibnamefont{Okamoto}}, \bibnamefont{and}
  \bibinfo{author}{\bibfnamefont{P.}~\bibnamefont{Mendels}},
  \bibinfo{journal}{Phys. Rev. Lett.} \textbf{\bibinfo{volume}{115}},
  \bibinfo{pages}{047201} (\bibinfo{year}{2015}),
  \urlprefix\url{http://link.aps.org/doi/10.1103/PhysRevLett.115.047201}.

\bibitem[{\citenamefont{Norman and Micklitz}(2010)}]{Norman10}
\bibinfo{author}{\bibfnamefont{M.~R.} \bibnamefont{Norman}} \bibnamefont{and}
  \bibinfo{author}{\bibfnamefont{T.}~\bibnamefont{Micklitz}},
  \bibinfo{journal}{Phys. Rev. B} \textbf{\bibinfo{volume}{81}},
  \bibinfo{pages}{024428} (\bibinfo{year}{2010}),
  \urlprefix\url{http://link.aps.org/doi/10.1103/PhysRevB.81.024428}.

\bibitem[{\citenamefont{Micklitz and Norman}(2010)}]{Micklitz10}
\bibinfo{author}{\bibfnamefont{T.}~\bibnamefont{Micklitz}} \bibnamefont{and}
  \bibinfo{author}{\bibfnamefont{M.~R.} \bibnamefont{Norman}},
  \bibinfo{journal}{Phys. Rev. B} \textbf{\bibinfo{volume}{81}},
  \bibinfo{pages}{174417} (\bibinfo{year}{2010}),
  \urlprefix\url{http://link.aps.org/doi/10.1103/PhysRevB.81.174417}.

\bibitem[{\citenamefont{Nakatsuji et~al.}(2005)\citenamefont{Nakatsuji, Nambu,
  Tonomura, Sakai, Jonas, Broholm, Tsunetsugu, Qiu, and Maeno}}]{Nakatsuji05}
\bibinfo{author}{\bibfnamefont{S.}~\bibnamefont{Nakatsuji}},
  \bibinfo{author}{\bibfnamefont{Y.}~\bibnamefont{Nambu}},
  \bibinfo{author}{\bibfnamefont{H.}~\bibnamefont{Tonomura}},
  \bibinfo{author}{\bibfnamefont{O.}~\bibnamefont{Sakai}},
  \bibinfo{author}{\bibfnamefont{S.}~\bibnamefont{Jonas}},
  \bibinfo{author}{\bibfnamefont{C.}~\bibnamefont{Broholm}},
  \bibinfo{author}{\bibfnamefont{H.}~\bibnamefont{Tsunetsugu}},
  \bibinfo{author}{\bibfnamefont{Y.}~\bibnamefont{Qiu}}, \bibnamefont{and}
  \bibinfo{author}{\bibfnamefont{Y.}~\bibnamefont{Maeno}},
  \bibinfo{journal}{Science} \textbf{\bibinfo{volume}{309}},
  \bibinfo{pages}{1697} (\bibinfo{year}{2005}),
  \eprint{http://www.sciencemag.org/content/309/5741/1697.full.pdf},
  \urlprefix\url{http://www.sciencemag.org/content/309/5741/1697.abstract}.

\bibitem[{\citenamefont{MacLaughlin et~al.}(2010)\citenamefont{MacLaughlin,
  Nambu, Ohta, Machida, Nakatsuji, and Bernal}}]{MacLaughlin10}
\bibinfo{author}{\bibfnamefont{D.~E.} \bibnamefont{MacLaughlin}},
  \bibinfo{author}{\bibfnamefont{Y.}~\bibnamefont{Nambu}},
  \bibinfo{author}{\bibfnamefont{Y.}~\bibnamefont{Ohta}},
  \bibinfo{author}{\bibfnamefont{Y.}~\bibnamefont{Machida}},
  \bibinfo{author}{\bibfnamefont{S.}~\bibnamefont{Nakatsuji}},
  \bibnamefont{and} \bibinfo{author}{\bibfnamefont{O.~O.}
  \bibnamefont{Bernal}}, \bibinfo{journal}{Journal of Physics: Conference
  Series} \textbf{\bibinfo{volume}{225}}, \bibinfo{pages}{012031}
  (\bibinfo{year}{2010}),
  \urlprefix\url{http://stacks.iop.org/1742-6596/225/i=1/a=012031}.

\bibitem[{\citenamefont{Stock et~al.}(2010)\citenamefont{Stock, Jonas, Broholm,
  Nakatsuji, Nambu, Onuma, Maeno, and Chung}}]{Stock10}
\bibinfo{author}{\bibfnamefont{C.}~\bibnamefont{Stock}},
  \bibinfo{author}{\bibfnamefont{S.}~\bibnamefont{Jonas}},
  \bibinfo{author}{\bibfnamefont{C.}~\bibnamefont{Broholm}},
  \bibinfo{author}{\bibfnamefont{S.}~\bibnamefont{Nakatsuji}},
  \bibinfo{author}{\bibfnamefont{Y.}~\bibnamefont{Nambu}},
  \bibinfo{author}{\bibfnamefont{K.}~\bibnamefont{Onuma}},
  \bibinfo{author}{\bibfnamefont{Y.}~\bibnamefont{Maeno}}, \bibnamefont{and}
  \bibinfo{author}{\bibfnamefont{J.-H.} \bibnamefont{Chung}},
  \bibinfo{journal}{Phys. Rev. Lett.} \textbf{\bibinfo{volume}{105}},
  \bibinfo{pages}{037402} (\bibinfo{year}{2010}),
  \urlprefix\url{http://link.aps.org/doi/10.1103/PhysRevLett.105.037402}.

\bibitem[{\citenamefont{Shimizu et~al.}(2003)\citenamefont{Shimizu, Miyagawa,
  Kanoda, Maesato, and Saito}}]{Shimizu03}
\bibinfo{author}{\bibfnamefont{Y.}~\bibnamefont{Shimizu}},
  \bibinfo{author}{\bibfnamefont{K.}~\bibnamefont{Miyagawa}},
  \bibinfo{author}{\bibfnamefont{K.}~\bibnamefont{Kanoda}},
  \bibinfo{author}{\bibfnamefont{M.}~\bibnamefont{Maesato}}, \bibnamefont{and}
  \bibinfo{author}{\bibfnamefont{G.}~\bibnamefont{Saito}},
  \bibinfo{journal}{Phys. Rev. Lett.} \textbf{\bibinfo{volume}{91}},
  \bibinfo{pages}{107001} (\bibinfo{year}{2003}),
  \urlprefix\url{http://link.aps.org/doi/10.1103/PhysRevLett.91.107001}.

\bibitem[{\citenamefont{Bramwell and Gingras}(2001)}]{Bramwell01}
\bibinfo{author}{\bibfnamefont{S.~T.} \bibnamefont{Bramwell}} \bibnamefont{and}
  \bibinfo{author}{\bibfnamefont{M.~J.~P.} \bibnamefont{Gingras}},
  \bibinfo{journal}{Science} \textbf{\bibinfo{volume}{294}},
  \bibinfo{pages}{1495} (\bibinfo{year}{2001}),
  \eprint{http://www.sciencemag.org/content/294/5546/1495.full.pdf},
  \urlprefix\url{http://www.sciencemag.org/content/294/5546/1495.abstract}.

\bibitem[{\citenamefont{Garanin and Canals}(1999)}]{Garanin99}
\bibinfo{author}{\bibfnamefont{D.~A.} \bibnamefont{Garanin}} \bibnamefont{and}
  \bibinfo{author}{\bibfnamefont{B.}~\bibnamefont{Canals}},
  \bibinfo{journal}{Phys. Rev. B} \textbf{\bibinfo{volume}{59}},
  \bibinfo{pages}{443} (\bibinfo{year}{1999}),
  \urlprefix\url{http://link.aps.org/doi/10.1103/PhysRevB.59.443}.

\bibitem[{\citenamefont{Canals and Garanin}(2001)}]{Garanin01}
\bibinfo{author}{\bibfnamefont{B.}~\bibnamefont{Canals}} \bibnamefont{and}
  \bibinfo{author}{\bibfnamefont{D.~A.} \bibnamefont{Garanin}},
  \bibinfo{journal}{Canadian Journal of Physics} \textbf{\bibinfo{volume}{79}},
  \bibinfo{pages}{1323} (\bibinfo{year}{2001}),
  \urlprefix\url{http://www.nrcresearchpress.com/doi/abs/10.1139/p01-101}.

\bibitem[{\citenamefont{Fennell et~al.}(2009)\citenamefont{Fennell, Deen,
  Wildes, Schmalzl, Prabhakaran, Boothroyd, Aldus, McMorrow, and
  Bramwell}}]{Fennell09}
\bibinfo{author}{\bibfnamefont{T.}~\bibnamefont{Fennell}},
  \bibinfo{author}{\bibfnamefont{P.~P.} \bibnamefont{Deen}},
  \bibinfo{author}{\bibfnamefont{A.~R.} \bibnamefont{Wildes}},
  \bibinfo{author}{\bibfnamefont{K.}~\bibnamefont{Schmalzl}},
  \bibinfo{author}{\bibfnamefont{D.}~\bibnamefont{Prabhakaran}},
  \bibinfo{author}{\bibfnamefont{A.~T.} \bibnamefont{Boothroyd}},
  \bibinfo{author}{\bibfnamefont{R.~J.} \bibnamefont{Aldus}},
  \bibinfo{author}{\bibfnamefont{D.~F.} \bibnamefont{McMorrow}},
  \bibnamefont{and} \bibinfo{author}{\bibfnamefont{S.~T.}
  \bibnamefont{Bramwell}}, \bibinfo{journal}{Science}
  \textbf{\bibinfo{volume}{326}}, \bibinfo{pages}{415} (\bibinfo{year}{2009}),
  \eprint{http://www.sciencemag.org/content/326/5951/415.full.pdf},
  \urlprefix\url{http://www.sciencemag.org/content/326/5951/415.abstract}.

\bibitem[{\citenamefont{Kresse and Furthm\"uller}(1996)}]{Kresse1996}
\bibinfo{author}{\bibfnamefont{G.}~\bibnamefont{Kresse}} \bibnamefont{and}
  \bibinfo{author}{\bibfnamefont{J.}~\bibnamefont{Furthm\"uller}},
  \bibinfo{journal}{Phys. Rev. B} \textbf{\bibinfo{volume}{54}},
  \bibinfo{pages}{11169} (\bibinfo{year}{1996}),
  \urlprefix\url{http://link.aps.org/doi/10.1103/PhysRevB.54.11169}.

\bibitem[{\citenamefont{Kresse and Joubert}(1999)}]{Kresse1999}
\bibinfo{author}{\bibfnamefont{G.}~\bibnamefont{Kresse}} \bibnamefont{and}
  \bibinfo{author}{\bibfnamefont{D.}~\bibnamefont{Joubert}},
  \bibinfo{journal}{Phys. Rev. B} \textbf{\bibinfo{volume}{59}},
  \bibinfo{pages}{1758} (\bibinfo{year}{1999}),
  \urlprefix\url{http://link.aps.org/doi/10.1103/PhysRevB.59.1758}.

\bibitem[{\citenamefont{Ozaki}({2011})}]{openmx}
\bibinfo{author}{\bibfnamefont{T.}~\bibnamefont{Ozaki}},
  \bibinfo{journal}{{http://www.openmx-square.org}}  (\bibinfo{year}{{2011}}).

\bibitem[{\citenamefont{Chen and Hermele}(2012)}]{Chen12}
\bibinfo{author}{\bibfnamefont{G.}~\bibnamefont{Chen}} \bibnamefont{and}
  \bibinfo{author}{\bibfnamefont{M.}~\bibnamefont{Hermele}},
  \bibinfo{journal}{Phys. Rev. B} \textbf{\bibinfo{volume}{86}},
  \bibinfo{pages}{235129} (\bibinfo{year}{2012}),
  \urlprefix\url{http://link.aps.org/doi/10.1103/PhysRevB.86.235129}.

\bibitem[{\citenamefont{Cox and Zawadowski}(1998)}]{Cox98}
\bibinfo{author}{\bibfnamefont{D.~L.} \bibnamefont{Cox}} \bibnamefont{and}
  \bibinfo{author}{\bibfnamefont{A.}~\bibnamefont{Zawadowski}},
  \bibinfo{journal}{Advances in Physics} \textbf{\bibinfo{volume}{47}},
  \bibinfo{pages}{599} (\bibinfo{year}{1998}).

\bibitem[{\citenamefont{Onoda and Tanaka}(2011)}]{Onoda11}
\bibinfo{author}{\bibfnamefont{S.}~\bibnamefont{Onoda}} \bibnamefont{and}
  \bibinfo{author}{\bibfnamefont{Y.}~\bibnamefont{Tanaka}},
  \bibinfo{journal}{Phys. Rev. B} \textbf{\bibinfo{volume}{83}},
  \bibinfo{pages}{094411} (\bibinfo{year}{2011}),
  \urlprefix\url{http://link.aps.org/doi/10.1103/PhysRevB.83.094411}.

\bibitem[{\citenamefont{Norman}(1995)}]{Norman1995}
\bibinfo{author}{\bibfnamefont{M.~R.} \bibnamefont{Norman}},
  \bibinfo{journal}{Physical Review B} \textbf{\bibinfo{volume}{52}},
  \bibinfo{pages}{1421} (\bibinfo{year}{1995}), ISSN \bibinfo{issn}{01631829},
  \eprint{9501111}.

\bibitem[{\citenamefont{Zhou et~al.}(2008)\citenamefont{Zhou, Wiebe, Janik,
  Balicas, Yo, Qiu, Copley, and Gardner}}]{Zhou08}
\bibinfo{author}{\bibfnamefont{H.~D.} \bibnamefont{Zhou}},
  \bibinfo{author}{\bibfnamefont{C.~R.} \bibnamefont{Wiebe}},
  \bibinfo{author}{\bibfnamefont{J.~A.} \bibnamefont{Janik}},
  \bibinfo{author}{\bibfnamefont{L.}~\bibnamefont{Balicas}},
  \bibinfo{author}{\bibfnamefont{Y.~J.} \bibnamefont{Yo}},
  \bibinfo{author}{\bibfnamefont{Y.}~\bibnamefont{Qiu}},
  \bibinfo{author}{\bibfnamefont{J.~R.~D.} \bibnamefont{Copley}},
  \bibnamefont{and} \bibinfo{author}{\bibfnamefont{J.~S.}
  \bibnamefont{Gardner}}, \bibinfo{journal}{Phys. Rev. Lett.}
  \textbf{\bibinfo{volume}{101}}, \bibinfo{pages}{227204}
  (\bibinfo{year}{2008}),
  \urlprefix\url{http://link.aps.org/doi/10.1103/PhysRevLett.101.227204}.

\bibitem[{\citenamefont{Gschneidner et~al.}({2007})\citenamefont{Gschneidner,
  Bunzli, and Pecharsky}}]{handbook}
\bibinfo{author}{\bibfnamefont{K.~A.} \bibnamefont{Gschneidner}},
  \bibinfo{author}{\bibfnamefont{J.-C.} \bibnamefont{Bunzli}},
  \bibnamefont{and}
  \bibinfo{author}{\bibfnamefont{V.}~\bibnamefont{Pecharsky}},
  \emph{\bibinfo{title}{{Handbook on the Physics and Chemistry of Rare Earth:
  Optical spectroscopy}}} (\bibinfo{publisher}{{Elsevier}},
  \bibinfo{year}{{2007}}).

\bibitem[{\citenamefont{Nakatsuji et~al.}(2006)\citenamefont{Nakatsuji,
  Machida, Maeno, Tayama, Sakakibara, Duijn, Balicas, Millican, Macaluso, and
  Chan}}]{Nakatsuji06}
\bibinfo{author}{\bibfnamefont{S.}~\bibnamefont{Nakatsuji}},
  \bibinfo{author}{\bibfnamefont{Y.}~\bibnamefont{Machida}},
  \bibinfo{author}{\bibfnamefont{Y.}~\bibnamefont{Maeno}},
  \bibinfo{author}{\bibfnamefont{T.}~\bibnamefont{Tayama}},
  \bibinfo{author}{\bibfnamefont{T.}~\bibnamefont{Sakakibara}},
  \bibinfo{author}{\bibfnamefont{J.~v.} \bibnamefont{Duijn}},
  \bibinfo{author}{\bibfnamefont{L.}~\bibnamefont{Balicas}},
  \bibinfo{author}{\bibfnamefont{J.~N.} \bibnamefont{Millican}},
  \bibinfo{author}{\bibfnamefont{R.~T.} \bibnamefont{Macaluso}},
  \bibnamefont{and} \bibinfo{author}{\bibfnamefont{J.~Y.} \bibnamefont{Chan}},
  \bibinfo{journal}{Phys. Rev. Lett.} \textbf{\bibinfo{volume}{96}},
  \bibinfo{pages}{087204} (\bibinfo{year}{2006}),
  \urlprefix\url{http://link.aps.org/doi/10.1103/PhysRevLett.96.087204}.

\bibitem[{\citenamefont{Sakurai}(1994)}]{Sakurai}
\bibinfo{author}{\bibfnamefont{J.~J.} \bibnamefont{Sakurai}},
  \emph{\bibinfo{title}{Modern Quantum Mechanics}}
  (\bibinfo{publisher}{Addison-Wesley}, \bibinfo{year}{1994}).

\bibitem[{\citenamefont{Pethick and Smith}(2008)}]{Pethick}
\bibinfo{author}{\bibfnamefont{C.~J.} \bibnamefont{Pethick}} \bibnamefont{and}
  \bibinfo{author}{\bibfnamefont{H.}~\bibnamefont{Smith}},
  \emph{\bibinfo{title}{Bose-Einstein Condensation in Dilute Gases}}
  (\bibinfo{publisher}{{Cambridge University Press}}, \bibinfo{year}{2008}).

\bibitem[{\citenamefont{Leggett}(1975)}]{Leggett75}
\bibinfo{author}{\bibfnamefont{A.~J.} \bibnamefont{Leggett}},
  \bibinfo{journal}{Rev. Mod. Phys.} \textbf{\bibinfo{volume}{47}},
  \bibinfo{pages}{331} (\bibinfo{year}{1975}),
  \urlprefix\url{http://link.aps.org/doi/10.1103/RevModPhys.47.331}.

\end{thebibliography}

\end{document}